\documentclass[journal,twoclumn]{IEEEtran}
\usepackage{multicol}
\usepackage{etoolbox}
\makeatletter
\patchcmd{\@makecaption}
  {\scshape}
  {}
  {}
  {}
\makeatletter
\patchcmd{\@makecaption}
  {\\}
  {.\ }
  {}
  {}
\makeatother

\usepackage{amsfonts}
\usepackage{amsmath}
\usepackage{mathrsfs}
\usepackage{mathtools}
\usepackage{amsfonts}
\usepackage{amssymb}
\usepackage{graphicx}
\usepackage{epsfig}
\usepackage{psfrag}{}
\usepackage{array}
\usepackage{cases}
\usepackage{eufrak}
\usepackage{cite,graphicx,amssymb,color}
\usepackage{algorithmic}
\usepackage{algorithm}
\usepackage{setspace}
\usepackage{subfigure}
\usepackage{bm}
\usepackage{multirow}
\usepackage{threeparttable}
\usepackage{array}
\usepackage{makecell}
\usepackage{soul}
\usepackage{float}
\usepackage{booktabs}
\newfloat{figtab}{htb}{fgtb}
\makeatletter
  \newcommand\figcaption{\def\@captype{figure}\caption}
  \newcommand\tabcaption{\def\@captype{table}\caption}
\makeatother

\newcommand{\tabincell}[2]{\begin{tabular}{@{}#1@{}}#2\end{tabular}}

\newtheorem{Thm}{Theorem}

\newtheorem{Prob}{Problem}
\newtheorem{Rem}{Remark}

\newtheorem{claim}{Claim}
\newtheorem{Proof}{Proof}

\IEEEoverridecommandlockouts

\begin{document}
\title{Adaptive Streaming of 360 Videos with Perfect, Imperfect, and Unknown FoV Viewing Probabilities in Wireless Networks}
\author{\IEEEauthorblockN{Lingzhi Zhao, Ying Cui,~\IEEEmembership{Member,~IEEE}, Zhi Liu,~\IEEEmembership{Senior Member,~IEEE}, Yunfei Zhang, and Sheng Yang,~\IEEEmembership{Member,~IEEE}}
\thanks{Manuscript received August 18, 2020; revised April 12, 2021; accepted June 21, 2021.
 The work of Y. Cui was supported in part by STCSM 18DZ2270700 and in part by the Natural Science Foundation of Shanghai under Grant 20ZR1425300.
 The work of Z. Liu was supported in part by JSPS KAKENHI under Grants 19H04092, 20H04174, and in part by ROIS NII Open Collaborative Research 2020 (20FA02), 2021 (21FA02).
This paper was presented in part at IEEE GLOBECOM 2020 \cite{GLOBECOM20}.
The associate editor coordinating the review of this paper and approving it for publication was Francesco De Natale. \textit{(Corresponding author: Ying Cui.)}

L. Zhao and Y. Cui are with the Department of Electronic Engineering, Shanghai Jiao Tong University, Shanghai 200240, China (e-mail: cuiying@sjtu.edu.cn).

Z. Liu is with
Graduate School of Informatics and Engineering, the University of Electro-Communications, Tokyo 182-8585, Japan.

Y. Zhang is with Tencent Technology Co., Ltd, Shenzhen, 518054, China.

S. Yang is with the Laboratory of Signals and Systems, CentraleSup\'elec-CNRS-Universit\'e Paris-Sud, 91192 Gif-sur-Yvette, France.
}}

\maketitle


\begin{abstract}
This paper investigates adaptive streaming of one or multiple tiled 360 videos from a multi-antenna base station (BS) to one or multiple single-antenna users, respectively, in a multi-carrier wireless system. We aim to maximize the video quality while keeping rebuffering time small via encoding rate adaptation at each group of pictures (GOP) and transmission adaptation at each (transmission) slot. To capture the impact of field-of-view (FoV) prediction, we consider three cases of FoV viewing probability distributions, i.e., perfect, imperfect, and unknown FoV viewing probability distributions, and use the average total utility, worst average total utility, and worst total utility as the respective performance metrics. In the single-user scenario, we optimize the encoding rates of the tiles, encoding rates of the FoVs, and transmission beamforming vectors for all subcarriers to maximize the total utility in each case. In the multi-user scenario, we adopt rate splitting with successive decoding and optimize the encoding rates of the tiles, encoding rates of the FoVs, rates of the common and private messages, and transmission beamforming vectors for all subcarriers to maximize the total utility in each case. Then, we separate the challenging optimization problem into multiple tractable problems in each scenario. In the single-user scenario, we obtain a globally optimal solution of each problem using transformation techniques and the Karush-Kuhn-Tucker (KKT) conditions. In the multi-user scenario, we obtain a KKT point of each problem using the concave-convex procedure (CCCP).
Finally, numerical results demonstrate that the proposed solutions achieve notable gains over existing schemes in all three cases. To the best of our knowledge, this is the first work revealing the impact of FoV prediction on the performance of adaptive streaming of tiled 360 videos.
\end{abstract}



\section{Introduction}
Virtual reality (VR) techniques can provide a better quality of experience (QoE) in interactive applications and have vast applications in entertainment, education, medicine, etc. It is predicted that the VR market will reach 87.97 billion USD by 2025 \cite{web}. 
A VR spherical video is generated by capturing a scene of interest in every direction simultaneously using omnidirectional cameras. A user can freely watch the scene of interest in any viewing direction at any time, hence enjoying an immersive viewing experience. A 360 video is generated by projecting a spherical video onto a rectangle\cite{IEEEProc19}. A 360 video is of a much larger size than a traditional video. At any moment, a user watching a 360 video is interested in only one field-of-view (FoV), the center of each is referred to as viewpoint.

Viewpoint or FoV prediction based on past viewpoint sequences and video features, e.g., saliency, has been widely studied \cite{ACM17,ATC16,SECON17,JSTSP20,ACM/MM18,CVPR18,TMM20}. In particular, some works \cite{ACM17,ATC16,SECON17,JSTSP20,CVPR18} predict the FoV that is most likely to be watched, and some works\cite{ACM/MM18,TMM20} predict a set of FoVs that may be viewed and the viewing probability distribution on this set. The tiling technique, which divides a 360 video into rectangular segments, referred to as tiles, enables flexible transmission of FoVs that are most likely to be viewed. It can reduce the communication resource while maintaining the QoE to a certain extent. Pre-encoding a tile into multiple representations with different quality levels allows quality adaptation according to users’ channel conditions. Therefore, adaptive streaming of tiled 360 videos based on FoV prediction results has received increasing attention.

In \cite{ACM17,TSP18,ICASSP19,BLTJ12,VR/AR18,WCL19,WCL18,TMM20L,JSTSP20}, the authors study adaptive streaming of a tiled 360 video to a single user\cite{ACM17,TSP18,ICASSP19,BLTJ12,VR/AR18} or multiple users\cite{WCL19,WCL18,TMM20L,JSTSP20} in wireless networks. In the multi-user scenario, our previous works\cite{WCL19,WCL18,TMM20L} consider the overlapping of the FoVs of different users and focus on exploiting natural\cite{WCL19,WCL18,TMM20L}, relative smoothness-enabled\cite{TMM20L}, and transcoding-enabled\cite{WCL18,TMM20L} multicast opportunities to improve transmission efficiency, whereas \cite{JSTSP20} does not exploit potential multicast opportunities.
In \cite{HUANG20191,IS19,TIP20}, the authors study adaptive streaming of multiple tiled 360 videos to multiple users, respectively, in wireless networks. Specifically, in \cite{WCL19,WCL18,ACM17,TSP18,ICASSP19,BLTJ12,VR/AR18,TIP20,TMM20L,JSTSP20,HUANG20191,IS19}, the authors optimize the quality level selection and communication resource allocation to maximize the total utility\cite{ICASSP19,VR/AR18,HUANG20191,IS19,TIP20}, minimize the total distortion\cite{ACM17,JSTSP20}, minimize the total transmission power \cite{WCL18,WCL19,TMM20L}, or minimize the bandwidth consumption\cite{BLTJ12,TSP18}. The obtained solutions in \cite{ACM17,TSP18,ICASSP19,BLTJ12,VR/AR18,HUANG20191,TIP20} are heuristic, the obtained solutions in \cite{TMM20L,JSTSP20,IS19} are locally optimal, and the obtained solutions in \cite{WCL19,WCL18} are globally optimal.

The existing works \cite{WCL19,WCL18,ACM17,TSP18,ICASSP19,BLTJ12,VR/AR18,TIP20,TMM20L,JSTSP20,HUANG20191,IS19} mainly consider two FoV transmission methods. In \cite{WCL19,WCL18,TMM20L,ACM17,TSP18,VR/AR18,TIP20,IS19,JSTSP20}, the authors transmit the set of tiles that cover the FoV which is most likely to be watched at a certain encoding rate and a safe margin at the same\cite{WCL19,WCL18,TMM20L} or a lower encoding rate\cite{ACM17,TSP18,VR/AR18,TIP20,IS19,JSTSP20,BLTJ12}. In \cite{ICASSP19,HUANG20191}, the authors transmit the set of tiles that cover all FoVs that may be viewed at different encoding rates determined according to the viewing probability distribution over these FoVs. 
In \cite{WCL19,WCL18,ACM17,TSP18,ICASSP19,BLTJ12,VR/AR18,TIP20,TMM20L,JSTSP20,HUANG20191,IS19}, the quality levels of adjacent tiles viewed by a user may vary significantly, leading to a poor viewing experience. To address such issue, in\cite{ACM17,TMM20L,JSTSP20,HUANG20191}, quality smoothness requirements are incorporated in the objective functions\cite{ACM17,JSTSP20,HUANG20191} or the constraints\cite{ICASSP19,TMM20L}.

There are three main limitations in the existing works on adaptive streaming of tiled 360 videos in wireless networks. Firstly, most existing works \cite{WCL19,WCL18,ACM17,TSP18,ICASSP19,BLTJ12,VR/AR18,TIP20,TMM20L,JSTSP20,HUANG20191,IS19} rely on the assumption of perfect FoV prediction. It is unknown how FoV prediction errors influence the performance of adaptive streaming of tiled 360 videos. Secondly, the transmission designs in \cite{WCL19,WCL18,TMM20L,JSTSP20,HUANG20191,IS19,TIP20} are based on orthogonal multiple access schemes which are less spectrum efficient. It is not clear how advanced nonorthogonal transmission schemes can improve the performance of adaptive wireless streaming of tiled 360 videos. Thirdly, \cite{WCL19,ACM17,TSP18,ICASSP19,BLTJ12,VR/AR18,TIP20,TMM20L,JSTSP20,HUANG20191,IS19} all consider single-antenna servers, which cannot exploit spatial degrees of freedom and hence cannot provide satisfactory performance for adaptive wireless streaming of tiled 360 videos. It is interesting to know how current multi-antenna base stations (BS) can improve performance.

In this paper, we would like to address the above limitations and questions. We investigate adaptive streaming of one or multiple tiled 360 videos from a multi-antenna BS to one or multiple single-antenna users, respectively, in a multi-carrier wireless system. The goal is to maximize the video quality while keeping the rebuffering time small. Our main contributions are summarized below.\footnote{This paper extends the results in the conference version \cite{GLOBECOM20}, which only considers a simpler version of the utility maximization in the multi-user scenario.}

\begin{itemize}
  \item To capture the impact of FoV prediction, we consider three cases of FoV viewing probability distributions, i.e., perfect, imperfect, and unknown FoV viewing probability distributions, and use the average total utility, worst average total utility, and worst total utility as the respective performance metrics.
  \item In the single-user scenario, we optimize the encoding rates of the tiles, encoding rates of the FoVs, and transmission beamforming vectors for all subcarriers to maximize the total utility in each case of FoV viewing probability distribution. Then, we separate the optimization problem into multiple tractable problems which can provide satisfactory performance. We obtain globally optimal solutions of the separate optimization problems using transformation techniques and the Karush-Kuhn-Tucker (KKT) conditions in each case. Besides, we characterize optimality properties in the three cases, which indicate the impact of FoV prediction.
  \item In the multi-user scenario, we adopt rate splitting (which partially decodes interference and partially treats interference as noise \cite{JSAC,TCOM16}) with successive decoding for efficient transmission. Note that rate splitting successfully bridges the two extreme strategies, i.e., Nonorthogonal Multiple Access (NOMA) and Space Division Multiple Access (SDMA), and improves the spectrum efficiency in serving multiple users. In each case of FoV viewing probability distributions, we optimize the encoding rates of the tiles, encoding rates of the FoVs, rates of the common and private messages, and transmission beamforming vectors for all subcarriers to maximize the total utility. Similarly, we separate the problem into multiple tractable problems which can achieve appealing performance. We obtain KKT points of the separate optimization problems using the concave-convex procedure (CCCP)\cite{TSP17}. Furthermore, we characterize optimality properties in the three cases. 
  \item Finally, we evaluate the quality, quality variation, and rebuffering time of the proposed solutions. Numerical results show substantial gains of the proposed solutions over existing schemes in all three cases and reveal the impact of FoV prediction and its accuracy on adaptive streaming of tiled 360 videos in wireless networks.
\end{itemize}

The key notation used in this paper is listed in Table \ref{table0}.

\begin{table}[t]
\caption{\small{KEY NOTAION}}
\begin{center}
\resizebox{9cm}{!}{
\begin{tabular}{|c|c|} 
\hline 
Notation & Description \\ \hline  
$\overline{I}$ &   number of viewpoints (FoVs) of a 360 video \\ \hline
$\mathcal{F}_{i}$ &  set of $F_{i}$ tiles fully or partially included in the $i$-th FoV   \\ \hline
$D_{l}$ &  encoding rate of the $l$-th representation   \\ \hline
$\mathcal{K}$ & set of $K$ user indices (video indices) \\ \hline
$\mathcal{I}_{k}$ &   set of indices of the $I_{k}$ FoVs of video $k$ that may be watched by user $k$  \\ \hline
$p_{i,k}$ &   probability that the $i$-th FoV of video $k$ is viewed by user $k$ \\ \hline
$\hat{p}_{i,k}$ &   estimated probability that the $i$-th FoV of video $k$ is viewed by user $k$ \\ \hline
$\Delta_{i,k}$ & estimation error of the probability that the $i$-th FoV of video $k$ is viewed by user $k$ \\ \hline
$\delta $ & tolerance for quality variation in an FoV \\ \hline
$M$ &   number of the antennas \\ \hline
$N$ & number of the subcarriers \\ \hline
$B$ & bandwidth of each subcarrier \\\hline
$R_{x,y,k}$ &   encoding rate of the $(x,y)$-th tile of video $k$ \\ \hline
$r_{i,k}$ &   encoding rate of the $i$-th FoV of video $k$ \\ \hline
$\mathbf{w}_{k,n}$ & transmission beamforming vector for user $k$ on subcarrier $n$ \\\hline
$d_{c,k}$ & transmission rate of the common part of the message for user $k$ \\ \hline
$d_{p,k}$ & transmission rate of the private part of the message for user $k$ \\ \hline
\end{tabular}
} \label{table0}
\end{center}
\end{table}
\section{System Model}\label{section2}
As illustrated in Fig. \ref{streaming}, we consider adaptive streaming of $K$ tiled 360 videos from the BS to $K$ users in a single-cell wireless network, respectively. The system has two time units, group of pictures (GOP) duration (usually 0.5-1 s) and (transmission) slot duration (usually 1-5 ms). Each GOP contains $T$ slots. Encoding rate adaptation is operated at the beginning of each GOP according to the FoV prediction results and channel statistics. In contrast, transmission adaptation is operated at the beginning of each slot according to the instantaneous channel conditions.
\subsection{Tiled 360 Videos}\label{s2_a}
The $K$ users are watching $K$ 360 videos. We consider tiling to enable flexible transmission of necessary FoVs of each 360 video. Specifically, each 360 video is divided into $X\times Y$ rectangular segments, referred to as tiles, where $X$ and $Y$ represent the numbers of segments in each column and each row, respectively. Define $\mathcal{X} \triangleq \{1,\ldots,X\}$ and $\mathcal{Y} \triangleq \{1,\ldots,Y\}$. The ($x,y$)-th tile refers to the tile in the $x$-th row and the $y$-th column, for all $x\in\mathcal{X}$ and $y\in\mathcal{Y}$. For each 360 video, consider $\overline{I}$ viewpoints (i.e., $\overline{I}$ FoVs). Denote $\overline{\mathcal{I}} \triangleq\{1,\ldots,\overline{I}\}$. For all $i\in\overline{\mathcal{I}}$, let $\mathcal{F}_{i}$ denote the set of $F_{i}$ tiles fully or partially included in the $i$-th FoV. Note that $ F_{i},i\in\overline{\mathcal{I}}$ can be different. Considering user heterogeneity (e.g., in cellular usage costs, display resolutions of devices, channel conditions, etc.), each tile is pre-encoded into $L$ representations corresponding to $L$ quality levels using High Efficiency Video Coding (HEVC), as in Dynamic Adaptive Streaming over HTTP (DASH). Let $\mathcal{L} \triangleq \{1,\ldots, L\}$ denote the set of quality levels. For all $l\in\mathcal{L}$, the $l$-th representation of each tile corresponds to the $l$-th lowest quality. For ease of exposition, assume that the encoding rates of the tiles with the same quality level are identical \cite{Zhang2013QoE,HUANG20191,IS19}.\footnote{Various tiles of the same quality may have different encoding rates due to their distinct spatial redundancies. The variation of encoding rates corresponding to the same quality is usually small in practice and hence is ignored in the existing literature \cite{Zhang2013QoE,HUANG20191,IS19} for tractability.} The encoding rate of the $l$-th representation of a tile is denoted by $D_{l}$ (in bits/s), where $D_{1} < \ldots < D_{L}$.

A user can freely switch views, when watching a 360 video. Assume that the FoV of each user does not change within one GOP. This paper focuses on one GOP unless otherwise specified. Let $\mathcal{K}\triangleq \{1,\ldots,K\}$ denote the set of user indices (video indices).
Let $\mathcal{I}_{k}$ represent the set of indices of the $I_{k}$ FoVs (corresponding to the considered GOP) of video $k\in\mathcal{K}$ that user $k$ may watch. Throughout the whole paper, we suppose that $\mathcal{I}_{k},k\in\mathcal{K}$ are known to the BS.\footnote{As the angular rate of a human’s head rotation is limited \cite{VR/AR17}, it is easy to predict possible FoVs that a user may watch \cite{ICASSP19,JSTSP20}.} $\mathcal{F}_{i},i\in\mathcal{I}_{k}$ may overlap, and user $k$ will watch only one of the $I_{k}$ FoVs. For all $k\in\mathcal{K}$ and $i\in\mathcal{I}_{k}$, let $p_{i,k}$ denote the probability that the $i$-th FoV of video $k$ is viewed by user $k$. Here, $p_{i,k} \geq 0, i\in\mathcal{I}_{k},k\in\mathcal{K}$, and $\sum_{i\in\mathcal{I}_{k}}p_{i,k} = 1,k\in\mathcal{K}.$ Denote $\mathbf{p}_{k} \triangleq (p_{i,k})_{i\in\mathcal{I}_{k}},k\in\mathcal{K}$. In the following, we consider three cases of FoV viewing probability distributions.

\begin{figure}[t]
\begin{center}
 {\resizebox{9cm}{!}{\includegraphics{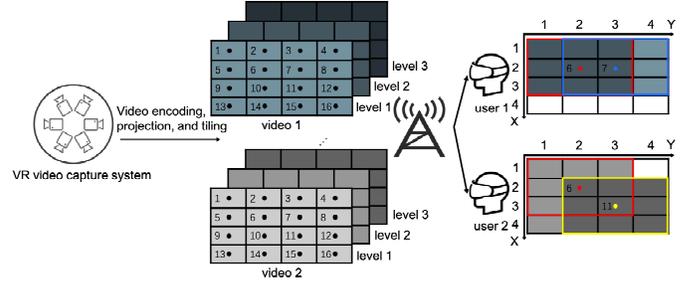}}}
\end{center}
   \caption{\small{System model. $X = 4$, $Y = 4$, $\overline{I} = 16$, $L = 3$, $K = 2$. $\mathcal{I}_{1} = \{6,7\}$, $\mathcal{F}_{6} = \{(1,1),(1,2),(1,3),(2,1),(2,2),(2,3),(3,1),(3,2),(3,3)\}$, $\mathcal{F}_{7} = \{(1,2),(1,3),(1,4),(2,2),(2,3),(2,4),(3,2),(3,3),(3,4)\}$, $r_{6,1} = D_{2}$, $r_{7,1} = D_{1}$, $R_{x,y,1} = D_{2},(x,y)\in\mathcal{F}_{6}$, $R_{x,y,1} = D_{1},(x,y)\in\mathcal{F}_{7}\backslash\mathcal{F}_{6}$. $\mathcal{I}_{2} = \{6,11\}$, $\mathcal{F}_{6} = \{(1,1),(1,2),(1,3),(2,1),(2,2),(2,3),(3,1),(3,2),(3,3)\}$, $\mathcal{F}_{11} = \{(2,2),(2,3),(2,4),(3,2),(3,3),(3,4),(4,2),(4,3),(4,4)\}$, $r_{6,2} = D_{2}$, $r_{11,2} = D_{3}$, $R_{x,y,2} = D_{3},(x,y)\in\mathcal{F}_{11}$, $R_{x,y,2} = D_{2},(x,y)\in\mathcal{F}_{6}\backslash\mathcal{F}_{11}$.}}
   \label{streaming}
\end{figure}

\textbf{Perfect FoV viewing probability distributions:} In this case, FoV viewing probability distributions have been estimated by some learning methods, and the estimation errors are negligible. That is, the BS knows the exact values of $\mathbf{p}_{k},k\in\mathcal{K}$\cite{SECON17,ICASSP19}.

\textbf{Imperfect FoV viewing probability distributions:} In this case, FoV viewing probability distributions have been estimated by some learning methods with certain estimation errors.\footnote{As estimation errors are inevitable in practice, the case of imperfect FoV viewing probability distributions is more practical than the case of perfect FoV viewing probability distributions.} For all $k\in\mathcal{K}$ and $i\in\mathcal{I}_{k}$, let $\hat{p}_{i,k}$ denote the estimated probability that the $i$-th FoV of video $k$ is viewed by user $k$, and let $\Delta_{i,k} \triangleq p_{i,k} - \hat{p}_{i,k}$ denote the corresponding estimation error. Here, $\hat{p}_{i,k} \geq 0, i\in\mathcal{I}_{k},k\in\mathcal{K}$, $\sum_{i\in\mathcal{I}_{k}}\hat{p}_{i,k} = 1,k\in\mathcal{K},$ $\sum\nolimits_{i\in\mathcal{I}_{k}}\Delta_{i,k} = 0$, and $|\Delta_{i,k}| \leq \varepsilon_{i,k}$ for some known $\varepsilon_{i,k} \in (0,1)$. Assume that the BS knows $\hat{p}_{i,k}$, $\varepsilon_{i,k},i\in\mathcal{I}_{k},k\in\mathcal{K}$ but does not know $\mathbf{p}_{k},k\in\mathcal{K}$. That is, the BS knows that the exact values of $\mathbf{p}_{k},k\in\mathcal{K}$ satisfy $\mathbf{p}_{k}\in\mathcal{P}_{k},k\in\mathcal{K}$, where
\begin{equation}
\mathcal{P}_{k}\triangleq\left\{\mathbf{p}_{k}~\Big|~\underline{p}_{i,k} \leq p_{i,k} \leq \overline{p}_{i,k},i\in\mathcal{I}_{k},\sum\limits_{i\in\mathcal{I}_{k}}p_{i,k} =1  \right\},k\in\mathcal{K},\nonumber
\end{equation}
with $\underline{p}_{i,k} \triangleq \max\{\hat{p}_{i,k}-\varepsilon_{i,k}, 0\}$ and $\overline{p}_{i,k} \triangleq \min\{\hat{p}_{i,k} + \varepsilon_{i,k}, 1\}$, $i \in \mathcal{I}_{k},k\in\mathcal{K}$.

\textbf{Unknown FoV viewing probability distributions:} In this case, the BS does not know any prior information about the exact values of $\mathbf{p}_{k},k\in\mathcal{K}$.

For all $k\in\mathcal{K}$, the $\overline{F}_{k}$ tiles in $\overline{\mathcal{F}}_{k}\triangleq\mathop\cup\nolimits_{i\in\mathcal{I}_{k}}\mathcal{F}_{i}$ may be transmitted to user $k$, where $\overline{F}_{k} \triangleq |\overline{\mathcal{F}}_{k}|$. Let $R_{x,y,k}$ (in bits/s) denote the encoding rate of the $(x,y)$-th tile (corresponding to the considered GOP) of video $k$, where
\begin{equation}
R_{x,y,k} \in\{0,D_{1},\ldots,D_{L}\},~(x,y)\in\overline{\mathcal{F}}_{k},~k\in\mathcal{K}.\label{tile_rate_max}
\end{equation}
Here, $R_{x,y,k} = 0$ indicates that the $(x,y)$-th tile of video $k$ will not be transmitted to user $k$, and $R_{x,y,k} = D_{l}$ indicates that the $l$-th representation of the $(x,y)$-th tile will be transmitted to user $k$. For all $k\in\mathcal{K}$, encoded (source coding) bits of different tiles in $\overline{\mathcal{F}}_{k}$ that will be transmitted to user $k$ are ``aggregated" into one message. The encoding rate of the aggregated message for user $k$ is $\sum\nolimits_{(x,y)\in\overline{\mathcal{F}}_{k}}R_{x,y,k}$.

To avoid degrading QoE, we consider a relative smoothness requirement for quality variation in an FoV \cite{TMM20L}:
\begin{equation}
r_{i,k} \leq R_{x,y,k} \leq r_{i,k} + \delta ,~(x,y)\in\mathcal{F}_{i},~i\in\mathcal{I}_{k},~k\in\mathcal{K},\label{rate_smooth}
\end{equation}
where
\begin{equation}
r_{i,k} \in\{0,D_{1},\ldots,D_{L}\},~i\in\mathcal{I}_{k},~k\in\mathcal{K}.\label{fov_rate_max}
\end{equation}
Here, $r_{i,k}$ (in bits/s) represents the minimum of the encoding rates of the tiles in the $i$-th FoV of video $k$, also referred to as the encoding rate of the $i$-th FoV of video $k$ and indicates the quality level of the $i$-th FoV for user $k$; and $\delta > 0$ is a small number representing the tolerance for quality variation in an FoV (note that the quality variation over tiles in one FoV is not visible if $\delta$ is small enough). An illustration example can be found in Fig. \ref{streaming}.

In this paper, we mainly focus on maximizing video quality while keeping the rebuffering time small.\footnote{Later, we shall see that the proposed approach can also achieve high video quality, low quality variation and short rebuffering time.} Toward the goal of maximizing video quality, we choose the following performance metrics. Let $U(r)$ denote the utility for an FoV with the encoding rate $r$. Here, $U(\cdot)$ can be any nonnegative, strictly increasing and strictly concave function,\footnote{Logarithmic functions satisfy the requirements on $U(\cdot)$. Besides, numerical results show that Peak Signal-to-Noise Ratio (PSNR) and Structural Similarity Index Measure (SSIM) also satisfy the requirements on $U(\cdot)$\cite{PSNR,SSIM}.} and $U(0) = 0$. Its monotonicity can capture the notion that perceptual quality increases with the encoding rate. Its concavity can capture the notion that the increase rate of perceptual quality decreases with the encoding rate. Let $Q^{(\phi)}(\mathbf{r}), \phi =$ pp, ip, and up denote the performance metrics in the three cases of FoV viewing probability distributions. In the case of perfect FoV viewing probability distributions (i.e., case-pp), we use the average total utility, $\sum_{k\in\mathcal{K}}\sum_{i\in\mathcal{I}_{k}}p_{i,k}U(r_{i,k})$, as the performance metric. In the case of imperfect FoV viewing probability distributions (i.e., case-ip), we use the worst (across all possible FoV viewing probability distributions) average total utility, $\sum_{k\in\mathcal{K}}\min_{\mathbf{p}_{k}\in\mathcal{P}_{k}} \sum_{i\in\mathcal{I}_{k}}p_{i,k}U(r_{i,k})$, as the performance metric. In the case of unknown FoV viewing probability distributions (i.e., case-up), we use the worst (across all possible FoVs) total utility, $\sum_{k\in\mathcal{K}} \min_{i\in\mathcal{I}_{k}}~U(r_{i,k})$, as the performance metric. Notice that performance metrics in the last two cases are commonly adopted in robust optimization to guarantee the worst-case performance when there is uncertainty about system (problem) parameters \cite{boyd2004convex}. Therefore, the performance metrics in the three cases of FoV viewing probability distributions can be written as
\begin{align}
Q^{(\phi)}(\mathbf{r}) = \left\{ \begin{array}{ll}
 \sum_{k\in\mathcal{K}}\sum_{i\in\mathcal{I}_{k}}p_{i,k}U(r_{i,k}), & \phi = \text{pp},\\
 \sum_{k\in\mathcal{K}}\min_{\mathbf{p}_{k}\in\mathcal{P}_{k}} \sum_{i\in\mathcal{I}_{k}}p_{i,k}U(r_{i,k}), & \phi = \text{ip},\\
 \sum_{k\in\mathcal{K}} \min_{i\in\mathcal{I}_{k}}~U(r_{i,k}), & \phi = \text{up}.
  \end{array} \right.\label{utility}
\end{align}

\subsection{Physical Layer Model}
The BS has $M$ antennas, and each user has one antenna. We consider a multi-carrier system. Let $N$ and $\mathcal{N} \triangleq\{1, \ldots, N\}$ denote the number of subcarriers and the set of subcarrier indices, respectively. The bandwidth of each subcarrier is $B$ (in Hz). We assume block fading, i.e., the channel on each subcarrier remains constant within each slot and changes in an independent and identically distributed (i.i.d.) manner over slots within one GOP. Let $\mathcal{T}\triangleq \{1,\cdots,T\}$ denote the set of the slots in the considered GOP. Let $\mathbf{h}_{k,n}^{H}(t) \in\mathbb{C}^{1\times M}$ denote the channel state on subcarrier $n$ between user $k$ and the BS at slot $t$. We assume that the channel state information is perfectly known at the BS and the users. Let $\mathbf{x}_{n}(t)\in\mathbb{C}^{M\times1}$ be a transmitted signal on subcarrier $n$ at slot $t$. The total average transmission power constraint at slot $t$ is given by: 
\begin{align}
\sum\nolimits_{n\in\mathcal{N}}\mathbb{E}[\|\mathbf{x}_{n}(t)\|^{2}_{2}] \leq P,~t\in\mathcal{T},\label{x_general}
\end{align}
where $P$ is the total transmission power budget. The received signal at user $k$ on subcarrier $n$ at slot $t$ is given by:
\begin{equation}
y_{k,n}(t) = \mathbf{h}_{k,n}^{H}(t)\mathbf{x}_{n}(t) + z_{k,n}(t),~k\in\mathcal{K},~n\in\mathcal{N},~t\in\mathcal{T},\label{received_signal}
\end{equation}
where $z_{k,n}(t) \sim \mathcal{CN}(0, \sigma^{2})$ represents the received Additive White Gaussian Noise (AWGN) at user $k$ on subcarrier $n$ at slot $t$, and $\sigma^{2}$ represents the noise power. In Section \ref{s3} and Section \ref{section3}, we shall consider the single-user scenario and multi-user scenario, respectively. The respective detailed physical layer models will be illustrated shortly.


\section{Adaptive Video Streaming In Single-user Scenario}\label{s3}
In this section, we consider the single-user scenario, i.e., $K = 1$. First, we illustrate the transmission scheme. Then, we formulate a utility maximization problem and separate it into multiple tractable problems for each case. Finally, we solve the problems. We omit index $k$ in the notations and expressions introduced in Section \ref{section2} in the single-user scenario for notation simplicity.
\subsection{Transmission Scheme}
The aggregated message for the user sent at slot $t$ is encoded (channel coding) into a codeword that spans over $N$ subcarriers. Let $s_{n}$ denote a symbol transmitted on the $n$-th subcarrier. Denote $\mathbf{s}\triangleq (s_{n})_{n\in\mathcal{N}}$ and assume that $\mathbb{E}[\mathbf{s}\mathbf{s}^{H}]= \mathbf{I}$. We consider linear precoding on each subcarrier. The transmitted signal on subcarrier $n$ at slot $t$ is given by:
\begin{equation}
\mathbf{x}_{n}(t) = \mathbf{w}_{n}(t)s_{n},~n\in\mathcal{N},~t\in\mathcal{T},\label{x_single_expectation}
\end{equation}
where $\mathbf{w}_{n}(t)\in\mathbb{C}^{M\times1}$ denotes the beamforming vector on subcarrier $n$ at slot $t$. Substituting \eqref{x_single_expectation} into \eqref{x_general}, we have the total power constraint at slot $t$:
\begin{equation}
\sum\nolimits_{n\in\mathcal{N}}\|\mathbf{w}_{n}(t)\|_{2}^{2} \leq P,~t\in\mathcal{T}.\label{single_power_allocation_constraint}
\end{equation}
Substituting \eqref{x_single_expectation} into \eqref{received_signal}, we can derive the Signal-to-Noise Ratios (SNR) on subcarrier $n$ at slot $t$, i.e., $\frac{|\mathbf{h}_{n}^{H}(t)\mathbf{w}_{n}(t)|^{2}}{\sigma^{2}}$. We consider Gaussian coding\cite{INFOR03,JSAC,TCOM16}. When the following encoding rate constraint for the GOP:
\begin{align}
&\sum\limits_{(x,y)\in\overline{\mathcal{F}}} R_{x,y} \leq \frac{1}{T}\sum_{t\in\mathcal{T}}\sum\limits_{n\in\mathcal{N}}B\log_{2}\left(1 + \frac{|\mathbf{h}_{n}^{H}(t)\mathbf{w}_{n}(t)|^{2}}{\sigma^{2}}\right)\label{single_successful_transmit}
\end{align}
is satisfied, rebuffering for the considered GOP can be avoided.\footnote{In this paper, we adopt a stronger requirement to avoid rebuffering in 360 video streaming due to frequent FoV switch.}
\subsection{Problem Formulation}
In the single-user scenario, the performance metrics in \eqref{utility} can be rewritten as
\begin{align}
Q^{(\phi)}(\mathbf{r}) = \left\{ \begin{array}{ll}
 \sum_{i\in\mathcal{I}}p_{i}U(r_{i}), & \phi = \text{pp},\\
 \min_{\mathbf{p}\in\mathcal{P}} \sum_{i\in\mathcal{I}}p_{i}U(r_{i}), & \phi = \text{ip},\\
 \min_{i\in\mathcal{I}}~U(r_{i}), & \phi = \text{up}.
  \end{array} \right.\label{utility_s}
\end{align}
Our goal is to maximize the video quality and avoid rebuffering meanwhile. Toward this end, we optimize the encoding rates of the tiles $\mathbf{R} \triangleq (R_{x,y})_{(x,y)\in\overline{\mathcal{F}}}$, encoding rates of the FoVs $\mathbf{r} \triangleq (r_{i})_{i\in\mathcal{I}}$, and transmission beamforming vectors $\mathbf{w}(t)\triangleq (\mathbf{w}_{n}(t))_{n\in\mathcal{N}},t\in\mathcal{T}$ to maximize the performance metrics in \eqref{utility_s} subject to the constraints in \eqref{tile_rate_max}, \eqref{rate_smooth}, \eqref{fov_rate_max}, \eqref{single_power_allocation_constraint}, \eqref{single_successful_transmit}. 
Note that $\mathbf{R}$ and $\mathbf{r}$ are discrete variables. For tractability, we consider a relaxed version of the discrete optimization problem, as in \cite{JSTSP20,HUANG20191}. That is, we replace the discrete constraints in \eqref{tile_rate_max} and \eqref{fov_rate_max} with the following continuous constraints:
\begin{align}
&0 \leq R_{x,y} \leq D_{L},~(x,y)\in\overline{\mathcal{F}},\label{single_tile_relax}\\
&0 \leq r_{i} \leq D_{L},~i\in\mathcal{I}.\label{single_fov_relax}
\end{align}
Therefore, we formulate the following optimization problem.

\begin{Prob}[Total Utility Maximization in Single-user Scenario]\label{single_case_general_new}For $\phi$ = \text{pp},~\text{ip},~\text{up},
\begin{align}\max_{\mathbf{R},\mathbf{r},\mathbf{w}(t),t\in\mathcal{T}}\quad &Q^{(\phi)}(\mathbf{r})\nonumber\\
    \mathrm{s.t.}\quad&\eqref{rate_smooth},~\eqref{single_power_allocation_constraint},~\eqref{single_successful_transmit},~\eqref{single_tile_relax},~\eqref{single_fov_relax}.\nonumber
\end{align}
\end{Prob}

Note that the optimal solution of Problem \ref{single_case_general_new} depends on $D_{L}$ and is not related to $D_{1},\ldots,D_{L-1}$. Define $\mathcal{D} \triangleq \{0,D_{1},\ldots,D_{L}\}.$ Based on any feasible solution of Problem \ref{single_case_general_new}, denoted by $(\mathbf{R}^{(\phi)},\mathbf{r}^{(\phi)})$, we can construct feasible discrete encoding rates of the tiles and the FoVs, denoted by $(\widetilde{\mathbf{R}}^{(\phi)},\widetilde{\mathbf{r}}^{(\phi)})$, where $\widetilde{R}_{x,y}^{(\phi)} =\max\{d\in\mathcal{D}|d\leq R^{(\phi)}_{x,y}\},(x,y)\in\overline{\mathcal{F}}$ and $\widetilde{r}_{i}^{(\phi)} = \max\{d\in\mathcal{D}|d\leq r^{(\phi)}_{i}\}, i\in\mathcal{I}$. Note that performance loss induced by solving Problem \ref{single_case_general_new} and constructing a feasible solution of the discrete problem based on the feasible solution of Problem \ref{single_case_general_new} is acceptable when $D_{2}-D_{1},\ldots,D_{L}-D_{L-1}$ are not large, which will be shown in Section \ref{simulation_single}.

The BS obtains the channel condition of each slot at the beginning of the slot and performs encoding rate adaptation at the beginning of each GOP, i.e., the beginning of the first slot in each GOP. Thus, in practice, Problem \ref{single_case_general_new}, an ideal formulation for the offline scenario, cannot be solved at the beginning of the first slot of the considered GOP without knowledge of the channel conditions of the subsequent slots of the GOP. To obtain a practical design, we separate Problem \ref{single_case_general_new} into $T$ optimization problems. One is for the encoding rate adaptation of the considered GOP and transmission adaptation at slot $1$ in the GOP, and the others are for transmission adaptation at slots $2,\ldots,T$ in the GOP.

Specifically, we introduce the encoding rate constraint for slot $1$:
\begin{align}
\sum\limits_{(x,y)\in\overline{\mathcal{F}}} R_{x,y} \leq \sum\limits_{n\in\mathcal{N}}B\log_{2}\left(1 + \frac{|\mathbf{h}_{n}^{H}(1)\mathbf{w}_{n}(1)|^{2}}{\sigma^{2}}\right).\label{single_successful_transmit_first}
\end{align}
The optimization problem for the encoding rate adaptation of the GOP and the transmission adaptation of slot $1$ is as follows.\footnote{In Problem \ref{single_case_general}, the constraint in \eqref{single_power_allocation_constraint} is only for slot $t=1$.}

\begin{Prob}[Total Utility Maximization at $t = 1$ in Single-user Scenario]\label{single_case_general}For $\phi$ = \text{pp},~\text{ip},~\text{up},
\begin{align}
&U^{(\phi)\star} \triangleq \max_{\mathbf{R},\mathbf{r},\mathbf{w}(1)}\quad Q^{(\phi)}(\mathbf{r})\nonumber\\
    &\mathrm{s.t.}\quad\eqref{rate_smooth},~\eqref{single_power_allocation_constraint},~\eqref{single_tile_relax},~\eqref{single_fov_relax},~\eqref{single_successful_transmit_first}.\nonumber
\end{align}
\end{Prob}

With the knowledge of the channel condition, $\mathbf{h}_{n}^{H}(1),n\in\mathcal{N}$, Problem \ref{single_case_general} can be solved at the beginning of slot $1$. The objective function of Problem \ref{single_case_general} is concave and the constraints in \eqref{rate_smooth},~\eqref{single_power_allocation_constraint},~\eqref{single_tile_relax},~\eqref{single_fov_relax} and \eqref{single_successful_transmit_first} are convex. Thus, Problem \ref{single_case_general} is convex with respect to (w.r.t.) ($\mathbf{R}, \mathbf{r},\mathbf{w}(1)$), and the optimal solution of Problem \ref{single_case_general} can be obtained. 

The optimization problem for the transmission adaptation of each slot $t = 2,\ldots,T$ is as follows.\footnote{In Problem \ref{single_infeasibility_min}, the constraint in \eqref{single_power_allocation_constraint} is only for the considered slot $t$.}

\begin{Prob}[Transmission Rate Maximization at $t = 2,\ldots,T$ in Single-user Scenario]\label{single_infeasibility_min}For all $t=2,\ldots,T$,
\begin{align}
\max_{\mathbf{w}(t)}\quad &\sum\nolimits_{n\in\mathcal{N}}B\log_{2}\left(1 + \frac{|\mathbf{h}_{n}^{H}(t)\mathbf{w}_{n}(t)|^{2}}{\sigma^{2}}\right) \nonumber\\
    &\mathrm{s.t.}~\eqref{single_power_allocation_constraint}.\nonumber
\end{align}
\end{Prob}

Note that the objective function of Problem \ref{single_infeasibility_min} and the constraints in \eqref{single_power_allocation_constraint} are convex. Thus, for all $t=2,\ldots,T$, Problem \ref{single_infeasibility_min} is convex w.r.t $\mathbf{w}(t)$, and hence can be solved optimally. Besides, note that Problem \ref{single_infeasibility_min} does not rely on $\phi$, indicating that the transmission adaptations of each subsequent slot in the three cases are identical.

\begin{Rem}[Interpretation of Separate Approach in Single-user Scenario]
The encoding rate constraint for slot $1$ in \eqref{single_successful_transmit_first} of Problem \ref{single_case_general} together with Problem \ref{single_infeasibility_min} for $t=2,\ldots,T$ is to reduce the infeasibility of the encoding rate constraint for the GOP in \eqref{single_successful_transmit} of Problem \ref{single_case_general_new}. In Problem \ref{single_case_general}, the encoding rate adaptation of the GOP relies only on the channel condition of slot $1$ rather than the channel conditions of all slots in the GOP. As the number of subcarriers in a practical system is usually large (e.g., $N = 128$ \cite{5G}), the average channel condition for one subcarrier at each slot does not change much over slots. Thus, the encoding rate adaptation for the GOP and the transmission adaptation for each slot offered by the separate approach are close to those offered by solving Problem \ref{single_case_general_new} and will not violate the encoding rate constraint for the GOP in \eqref{single_successful_transmit} that much. The violation if exists yields rebuffering. Later in Section \ref{simulation_single}, we shall see that with the separate approach, we can improve video quality while keeping the rebuffering time small, i.e., approximately achieve the goal of Problem \ref{single_case_general_new} (which cannot be solved unless in the ideal offline scenario).
\end{Rem}

\subsection{Solution}\label{single_solution}
\subsubsection{Optimal Solution of Problem \ref{single_case_general}}
As $Q^{(\phi)}(\mathbf{r})$, $\phi$ = ip, up, are non-differentiable, standard convex optimization methods cannot be used for solving Problem \ref{single_case_general} with $\phi$ = ip and up. Furthermore, an analytical solution is usually more computationally efficient than a numerical solution. In this part, we develop an efficient algorithm for solving Problem \ref{single_case_general}, by exploring its structural properties. Specifically, we equivalently transform Problem \ref{single_case_general} into the following problem.
\begin{Prob}[Equivalent Problem of Problem \ref{single_case_general}]\label{single_case_general_equal} For $\phi$ = \text{pp},~\text{ip},~\text{up},
\begin{align}
\max_{\mathbf{R},\mathbf{r}}\quad &Q^{(\phi)}(\mathbf{r})\nonumber\\
    \mathrm{s.t.}\quad&\eqref{rate_smooth},~\eqref{single_tile_relax},~\eqref{single_fov_relax},\nonumber\\
    &\sum\nolimits_{(x,y)\in\overline{\mathcal{F}}}R_{x,y} \leq C^{\dag}(1).\label{single_case_general_equal_constraint}
\end{align}
Let ($\mathbf{R}^{(\phi)\dag}$, $\mathbf{r}^{(\phi)\dag}$) denote an optimal solution of Problem \ref{single_case_general_equal}. $C^{\dag}(1)$ represents the transmission rate and is given by the following problem.
\end{Prob}
\begin{align}
C^{\dag}(t) \triangleq \max_{\mathbf{v}(t)}\quad&\sum\nolimits_{n\in\mathcal{N}}B\log_{2}\left(1 + \frac{\|\mathbf{h}_{n}(t)\|_{2}^{2}v_{n}(t)}{\sigma^{2}}\right)\label{water}\\
    \mathrm{s.t.}\quad&\sum\nolimits_{n\in\mathcal{N}}v_{n}(t)\leq P,\nonumber\\
    &v_{n}(t) \geq 0,~n\in\mathcal{N}\nonumber,
\end{align}
where $\mathbf{v}(t)\triangleq (v_{n}(t))_{n\in\mathcal{N}}$. Let $\mathbf{v}^{\dag}(t)$ denote an optimal solution of the problem in \eqref{water}. 

Note that $v_{n}(1),n\in\mathcal{N}$ in the problem in \eqref{water} can be viewed as the power allocated on the $n$-th subcarrier at slot 1. The equivalence between Problem \ref{single_case_general} and Problem \ref{single_case_general_equal} is summarized below.
\begin{Thm}[Equivalence between Problem \ref{single_case_general} and Problem \ref{single_case_general_equal}]\label{lemma_decouple}
For $\phi$ = pp, ip, up, an optimal solution of Problem \ref{single_case_general} is given by ($\mathbf{R}^{(\phi)\dag}, \mathbf{r}^{(\phi)\dag}, \mathbf{w}(1)^{\star}$), where ($\mathbf{R}^{(\phi)\dag},\mathbf{r}^{(\phi)\dag}$) is an optimal solution of Problem \ref{single_case_general_equal}, and 
\begin{align}
\mathbf{w}_{n}(1)^{\star} =\frac{\mathbf{h}_{n}(1)}{\|\mathbf{h}_{n}(1)\|_{2}}\sqrt{v^{\dag}_{n}(1)},~n\in\mathcal{N}.\label{bf2pwr}
\end{align}
\end{Thm}
\begin{Proof}
Please refer to Appendix A.
\end{Proof}

By Theorem~\ref{lemma_decouple}, we know that the optimal beamforming vectors in the three cases of FoV viewing probability distributions share the same form, which correspond to the Maximum Ratio Transmission (MRT) beamfomers\cite{MRT}. According to Theorem~\ref{lemma_decouple}, we can obtain an optimal solution of Problem \ref{single_case_general} by solving Problem \ref{single_case_general_equal}. 

First, we solve the convex problem in \eqref{water}. By the KKT conditions, we can obtain its semi-closed form optimal solution and optimal value:
\begin{align}
&v^{\dag}_{n}(1) = \max\Big\{0, \frac{1}{\rho^{\dag}\ln2} - \frac{\sigma^{2}}{\|\mathbf{h}_{n}(1)\|_{2}^{2}}\Big\},~n\in\mathcal{N},\label{water_solution}\\
&C^{\dag}(1) = \nonumber\\
&\sum\limits_{n\in\mathcal{N}}B\log_{2}\left(1 + \frac{\|\mathbf{h}_{n}(1)\|_{2}^{2}\max\Big\{0, \frac{1}{\rho^{\dag}\ln2} - \frac{\sigma^{2}}{\|\mathbf{h}_{n}(1)\|_{2}^{2}}\Big\}}{\sigma^{2}}\right),\label{optimal_C}
\end{align}
where $\rho^{\dag}$ satisfies $\sum_{n\in\mathcal{N}}\max\{0, \frac{1}{\rho^{\dag}\ln2} - \frac{\sigma^{2}}{\|\mathbf{h}_{n}(t)\|_{2}^{2}}\} = P$. Note that $\rho^{\dag}$ can be obtained by the bisection method. The computational complexity for calculating $\mathbf{v}^{\dag}(1)$ is $\mathcal{O}(MN)$. The optimal solution in \eqref{water_solution} has a water-filling structure with $\frac{1}{\rho^{\dag}\ln2}$ being the water level. The subcarrier with a higher channel gain will be allocated more power. Based on \eqref{bf2pwr} and \eqref{water_solution}, we can obtain: 
\begin{align}
&\mathbf{w}_{n}(1)^{\star} = \frac{\mathbf{h}_{n}(1)}{\|\mathbf{h}_{n}(1)\|_{2}}\sqrt{\max\Big\{0, \frac{1}{\rho^{\dag}\ln2} - \frac{\sigma^{2}}{\|\mathbf{h}_{n}(1)\|_{2}^{2}}\Big\}}.\nonumber
\end{align}

Next, we solve Problem \ref{single_case_general_equal} for each of the three cases of FoV viewing probability distributions given $C^{\dag}(1)$ in \eqref{optimal_C}.

\textbf{Perfect FoV viewing probability distribution:} As $Q^{(\text{pp})}(\mathbf{r})$ is differentiable, Problem \ref{single_case_general_equal} is much easier to handle.

\textbf{Imperfect FoV viewing probability distribution:}
Problem \ref{single_case_general_equal} with $\phi$ = ip is a max-min problem w.r.t. ($\mathbf{R},\mathbf{r},\mathbf{p}$). Note that max-min problems are in general very challenging. Furthermore, $Q^{(\text{ip})}(\mathbf{r})$ is non-differentiable. We transform Problem \ref{single_case_general_equal} into an equivalent convex problem with a differentiable objective function by replacing the inner problem with its dual problem\cite{Ye2019Optimal}.\footnote{Our previous work \cite{Ye2019Optimal} considers optimal random caching designs for perfect, imperfect, and unknown file popularity distributions in a large-scale multi-tier wireless network. It motivates us to study imperfect FoV prediction.} 
\begin{Prob}[Equivalent Problem of Problem \ref{single_case_general_equal} with $\phi$ = ip]\label{single_case_two_equal}
\begin{align}
\max_{\substack{\mathbf{R},\mathbf{r},\bm{\lambda} \succeq 0,\\ \bm{\tau} \succeq 0, \gamma}}\quad&\sum\nolimits_{i\in\mathcal{I}}(\tau_{i}\underline{p}_{i} - \lambda_{i}\overline{p}_{i}) - \gamma \nonumber\\
    \mathrm{s.t.}\quad&\eqref{rate_smooth},~\eqref{single_tile_relax},~\eqref{single_fov_relax},~\eqref{single_case_general_equal_constraint},\nonumber\\
    &U(r_{i}) + \lambda_{i} - \tau_{i} + \gamma \geq 0, ~i\in\mathcal{I},\label{single_dual_constraint}
\end{align}
where $\bm{\lambda}\triangleq (\lambda_{i})_{i\in\mathcal{I}}$, $\bm{\tau}\triangleq (\tau_{i})_{i\in\mathcal{I}}$. Let $(\mathbf{R}^{(\text{ip})\dag}, \mathbf{r}^{(\text{ip})\dag}, \bm{\lambda}^{(\text{ip})\dag}, \bm{\tau}^{(\text{ip})\dag}, \gamma^{(\text{ip})\dag})$ denote an optimal solution of Problem \ref{single_case_two_equal}.
\end{Prob}

Note that $\lambda_{i}$, $\tau_{i}$, and $\gamma$ are dual variables of the inner problem, corresponding to $p_{i} \leq \overline{p}_{i}$, $p_{i} \geq \underline{p}_{i}$ and $\sum_{i\in\mathcal{I}}p_{i} = 1$, respectively. The equivalence between Problem \ref{single_case_general_equal} with $\phi$ = ip and Problem \ref{single_case_two_equal} is summarized below.
\begin{Thm}[Equivalence between Problem \ref{single_case_general_equal} with $\phi$ = ip and Problem \ref{single_case_two_equal}]\label{lemma_single_case2}
$(\mathbf{R}^{(\text{ip})\dag}, \mathbf{r}^{(\text{ip})\dag})$ is an optimal solution of Problem \ref{single_case_general_equal} with $\phi$ = ip.
\end{Thm}
\begin{Proof}
Please refer to Appendix B.
\end{Proof}

By Theorem \ref{lemma_single_case2}, we can solve Problem \ref{single_case_general_equal} with $\phi$ = ip by solving Problem \ref{single_case_two_equal}.

\textbf{Unknown FoV viewing probability distribution:}
$Q^{(\text{up})}(\mathbf{r})$ is non-differentiable. We can cast Problem \ref{single_case_general_equal} with $\phi$ = up in hypograph form as follows\cite[pp. 134]{boyd2004convex}. 
\begin{Prob}[Equivalent Problem of Problem \ref{single_case_general_equal} with $\phi$ = up]\label{single_case_three_equal}
\begin{align}
\max_{\mathbf{R},\mathbf{r},y}\quad &y\nonumber\\
    \mathrm{s.t.}\quad&\eqref{rate_smooth},~\eqref{single_tile_relax},~\eqref{single_fov_relax},~\eqref{single_case_general_equal_constraint},\nonumber\\
    &y \leq U(r_{i}), ~i\in\mathcal{I}.\label{single_epigraph_p3}
\end{align}
Let $(\mathbf{R}^{(\text{up})\dag}, \mathbf{r}^{(\text{up})\dag}, y^{(\text{up})\dag})$ denote an optimal solution of Problem \ref{single_case_three_equal}.
\end{Prob}

It is clear that $(\mathbf{R}^{(\text{up})\dag}, \mathbf{r}^{(\text{up})\dag}, y^{(\text{up})\dag})$ is an optimal solution of Problem \ref{single_case_general_equal} with $\phi$ = up. Thus, we can solve Problem \ref{single_case_general_equal} with $\phi$ = up by solving Problem \ref{single_case_three_equal}. Note that Problem \ref{single_case_general_equal} with $\phi$ = pp, Problem \ref{single_case_two_equal} and Problem \ref{single_case_three_equal} are convex with differentiable objective functions and constraint functions and can be solved efficiently using standard convex optimization methods such as interior-point methods\cite{boyd2004convex}. When an interior point method is applied, the computational complexity for solving Problem \ref{single_case_general_equal} with $\phi$ = pp, Problem \ref{single_case_two_equal}, and Problem \ref{single_case_three_equal} is $\mathcal{O}(\overline{F}^{2}S^{1.5})$, where $S \triangleq \sum_{i\in\mathcal{I}}F_{i}$. The details for obtaining a globally optimal solution of Problem \ref{single_case_general} with $\phi$ = pp, ip, up are summarized in Algorithm~\ref{alg:single}.

\begin{algorithm}[t]\small
  \setstretch{1.35}
    \caption{Obtaining an Optimal Solution of Problem \ref{single_case_general}}
\begin{small}
        \begin{algorithmic}[1]
          \STATE \quad Compute $\rho^{\dag}$ by solving $\sum_{n\in\mathcal{N}}\max\{0, \frac{1}{\rho^{\dag}\ln2} - \frac{\sigma^{2}}{\|\mathbf{h}_{n}(1)\|_{2}^{2}}\} = P$ via bisection search;
          \STATE \quad Compute $v^{\dag}_{n}(1),n\in\mathcal{N}$ according to (\ref{water_solution});
           \STATE \quad Obtain an optimal solution of Problem~\ref{single_case_general_equal} with $\phi$ = pp, Problem~\ref{single_case_two_equal} or Problem~\ref{single_case_three_equal} using an interior point method.
    \end{algorithmic}\label{alg:single}
    \end{small}
\end{algorithm}

Finally, we analyze optimality properties of Problem \ref{single_case_general}. For all $i\in\mathcal{I},$ define $\mathcal{T}_{i} \triangleq \mathcal{F}_{i}\backslash\overline{\mathcal{F}}$. Note that $\mathcal{T}_{i} \cap \mathcal{T}_{j} = \emptyset$, for all $i,j\in\mathcal{I},i\not=j$.

\begin{Thm}[Optimality Properties of Problem \ref{single_case_general}]\label{single_theorem}(\romannumeral1) For $\phi$ = pp, ip, up, $R^{(\phi)\star}_{x,y} = \max\limits_{i\in\mathcal{I}: (x,y)\in\mathcal{F}_{i}}r^{(\phi)\star}_{i},(x,y)\in\overline{\mathcal{F}}$. (\romannumeral2) For $\phi$ = pp and for all $i,j\in\mathcal{I},i\not=j,$ if $p_{i} \leq p_{j}$ and $|\mathcal{T}_{i}| > |\mathcal{T}_{j}| > 1$, then $r^{(\text{pp})\star}_{i} \leq r^{(\text{pp})\star}_{j}$. For $\phi$ = ip and for all $i,j\in\mathcal{I},i\not=j,$ if $\overline{p}_{i} \leq \underline{p}_{j}$ and $|\mathcal{T}_{i}| > |\mathcal{T}_{j}| > 1$, then $r^{(\text{ip})\star}_{i} \leq r^{(\text{ip})\star}_{j}$. For $\phi$ = up, $r^{\text{(up)}\star}_{i},i\in\mathcal{I}$ are identical. (\romannumeral3) $U^{(\text{pp})\star} \geq U^{(\text{ip})\star} \geq U^{(\text{up})\star}$, where $U^{(\text{pp})\star}$, $U^{(\text{ip})\star}$, and $U^{(\text{up})\star}$ are the optimal values of Problem \ref{single_case_general} with $\phi$ = pp, ip and up, respectively.
\end{Thm}
\begin{Proof}
Please refer to Appendix C.
\end{Proof}

Statement (\romannumeral1) of Theorem \ref{single_theorem} indicates that in each case, for all $(x,y)\in\overline{\mathcal{F}}$, the first inequality in \eqref{rate_smooth} for at least one FoV that covers the $(x,y)$-th tile is active at an optimal solution. Statement (\romannumeral2) of Theorem \ref{single_theorem} indicates that in the cases of perfect and imperfect FoV viewing probability distributions, an FoV with a higher viewing probability has a higher encoding rate; and in the case of unknown FoV viewing probability distribution, the encoding rates of all FoVs in $\mathcal{I}$ are identical, as they are treated the same. Statement (\romannumeral3) of Theorem \ref{single_theorem} shows the relationship among the optimal values of Problem \ref{single_case_general} for the three cases.

\subsubsection{Optimal Solution of Problem \ref{single_infeasibility_min}}By replacing $\mathbf{w}_{n}(t)$ with $\frac{\mathbf{h}_{n}(t)}{\|\mathbf{h}_{n}(t)\|_{2}}\sqrt{v_{n}(t)}$, Problem \ref{single_infeasibility_min} can be equivalently converted into the problem in \eqref{water}. For all $t = 2,\ldots,T$, we can obtain an optimal solution of Problem \ref{single_infeasibility_min} by \eqref{water_solution}: 
\begin{align}
\mathbf{w}_{n}(t)^{\star} = \frac{\mathbf{h}_{n}(t)}{\|\mathbf{h}_{n}(t)\|_{2}}\sqrt{\max\Big\{0, \frac{1}{\rho^{\dag}\ln2} - \frac{\sigma^{2}}{\|\mathbf{h}_{n}(t)\|_{2}^{2}}\Big\}}.
\end{align}


\section{Adaptive Video Streaming In Multi-user Scenario}\label{section3}
In this section, we consider the multi-user scenario, i.e., $K>1$. First, we elaborate on the transmission scheme. Then, we formulate a utility maximization problem and separate it into multiple tractable problems for each case. Finally, we solve the problems.
\subsection{Transmission Scheme}
We consider a rate splitting scheme \cite{JSAC,TCOM16}. Specifically, for all $k\in\mathcal{K}$, the aggregated message for user $k$ sent at slot $t$ is split into a common part of rate $d_{c,k}(t)$ and a private part of rate $d_{p,k}(t)$. When the encoding rate constraints for the GOP:
\begin{equation}
\sum\nolimits_{(x,y)\in\overline{\mathcal{F}}_{k}}R_{x,y,k} \leq \frac{1}{T}\sum_{t\in\mathcal{T}}\left(d_{c,k}(t) + d_{p,k}(t)\right),~k\in\mathcal{K}.\label{sum_rate_constraint}
\end{equation}
are satisfied, rebuffering for the considered GOP can be avoided.

Further, the common parts of the messages of the $K$ users are combined into a common message of rate $\sum\nolimits_{k\in\mathcal{K}}d_{c,k}(t)$. The private part of user $k$'s message is also referred to as user $k$'s private message. The common message and the $K$ users' private messages are then encoded (channel coding) into codewords that span over $N$ subcarriers, respectively. Let $s_{c,n}$ and $s_{k,n}$ denote a symbol of the common message and a symbol of user $k$'s private message, which are transmitted on the $n$-th subcarrier at slot $t$, respectively. For notation simplicity, define $\overline{\mathcal{K}} \triangleq \mathcal{K} \cup \{c\}$. Let $\mathbf{s}_{n} \triangleq (s_{k,n})_{k\in\overline{\mathcal{K}}}$ and assume that $\mathbb{E}[\mathbf{s}_{n}\mathbf{s}^{H}_{n}] = \mathbf{I},$ $n\in\mathcal{N}$. We consider linear precoding on each subcarrier. The transmitted signal on subcarrier $n$ at slot $t$ is given by:
\begin{equation}
\mathbf{x}_{n}(t) = \mathbf{w}_{c,n}(t)s_{c,n} + \sum\limits_{k\in\mathcal{K}}\mathbf{w}_{k,n}(t)s_{k,n},~n\in\mathcal{N},t\in\mathcal{T},\label{x_multi}
\end{equation}
where $\mathbf{w}_{c,n}(t)\in\mathbb{C}^{M\times1}$ and $\mathbf{w}_{k,n}(t)\in\mathbb{C}^{M\times1}$ are the common beamforming vector on subcarrier $n$ at slot $t$ and the private beamforming vector for user $k$ on subcarrier $n$ at slot $t$, respectively. Substituting \eqref{x_multi} into \eqref{x_general}, we have the total transmission power constraint:
\begin{equation}
\sum\nolimits_{n\in\mathcal{N}}\left(\|\mathbf{w}_{c,n}(t)\|_{2}^{2} + \sum\nolimits_{k\in\mathcal{K}}\|\mathbf{w}_{k,n}(t)\|_{2}^{2}\right) \leq P,~t\in\mathcal{T}.\label{rs_power_constraint}
\end{equation}

We consider successive decoding at each user. Specifically, the decoding procedure for user $k\in\mathcal{K}$ is as follows. First, user $k$ decodes the common message by treating the interference from the $K$ users' private messages on each subcarrier as noise. After successfully decoding and removing the common message, user $k$ decodes his private message by treating the interference from the remaining $K-1$ users' private messages on each subcarrier as noise. Substituting \eqref{x_multi} into \eqref{received_signal}, we can derive the Signal to Interference plus Noise Ratios (SINRs) of the common message and user $k$'s private message on subcarrier $n$ at slot $t$, i.e., $\frac{|\mathbf{h}_{k,n}^{H}(t)\mathbf{w}_{c,n}(t)|^{2}}{\sum\nolimits_{j\in\mathcal{K}}|\mathbf{h}_{{k,n}}^{H}(t)\mathbf{w}_{j,n}(t)|^{2} + \sigma^{2}}$ and $\frac{|\mathbf{h}_{{k,n}}^{H}(t)\mathbf{w}_{{k,n}}(t)|^{2}}{\sum\nolimits_{j\in\mathcal{K},j\not=k}|\mathbf{h}_{{k,n}}^{H}(t)\mathbf{w}_{j,n}(t)|^{2} + \sigma^{2}}$, respectively. We consider Gaussian coding\cite{INFOR03,JSAC,TCOM16}. We have the following transmission rate constraints:
\begin{align}
&\sum\nolimits_{k\in\mathcal{K}}d_{c,k}(t) \nonumber\\
& \leq \sum\limits_{n\in\mathcal{N}}B{\rm log}_2\left(1+\frac{|\mathbf{h}_{{k,n}}^{H}(t)\mathbf{w}_{c,n}(t)|^{2}}{\sum\nolimits_{j\in\mathcal{K}}|\mathbf{h}_{{k,n}}^{H}(t)\mathbf{w}_{j,n}(t)|^{2} + \sigma^{2}}\right),\nonumber\\
&~~~~~~~~~~~~~~~~~~~~~~~~~~~~~~~~~~~~~~~~~~~k\in\mathcal{K},~t\in\mathcal{T},\label{conmon_rate_constraint}
\end{align}
\begin{align}
&d_{p,k}(t) \nonumber\\
&\leq\sum\limits_{n\in\mathcal{N}}B{\rm log}_2\left(1+\frac{|\mathbf{h}_{{k,n}}^{H}(t)\mathbf{w}_{{k,n}}(t)|^{2}}{\sum\nolimits_{j\in\mathcal{K},j\not=k}|\mathbf{h}_{{k,n}}^{H}(t)\mathbf{w}_{j,n}(t)|^{2} + \sigma^{2}}\right),\nonumber\\
&~~~~~~~~~~~~~~~~~~~~~~~~~~~~~~~~~~~~~~~~~~~~~~k\in\mathcal{K},~t\in\mathcal{T}.\label{private_rate_constraint}
\end{align}

\subsection{Problem Formulation}\label{multiuser_problem_formulation}
We aim to maximize the video quality and avoid rebuffering meanwhile. Toward this end, in the multi-user scenario,
we optimize the encoding rates of the tiles $\mathbf{R} \triangleq (R_{x,y,k})_{(x,y)\in\overline{\mathcal{F}}_{k},k\in\mathcal{K}}$, encoding rates of the FoVs $\mathbf{r} \triangleq (r_{i,k})_{i\in\mathcal{I}_{k},k\in\mathcal{K}}$, rates of the common and private messages $\mathbf{d}(t)\triangleq (d_{c,k}(t),d_{p,k}(t))_{k\in\mathcal{K}},t\in\mathcal{T}$, and transmission beamforming vectors $\mathbf{w}(t)\triangleq (\mathbf{w}_{k,n}(t))_{k\in\overline{\mathcal{K}},n\in\mathcal{N}},t\in\mathcal{T}$ to maximize the performance metrics in \eqref{utility} subject to the constraints in \eqref{tile_rate_max}, \eqref{rate_smooth}, \eqref{fov_rate_max}, \eqref{sum_rate_constraint}, \eqref{rs_power_constraint}, \eqref{conmon_rate_constraint}, \eqref{private_rate_constraint}. Similarly, for tractability, we replace the discrete constraints in \eqref{tile_rate_max} and \eqref{fov_rate_max} with the following continuous constraints:
\begin{align}
&0 \leq R_{x,y,k} \leq D_{L},~(x,y)\in\overline{\mathcal{F}}_{k},~k\in\mathcal{K},\label{multi_tile_relax}\\
&0 \leq r_{i,k} \leq D_{L},~i\in\mathcal{I}_{k},~k\in\mathcal{K},\label{multi_fov_relax}
\end{align}
and consider the following relaxed version of the original discrete optimization problem.
\begin{Prob}[Total Utility Maximization in Multi-user Scenario]\label{rs_case_general_new} For $\phi$ = \text{pp},~\text{ip},~\text{up},
\begin{align}
\max_{\mathbf{R},\mathbf{r},\mathbf{d}(t),\mathbf{w}(t),t\in\mathcal{T}}\quad&Q^{(\phi)}(\mathbf{r})\nonumber\\
    \mathrm{s.t.}\quad&\eqref{rate_smooth},~\eqref{sum_rate_constraint},~\eqref{rs_power_constraint},~\eqref{conmon_rate_constraint},~\eqref{private_rate_constraint},~\eqref{multi_tile_relax},~\eqref{multi_fov_relax}.\nonumber
\end{align}
\end{Prob}

Analogously, based on any feasible solution of Problem \ref{rs_case_general_new}, denoted by $(\mathbf{R}^{(\phi)},\mathbf{r}^{(\phi)})$, we can construct feasible discrete encoding rates of the tiles and the FoVs, denoted by $(\widetilde{\mathbf{R}}^{(\phi)},\widetilde{\mathbf{r}}^{(\phi)})$, where $\widetilde{R}_{x,y,k}^{(\phi)} =\max\{d\in\mathcal{D}|d\leq R^{(\phi)}_{x,y,k}\},(x,y)\in\overline{\mathcal{F}}_{k}, k\in\mathcal{K}$ and $\widetilde{r}_{i,k}^{(\phi)} = \max\{d\in\mathcal{D}|d\leq r^{(\phi)}_{i,k}\}$, $i\in\mathcal{I}_{k},k\in\mathcal{K}$. The performance loss induced by the continuous relaxation is acceptable when $D_{2}-D_{1},\ldots,D_{L}-D_{L-1}$ are small, which will be shown in Section \ref{simulation_multi}.

Similarly, in practice, Problem \ref{rs_case_general_new} cannot be solved at the beginning of the first slot of the considered GOP, and hence we also separate Problem \ref{rs_case_general_new} into $T$ optimization problems, as in the single-user scenario.
In particular, we introduce the encoding rate constraints for slot $1$:
\begin{align}
\sum\nolimits_{(x,y)\in\overline{\mathcal{F}}_{k}}R_{x,y,k} \leq d_{c,k}(1) + d_{p,k}(1),~k\in\mathcal{K}.\label{sum_rate_constraint_first}
\end{align}
The optimization problem for the encoding rate adaptation of the GOP and the transmission adaptation of slot 1 is as follows.\footnote{In Problem \ref{rs_case_general}, the constraints in \eqref{rs_power_constraint},~\eqref{conmon_rate_constraint},~\eqref{private_rate_constraint} are only for slot $1$.}

\begin{Prob}[Total Utility Maximization at $t= 1$ in Multi-user Scenario]\label{rs_case_general}For $\phi$ = \text{pp},~\text{ip},~\text{up},
\begin{align}
U^{(\phi)\star} \triangleq & \max_{\mathbf{R},\mathbf{r},\mathbf{d}(1),\mathbf{w}(1)}\quad Q^{(\phi)}(\mathbf{r})\nonumber\\
    \mathrm{s.t.}\quad&\eqref{rate_smooth},~\eqref{rs_power_constraint},~\eqref{conmon_rate_constraint},~\eqref{private_rate_constraint},~\eqref{multi_tile_relax},~\eqref{multi_fov_relax},~\eqref{sum_rate_constraint_first}.\nonumber
\end{align}
Let $(\mathbf{R}^{(\phi)\star},\mathbf{r}^{(\phi)\star},\mathbf{d}(1)^{(\phi)\star},\mathbf{w}(1)^{(\phi)\star})$ denote an optimal solution of Problem \ref{rs_case_general} for $\phi$ = pp, ip, up.
\end{Prob}

With the knowledge of the channel condition, $\mathbf{h}^{H}_{k,n}(1),k\in\mathcal{K},n\in\mathcal{N}$, Problem \ref{rs_case_general} can be solved at the beginning of slot $1$. The objective function in \eqref{utility} is concave, the constraints in \eqref{rate_smooth}, \eqref{rs_power_constraint}, \eqref{multi_tile_relax}, \eqref{multi_fov_relax}, and \eqref{sum_rate_constraint_first} are convex, and the constraints in \eqref{conmon_rate_constraint} and \eqref{private_rate_constraint} are nonconvex. Therefore, Problem \ref{rs_case_general} is a nonconvex problem. There are generally no effective methods for solving a nonconvex problem optimally. The goal of solving a nonconvex problem is usually to design an iterative algorithm to obtain a stationary point or a KKT point (which satisfies necessary conditions for optimality if strong duality holds)\cite{nonlinear}. Let $(\mathbf{R}^{(\phi)\dag},\mathbf{r}^{(\phi)\dag},\mathbf{d}(1)^{(\phi)\dag},\mathbf{w}(1)^{(\phi)\dag})$ denote a KKT point of Problem \ref{rs_case_general}. The method for obtaining it will be introduced in Section \ref{multiuser_solution}.

Based on $\mathbf{R}^{(\phi)\dag}$, the optimization problem for the transmission adaptation of each subsequent slot is as follows.\footnote{In Problem \ref{multi_infeasibility_min}, the constraints in \eqref{rs_power_constraint},~\eqref{conmon_rate_constraint},~\eqref{private_rate_constraint} are only for the considered slot $t$.}

\begin{Prob}[Sum of Infeasibilities Minimization at $t = 2,\ldots,T$ in Multi-user Scenario]\label{multi_infeasibility_min}For all $t=2,\ldots,T$ and for $\phi$ = pp, ip, up,
\begin{align}
&\min_{\mathbf{d}(t),\mathbf{w}(t),\mathbf{s}(t)\succeq 0}\quad \sum_{k\in\mathcal{K}}s_{k}(t)\nonumber\\
    &\mathrm{s.t.}~\eqref{rs_power_constraint},~\eqref{conmon_rate_constraint},~\eqref{private_rate_constraint},\nonumber\\
    &\sum\limits_{(x,y)\in\overline{\mathcal{F}}_{k}}R^{(\phi)\dag}_{x,y,k}- \left(d_{c,k}(t) + d_{p,k}(t)\right)\leq s_{k}(t),~k\in\mathcal{K},\label{multi_vio_constraint}
\end{align}
where $\mathbf{s}(t) \triangleq (s_{k}(t))_{k\in\mathcal{K}}$. 
\end{Prob}

Note that Problem \ref{multi_infeasibility_min} relies on $\mathbf{R}^{(\phi)\dag}$, implying that the transmission adaptations of each subsequent slot in the three cases are different. This differs from the single-user scenario. In Problem \ref{multi_infeasibility_min}, $\mathbf{s}(t)$ can be interpreted as upper bounds on the maximum infeasibilities of the encoding rate constraints for slot $t$:
\begin{align}
\sum\nolimits_{(x,y)\in\overline{\mathcal{F}}_{k}}R^{(\phi)}_{x,y,k} \leq d_{c,k}(t) + d_{p,k}(t),~k\in\mathcal{K}\label{multi_vio_constraint_new}.
\end{align}
The goal of Problem \ref{multi_infeasibility_min} is to drive the sum of infeasibilities of \eqref{multi_vio_constraint_new} to zero \cite[pp. 580]{boyd2004convex}. Note that the objective function of Problem \ref{multi_infeasibility_min} and the constraints in \eqref{multi_vio_constraint} are linear, the constraint in \eqref{rs_power_constraint} is convex, and the constraints in \eqref{conmon_rate_constraint} and \eqref{private_rate_constraint} are nonconvex. Thus, Problem \ref{multi_infeasibility_min} is nonconvex.

\begin{figure*}[ht]
\small{\begin{align}
&L_{{k,n}}(\mathbf{w}_{n}(1),u_{c,n}(1);\tilde{\mathbf{w}}_{n}(1),\tilde{u}_{c,n}(1)) \triangleq 
\sum\limits_{j\in\mathcal{K}}|\mathbf{h}_{{k,n}}^{H}(1)\mathbf{w}_{j,n}(1)|^{2} + \sigma^{2} - \frac{2\Re\{\sum\nolimits_{j\in\overline{\mathcal{K}}}\tilde{\mathbf{w}}_{j,n}^{H}(1)\mathbf{h}_{{k,n}}(1)\mathbf{h}_{{k,n}}^{H}(1)\mathbf{w}_{j,n}(1)\} + 2\sigma^{2}}{\tilde{u}_{c,n}(1)} \nonumber\\
&~~~~~~~~~~~~~~~~~~~~~~~~~~~~~~~~~~~~~~~~~~~~~~~~~~~~~~~~~~~~~~~~~~~~~~~~~~~~~~~~~+\frac{\left(\sum\nolimits_{j\in\overline{\mathcal{K}}}|\mathbf{h}_{{k,n}}^{H}(1)\tilde{\mathbf{w}}_{j,n}(1)|^{2} + \sigma^{2}\right)u_{c,n}(1)}{(\tilde{u}_{c,n}(1))^{2}},~k\in\mathcal{K},~n\in\mathcal{N},\label{dc_approximate_common}\\
&G_{{k,n}}(\mathbf{w}_{n}(1),u_{k,n}(1);\tilde{\mathbf{w}}_{n}(1),\tilde{u}_{k,n}(1)) \triangleq 
\sum\limits_{\substack{j\in\mathcal{K},j\not=k}}|\mathbf{h}_{{k,n}}^{H}(1)\mathbf{w}_{j,n}(1)|^{2} + \sigma^{2} - \frac{2\Re\{\sum\nolimits_{j\in\mathcal{K}}\tilde{\mathbf{w}}_{j,n}^{H}(1)\mathbf{h}_{{k,n}}(1)\mathbf{h}_{{k,n}}^{H}(1)\mathbf{w}_{j,n}(1)\} + 2\sigma^{2}}{\tilde{u}_{k,n}(1)} \nonumber\\
&~~~~~~~~~~~~~~~~~~~~~~~~~~~~~~~~~~~~~~~~~~~~~~~~~~~~~~~~~~~~~~~~~~~~~~~~~~~~~~~~+\frac{\left(\sum\nolimits_{j\in\mathcal{K}}|\mathbf{h}_{{k,n}}^{H}(1)\tilde{\mathbf{w}}_{j,n}(1)|^{2} + \sigma^{2}\right)u_{k,n}(1)}{(\tilde{u}_{k,n}(1))^{2}},~k\in\mathcal{K},~n\in\mathcal{N}.\label{dc_approximate_private}
\end{align}
}
\hrulefill
\end{figure*}

\begin{Rem}[Interpretation of Separate Approach in Multi-user Scenario]The encoding rate constraints for slot $1$ in \eqref{sum_rate_constraint_first} of Problem \ref{rs_case_general} together with Problem \ref{multi_infeasibility_min} for $t=2,\ldots,T$ are to reduce the sum of infeasibilities of the encoding rate constraints for the GOP in \eqref{sum_rate_constraint} of Problem \ref{rs_case_general_new}. As illustrated in Remark 1, when $N$ is large (which is usually the case in practice), the encoding rate adaptation of the GOP and the transmission adaptation of each slot offered by the separate approach are close to those offered by solving Problem \ref{rs_case_general_new} and will not violate the encoding rate constraints for each user $k$ in \eqref{sum_rate_constraint} that much. The violation if exists yields rebuffering for each user $k$. Later in Section \ref{simulation_multi}, we shall see that with the separate approach, we can improve video quality while keeping the rebuffering time small, i.e., approximately achieving the goal of Problem \ref{rs_case_general_new}.
\end{Rem}

\subsection{Solution}\label{multiuser_solution}
\subsubsection{KKT Point of Problem \ref{rs_case_general}}Problem \ref{rs_case_general} is noncovex due to the nonconvexities of the constraints in \eqref{conmon_rate_constraint} and \eqref{private_rate_constraint}. Besides, Problem \ref{rs_case_general} with $\phi$ = ip and Problem \ref{rs_case_general} with $\phi$ = up have non-differentiable objective functions. 
Although it is difficult to obtain a globally optimal solution of the nonconvex problem in Problem \ref{rs_case_general}, we can characterize its optimality properties. For all $i\in\mathcal{I}_{k},k\in\mathcal{K},$ define $\mathcal{T}_{i,k} \triangleq \mathcal{F}_{i}\backslash\overline{\mathcal{F}}_{k}$. Note that $\mathcal{T}_{i,k} \cap \mathcal{T}_{j,k} = \emptyset,$ for all $i,j\in\mathcal{I}_{k},i\not=j,k\in\mathcal{K}$.
\begin{Thm}[Optimality Properties of Problem \ref{rs_case_general}]\label{theorem}
(\romannumeral1) For $\phi$ = pp, ip, up, $R^{(\phi)\star}_{x,y,k} = \max\limits_{i\in\mathcal{I}_{k}: (x,y)\in\mathcal{F}_{i}}r^{(\phi)\star}_{i,k},(x,y)\in\overline{\mathcal{F}}_{k},~k\in\mathcal{K}$. (\romannumeral2) For $\phi$ = pp and for all $i,j\in\mathcal{I}_{k},i\not=j,k\in\mathcal{K}$, if $p_{i,k} \leq p_{j,k}$ and $|\mathcal{T}_{i,k}| > |\mathcal{T}_{j,k}| > 1$, then $r^{(\text{pp})\star}_{i,k} \leq r^{(\text{pp})\star}_{j,k}$. For $\phi$ = ip and for all $i,j\in\mathcal{I}_{k},i\not=j,k\in\mathcal{K}$, if $\overline{p}_{i,k} \leq \underline{p}_{j,k}$ and $|\mathcal{T}_{i,k}| > |\mathcal{T}_{j,k}| > 1$, then $r^{(\text{ip})\star}_{i,k} \leq r^{(\text{ip})\star}_{j,k}$. For $\phi$ = up and for all $k\in\mathcal{K}$, $r^{\text{(up)}\star}_{i,k},i\in\mathcal{I}_{k}$ are identical. (\romannumeral3) $U^{(\text{pp})\star} \geq U^{(\text{ip})\star} \geq U^{(\text{up})\star}$.
\end{Thm}

\begin{Proof}
The proof of Theorem \ref{theorem} is similar to that of Theorem \ref{single_theorem}, and is omitted due to page limitation.
\end{Proof}

Theorem \ref{theorem} extends Theorem \ref{single_theorem} to the multi-user scenario and can be interpreted similarly.
In the following, we obtain a KKT point of Problem \ref{rs_case_general} using CCCP, which can exploit the partial convexity and usually converges faster than conventional gradient methods.
First, we address the challenge caused by the nonconvexities of the constraints in \eqref{conmon_rate_constraint} and \eqref{private_rate_constraint}. By introducing auxiliary variables and extra constraints, we can equivalently transform Problem \ref{rs_case_general} into the following problem.
\begin{Prob}[Equivalent Problem of Problem \ref{rs_case_general}]\label{rs_case_general_equal}For $\phi$ = \text{pp},~\text{ip},~\text{up},
\begin{align}
&\max_{\mathbf{R},\mathbf{r},\mathbf{d}(1),\mathbf{e}(1),\mathbf{u}(1),\mathbf{w}(1)}\quad Q^{(\phi)}(\mathbf{r})\nonumber\\
    &\mathrm{s.t.}\quad\eqref{rate_smooth},~\eqref{sum_rate_constraint},~\eqref{rs_power_constraint},~\eqref{multi_tile_relax},~\eqref{multi_fov_relax},\nonumber\\
    &\sum\nolimits_{k\in\mathcal{K}}d_{c,k}(1) \leq \sum\nolimits_{n\in\mathcal{N}}e_{c,n}(1),\label{common_sum_rate}\\
    &d_{p,k}(1) \leq \sum\nolimits_{n\in\mathcal{N}}e_{k,n}(1).~k\in\mathcal{K},\label{private_sum_rate}\\
    &\sum\limits_{j\in\mathcal{K}}|\mathbf{h}_{{k,n}}^{H}(1)\mathbf{w}_{j,n}(1)|^{2} + \sigma^{2}  \nonumber\\
    &-\frac{\sum\limits_{j\in\overline{\mathcal{K}}}|\mathbf{h}_{{k,n}}^{H}(1)\mathbf{w}_{j,n}(1)|^{2} + \sigma^{2}}{u_{c,n}(1)} \leq 0,~k\in\mathcal{K},~n\in\mathcal{N}, \label{dc_common}
      \end{align}
    \end{Prob}
  \begin{align}
    &\sum\limits_{j\in\mathcal{K},j\not=k}|\mathbf{h}_{{k,n}}^{H}(1)\mathbf{w}_{j,n}(1)|^{2} + \sigma^{2} \nonumber\\
    &-\frac{\sum\limits_{j\in\mathcal{K}}|\mathbf{h}_{{k,n}}^{H}(1)\mathbf{w}_{j,n}(1)|^{2} + \sigma^{2}}{u_{{k,n}}(1)} \leq 0,~k\in\mathcal{K},~n\in\mathcal{N},\label{dc_private}\\
    &2^{\frac{e_{k,n}(1)}{B}} \leq u_{{k,n}}(1),~k\in\overline{\mathcal{K}},~n\in\mathcal{N},\label{dc_variable_private}
\end{align}
where $\mathbf{e}(1)\triangleq (e_{k,n}(1))_{k\in\overline{\mathcal{K}},n\in\mathcal{N}}$, $\mathbf{u}(1)\triangleq (u_{k,n}(1))_{k\in\overline{\mathcal{K}},n\in\mathcal{N}}$. Let $(\mathbf{R}^{(\phi)\dag},\mathbf{r}^{(\phi)\dag},\mathbf{d}^{(\phi)\dag}(1),\mathbf{e}^{(\phi)\dag}(1),\mathbf{u}^{(\phi)\dag}(1),\mathbf{w}^{(\phi)\dag}(1))$ denote an optimal solution of Problem \ref{rs_case_general_equal}.

Note that ($\mathbf{e}(1),\mathbf{u}(1)$) are auxiliary variables, and \eqref{common_sum_rate}, \eqref{private_sum_rate}, \eqref{dc_common}, \eqref{dc_private}, \eqref{dc_variable_private} are extra constraints. By contradiction, we can easily show that the constraints in \eqref{dc_common}, \eqref{dc_private}, and \eqref{dc_variable_private} are active at an optimal solution. Therefore, it is obvious that Problem \ref{rs_case_general} and Problem \ref{rs_case_general_equal} are equivalent. Furthermore, notice that the constraints in \eqref{common_sum_rate} and \eqref{private_sum_rate} are convex w.r.t $(\mathbf{d}(1),\mathbf{e}(1))$, the constraints in \eqref{dc_variable_private} are convex w.r.t $(\mathbf{e}(1),\mathbf{u}(1))$, and each constraint function in (\ref{dc_common}) and (\ref{dc_private}) can be regarded as a difference of two convex functions w.r.t. $(\mathbf{w}(1),\mathbf{u}(1))$. Therefore, Problem \ref{rs_case_general_equal} is a difference of convex functions (DC) programming (one type of nonconvex problems). A KKT point of Problem \ref{rs_case_general_equal} can be obtained by CCCP\cite{TSP17}. The main idea is to solve a sequence of successively refined approximate convex problems, each of which is obtained by linearizing $\frac{\sum\nolimits_{j\in\overline{\mathcal{K}}}|\mathbf{h}_{{k,n}}^{H}(1)\mathbf{w}_{j,n}(1)|^{2} + \sigma^{2}}{u_{c,n}(1)}$ and $\frac{\sum\nolimits_{j\in\mathcal{K}}|\mathbf{h}_{{k,n}}^{H}(1)\mathbf{w}_{j,n}(1)|^{2} + \sigma^{2}}{u_{{k,n}}(1)}$ in \eqref{dc_common} and \eqref{dc_private}, respectively, and preserving the remaining convexity of Problem \ref{rs_case_general_equal}. Specifically, the convex approximations of the constraints in \eqref{dc_common} and \eqref{dc_private} at $(\tilde{\mathbf{w}}_{n}(1),\tilde{\mathbf{u}}(1))$ are given by:
\begin{align}
L_{{k,n}}(\mathbf{w}_{n}(1),u_{c,n}(1);\tilde{\mathbf{w}}_{n}(1),\tilde{u}_{c,n}(1)) \leq 0,~k\in\mathcal{K},~n\in\mathcal{N},\\
G_{{k,n}}(\mathbf{w}_{n}(1),u_{k,n}(1);\tilde{\mathbf{w}}_{n}(1),\tilde{u}_{k,n}(1)) \leq 0,~k\in\mathcal{K},~n\in\mathcal{N},
\end{align}
where $L_{{k,n}}(\mathbf{w}_{n}(1),u_{c,n}(1);\tilde{\mathbf{w}}_{n}(1),\tilde{u}_{c,n}(1))$, $G_{{k,n}}(\mathbf{w}_{n}(1),u_{k,n}(1);\tilde{\mathbf{w}}_{n}(1),\tilde{u}_{k,n}(1))$ are given by \eqref{dc_approximate_common} and \eqref{dc_approximate_private}, respectively, as shown at the top of this page.

In the following, we present the approximated convex problem at each iteration for each of the three cases of FoV viewing probability distributions.

\textbf{Perfect FoV viewing probability distributions:} The approximate convex problem of Problem \ref{rs_case_general_equal} with $\phi$ = pp at the $i$-th iteration is given by:
\begin{Prob}[Convex Approximation of Problem \ref{rs_case_general_equal} with $\phi$ = pp at $i$-th Iteration]\label{rs_one_DC_convex}
\begin{align}
&\max_{\mathbf{R},\mathbf{r},\mathbf{d}(1),\mathbf{e}(1),\mathbf{u}(1),\mathbf{w}(1)}\quad Q^{(\text{pp})}(\mathbf{r}) \nonumber\\
    &\mathrm{s.t.}\quad \eqref{rate_smooth},~\eqref{sum_rate_constraint},~\eqref{rs_power_constraint},~\eqref{multi_tile_relax},~\eqref{multi_fov_relax},~\eqref{common_sum_rate},~\eqref{private_sum_rate},~\eqref{dc_variable_private},\nonumber\\
    &L_{{k,n}}(\mathbf{w}_{n}(1),u_{c,n}(1);\mathbf{w}_{n}^{(\text{pp})(i-1)}(1),u_{c,n}^{(\text{pp})(i-1)}(1)) \leq 0,\nonumber\\
    &~~~~~~~~~~~~~~~~~~~~~~~~~~~~~~~~~~~~~~~~~~~k\in\mathcal{K},~n\in\mathcal{N},\nonumber\\
  &G_{{k,n}}(\mathbf{w}_{n}(1),u_{k,n}(1);\mathbf{w}_{n}^{(\text{pp})(i-1)}(1),u_{k,n}^{(\text{pp})(i-1)}(1)) \leq 0,\nonumber\\
  &~~~~~~~~~~~~~~~~~~~~~~~~~~~~~~~~~~~~~~~~~~~k\in\mathcal{K},~n\in\mathcal{N}.\nonumber
\end{align}
Let $(\mathbf{R}^{(\text{pp})(i)},\mathbf{r}^{(\text{pp})(i)},\mathbf{d}^{(\text{pp})(i)}(1),\mathbf{e}^{(\text{pp})(i)}(1),\mathbf{u}^{(\text{pp})(i)}(1),\\\mathbf{w}^{(\text{pp})(i)}(1))$ denote an optimal solution of Problem \ref{rs_one_DC_convex}.
\end{Prob}

\textbf{Imperfect FoV viewing probability distributions:}
$Q^{(\text{ip})}(\mathbf{r})$ is non-differentiable, and Problem \ref{rs_case_general_equal} with $\phi$ = ip is a max-min problem w.r.t. $(\mathbf{R},\mathbf{r},\mathbf{d}(1),\mathbf{e}(1),\mathbf{u}(1),\mathbf{w}(1),\mathbf{p})$. As in the single-user scenario, we transform Problem \ref{rs_case_general_equal} with $\phi$ = ip into an equivalent DC programming with a differentiable objective function by replacing the inner problem with its dual problem\cite{Ye2019Optimal}. 

\begin{Prob}[Equivalent Problem of Problem \ref{rs_case_general_equal} with $\phi$ = ip]\label{rs_case_two_equal}
\begin{align}
&\max_{\substack{\mathbf{R},\mathbf{r},\mathbf{d}(1),\mathbf{e}(1),\mathbf{u}(1),\mathbf{w}(1),\\\bm{\lambda} \succeq 0, \bm{\tau} \succeq 0, \bm{\gamma}}}\sum\limits_{k\in\mathcal{K}}\left(\sum\limits_{i\in\mathcal{I}_{k}}(\tau_{i,k}\underline{p}_{i,k} - \lambda_{i,k}\overline{p}_{i,k}) - \gamma_{k}\right) \nonumber\\
    &\mathrm{s.t.}~\eqref{rate_smooth},~\eqref{sum_rate_constraint},~\eqref{rs_power_constraint},~\eqref{multi_tile_relax},~\eqref{multi_fov_relax},~\eqref{common_sum_rate},~\eqref{private_sum_rate},~\eqref{dc_common},~\eqref{dc_private},~\eqref{dc_variable_private},\nonumber\\
    &U(r_{i,k}) + \lambda_{i,k} - \tau_{i,k} + \gamma_{k} \geq 0, ~i\in\mathcal{I}_{k},~k\in\mathcal{K},\label{dual_constraint}
\end{align}
where $\bm{\lambda}\triangleq (\lambda_{i,k})_{i\in\mathcal{I}_{k},k\in\mathcal{K}}$, $\bm{\tau}\triangleq (\tau_{i,k})_{i\in\mathcal{I}_{k},k\in\mathcal{K}}$, and $\bm{\gamma}\triangleq (\gamma_{k})_{k\in\mathcal{K}}$. Let $(\mathbf{R}^{(\text{ip})\dag},\mathbf{r}^{(\text{ip})\dag},\mathbf{d}^{(\text{ip})\dag}(1),\mathbf{e}^{(\text{ip})\dag}(1),\mathbf{u}^{(\text{ip})\dag}(1),\mathbf{w}^{(\text{ip})\dag}(1),\bm{\lambda}^{(\text{ip})\dag},\\\bm{\tau}^{(\text{ip})\dag},\bm{\gamma}^{(\text{ip})\dag})$ denote an optimal solution of Problem \ref{rs_case_two_equal}.
\end{Prob}

Note that $\lambda_{i,k}$, $\tau_{i,k}$, and $\gamma_{i,k}$ are dual variables of the inner problem, corresponding to $p_{i,k} \leq \overline{p}_{i,k}$, $p_{i,k} \geq \underline{p}_{i,k}$, and $\sum_{i\in\mathcal{I}_{k}}p_{i,k} = 1$, respectively. The equivalence between Problem \ref{rs_case_general_equal} with $\phi$ = ip and Problem \ref{rs_case_two_equal} is summarized below. 
\begin{Thm}[Equivalence between Problem \ref{rs_case_general_equal} with $\phi$ = ip and Problem \ref{rs_case_two_equal}]\label{lemma_rs_case2}$(\mathbf{R}^{(\text{ip})\dag},\mathbf{r}^{(\text{ip})\dag},\mathbf{d}^{(\text{ip})\dag}(1),\mathbf{e}^{(\text{ip})\dag}(1),\mathbf{u}^{(\text{ip})\dag}(1),\mathbf{w}^{(\text{ip})\dag}(1))$ is an optimal solution of Problem \ref{rs_case_general_equal} with $\phi$ = ip.
\end{Thm}
\begin{Proof}
The proof of Theorem \ref{lemma_rs_case2} is similar to that of Theorem \ref{lemma_single_case2}, and is omitted due to page limitation.
\end{Proof}

The approximate convex problem of Problem \ref{rs_case_two_equal} at the $i$-th iteration is given by:
\begin{Prob}[Convex Approximation of Problem \ref{rs_case_two_equal} at $i$-th Iteration]\label{rs_two_DC_convex}
\begin{align}
&\max_{\substack{\mathbf{R},\mathbf{r},\mathbf{d}(1),\mathbf{e}(1),\mathbf{u}(1),\mathbf{w}(1),\\\bm{\lambda} \succeq 0, \bm{\tau} \succeq 0, \bm{\gamma}}}\sum\limits_{k\in\mathcal{K}}\left(\sum\limits_{i\in\mathcal{I}_{k}}(\tau_{i,k}\underline{p}_{i,k} - \lambda_{i,k}\overline{p}_{i,k}) - \gamma_{k}\right) \nonumber\\
    &\mathrm{s.t.}\quad\eqref{rate_smooth},~\eqref{sum_rate_constraint},~\eqref{rs_power_constraint},~\eqref{multi_tile_relax},~\eqref{multi_fov_relax},~\eqref{common_sum_rate},~\eqref{private_sum_rate},~\eqref{dc_variable_private},~\eqref{dual_constraint},\nonumber\\
    &L_{{k,n}}(\mathbf{w}_{n}(1),u_{c,n}(1);\mathbf{w}_{n}^{(\text{ip})(i-1)}(1),u_{c,n}^{(\text{ip})(i-1)}(1)) \leq 0,\nonumber\\
    &~~~~~~~~~~~~~~~~~~~~~~~~~~~~~~~~~~~~~~~~~~~~k\in\mathcal{K},~n\in\mathcal{N},\nonumber\\
  &G_{{k,n}}(\mathbf{w}_{n}(1),u_{k,n}(1);\mathbf{w}_{n}^{(\text{ip})(i-1)}(1),u_{k,n}^{(\text{ip})(i-1)}(1)) \leq 0,\nonumber\\
  &~~~~~~~~~~~~~~~~~~~~~~~~~~~~~~~~~~~~~~~~~~~~k\in\mathcal{K},~n\in\mathcal{N}\nonumber.
\end{align}
Let $(\mathbf{R}^{(\text{ip})(i)},\mathbf{r}^{(\text{ip})(i)},\mathbf{d}^{(\text{ip})(i)}(1),\mathbf{e}^{(\text{ip})(i)}(1),\mathbf{u}^{(\text{ip})(i)}(1),\\\mathbf{w}^{(\text{ip})(i)}(1),\bm{\lambda}^{(\text{ip})(i)},\bm{\tau}^{(\text{ip})(i)},\bm{\gamma}^{(\text{ip})(i)})$ denote an optimal solution of Problem \ref{rs_two_DC_convex}.
\end{Prob}

\textbf{Unknown FoV viewing probability distributions:}
$Q_{\text{up}}(\mathbf{r})$ is non-differentiable. As in the single-user scenario, we can cast Problem \ref{rs_case_general_equal} with $\phi$ = up in hypograph form as:
\begin{Prob}[Equivalent Problem of Problem \ref{rs_case_general_equal} with $\phi$ = up]\label{rs_case_three_equal}
\begin{align}
&\max_{\substack{\mathbf{R},\mathbf{r},\mathbf{d}(1),\mathbf{e}(1),\\\mathbf{u}(1),\mathbf{w}(1), \mathbf{y} \succeq 0}}\sum\nolimits_{k\in\mathcal{K}}y_{k}\nonumber\\
    &\mathrm{s.t.}~\eqref{rate_smooth},~\eqref{sum_rate_constraint},~\eqref{rs_power_constraint},~\eqref{multi_tile_relax},~\eqref{multi_fov_relax},~\eqref{common_sum_rate},~\eqref{private_sum_rate},~\eqref{dc_common},~\eqref{dc_private},~\eqref{dc_variable_private},\nonumber\\
    &y_{k} \leq U(r_{i,k}), ~i\in\mathcal{I}_{k},~k\in\mathcal{K},\label{epigraph_p3}
\end{align}
where $\mathbf{y} \triangleq (y_{k})_{k\in\mathcal{K}}$.
\end{Prob}

The approximate convex problem of Problem \ref{rs_case_three_equal} at the $i$-th iteration is given by:
\begin{Prob}[Convex Approximation of Problem \ref{rs_case_three_equal} at $i$-th Iteration]\label{rs_three_DC_convex}
\begin{align}
&\max_{\substack{\mathbf{R},\mathbf{r},\mathbf{d}(1),\mathbf{e}(1),\\\mathbf{u}(1),\mathbf{w}(1),\mathbf{y} \succeq 0}}\sum\nolimits_{k\in\mathcal{K}}y_{k}\nonumber\\
    &\mathrm{s.t.}\quad\eqref{rate_smooth},~\eqref{sum_rate_constraint},~\eqref{rs_power_constraint},~\eqref{multi_tile_relax},~\eqref{multi_fov_relax},~\eqref{common_sum_rate},~\eqref{private_sum_rate},~\eqref{dc_variable_private},~\eqref{epigraph_p3},\nonumber\\
    &L_{{k,n}}(\mathbf{w}_{n}(1),u_{c,n}(1);\mathbf{w}_{n}^{(\text{up})(i-1)}(1),u_{c,n}^{(\text{up})(i-1)}(1)) \leq 0,\nonumber\\
    &~~~~~~~~~~~~~~~~~~~~~~~~~~~~~~~~~~~~~~~~~~~~~k\in\mathcal{K},~n\in\mathcal{N},\nonumber\\
  &G_{{k,n}}(\mathbf{w}_{n}(1),u_{k,n}(1);\mathbf{w}_{n}^{(\text{up})(i-1)}(1),u_{k,n}^{(\text{up})(i-1)}(1)) \leq 0,\nonumber\\
  &~~~~~~~~~~~~~~~~~~~~~~~~~~~~~~~~~~~~~~~~~~~~~k\in\mathcal{K},~n\in\mathcal{N}.\nonumber
\end{align}
Let $(\mathbf{R}^{(\text{up})(i)},\mathbf{r}^{(\text{up})(i)},\mathbf{d}^{(\text{up})(i)}(1),\mathbf{e}^{(\text{up})(i)}(1),\mathbf{u}^{(\text{up})(i)}(1),\\\mathbf{w}^{(\text{up})(i)}(1),\mathbf{y}^{(\text{up})(i)})$ denote an optimal solution of Problem \ref{rs_three_DC_convex}.
\end{Prob}

Note that Problem \ref{rs_one_DC_convex}, Problem \ref{rs_two_DC_convex}, and Problem \ref{rs_three_DC_convex} are convex with differentiable objective functions and constraint functions and can be solved efficiently using standard convex optimization methods \cite{boyd2004convex}. The details of CCCP for obtaining a KKT point of Problem \ref{rs_case_general_equal} with $\phi$ = pp, ip, up are summarized in Algorithm~\ref{alg:DC}. 

\begin{algorithm}[t]\small
  \setstretch{1.35}
    \caption{Obtaining a KKT Point of Problem \ref{rs_case_general_equal}}
\begin{small}
        \begin{algorithmic}[1]
           \STATE \textbf{initialization}: Choose any feasible point $\mathbf{X}^{(\phi)(0)}$ of Problem~\ref{rs_one_DC_convex}, Problem~\ref{rs_two_DC_convex}, or Problem~\ref{rs_three_DC_convex} and set $i = 1$.\\
           \STATE \textbf{repeat}
           \STATE \quad Obtain an optimal solution $\mathbf{X}^{(\phi)(i)}$ of Problem~\ref{rs_one_DC_convex}, Problem~\ref{rs_two_DC_convex}, or Problem~\ref{rs_three_DC_convex} with an interior point method.
           \STATE\quad Set $i:=i+1$
           \STATE \textbf{until} the convergence criterion $\|\mathbf{X}^{(\phi)(i)} - \mathbf{X}^{(\phi)(i-1)}\|_{2} \leq \epsilon$ is met.
    \end{algorithmic}\label{alg:DC}
    \end{small}
\end{algorithm}

\begin{claim}\label{claim_DC}
As $i \rightarrow \infty$, $\mathbf{X}^{(\phi)(i)}$ obtained by Algorithm~\ref{alg:DC} converges to a KKT point of Problem \ref{rs_case_general_equal} for all $\phi$ = pp, ip, up \cite{TSP17}.
\end{claim}

\begin{Proof}
We have shown that Problem \ref{rs_case_general_equal} is DC programming for $\phi$ = pp, ip, up, and we propose to solve it with CCCP. It has been validated in \cite{TSP17} that solving DC programming through CCCP always returns a KKT point.
\end{Proof}

By \cite{facchinei2017ghost}, we know that the number of iterations of Algorithm~\ref{alg:DC} does not scale with the problem size. Thus, the computational complexity order for Algorithm~\ref{alg:DC} is the same as that for solving Problem \ref{rs_one_DC_convex}, Problem \ref{rs_two_DC_convex}, or Problem \ref{rs_three_DC_convex} in Step 3. When an interior point method is applied, the computational complexity for solving Problem \ref{rs_one_DC_convex}, Problem \ref{rs_two_DC_convex}, or Problem \ref{rs_three_DC_convex} is $\mathcal{O}(K^{3.5}N^{3.5})$.
In practice, we can run Algorithm~\ref{alg:DC} multiple times with different feasible initial points to obtain multiple KKT points and choose the KKT point with the best objective value as a suboptimal solution. 

\subsubsection{KKT Point of Problem \ref{multi_infeasibility_min}} Problem \ref{multi_infeasibility_min} is nonconvex due to the nonconvexities of the constraints in \eqref{conmon_rate_constraint} and \eqref{private_rate_constraint}. We address the challenge for solving Problem \ref{multi_infeasibility_min} caused by the nonconvexities of the constraints in \eqref{conmon_rate_constraint} and \eqref{private_rate_constraint} using the same method for solving Problem \ref{rs_case_general}. Specifically, for all $t=2,\ldots,T$, by introducing auxiliary variables $(\mathbf{e}(t),\mathbf{u}(t))$ and extra constraints \eqref{common_sum_rate}, \eqref{private_sum_rate}, \eqref{dc_common}, \eqref{dc_private} and \eqref{dc_variable_private}, we can convert Problem \ref{multi_infeasibility_min} into a DC programming and solve it using CCCP in each case. The details are omitted due to page limitation.


\begin{figure}[t]
\begin{center}
 {\resizebox{5cm}{!}{\includegraphics{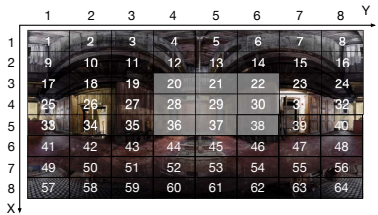}}}
\end{center}
   \caption{\small{Illustration of tiles, viewpoints and FoVs in a 360 video. The grey area represents FoV 29 which is centered at viewpoint 29.}}
   \label{example}
\end{figure}
\section{Numerical Results}
\subsection{Simulation Setup}
\begin{figure*}[t]
\begin{center}
 \subfigure[\small{Total utility versus $L$ at $\varepsilon$ = 0.4, $M$ = 8, $P$ = 30 dBm.}]
 {\resizebox{4.3cm}{!}{\includegraphics{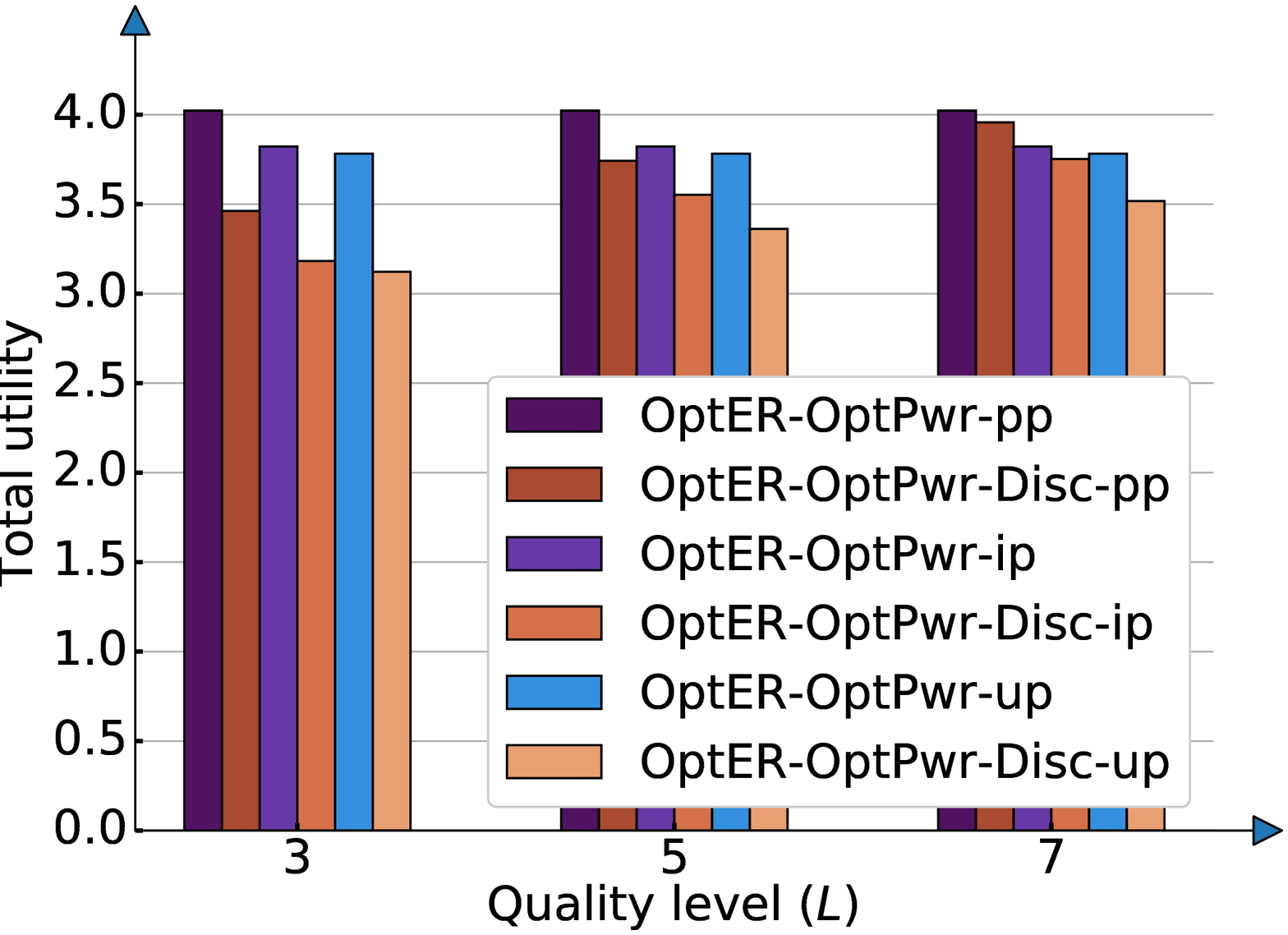}}}
  \subfigure[\small{Worst average total utility versus $\varepsilon$ at $M$ = 8, $P$ = 30 dBm.}]
 {\resizebox{4.3cm}{!}{\includegraphics{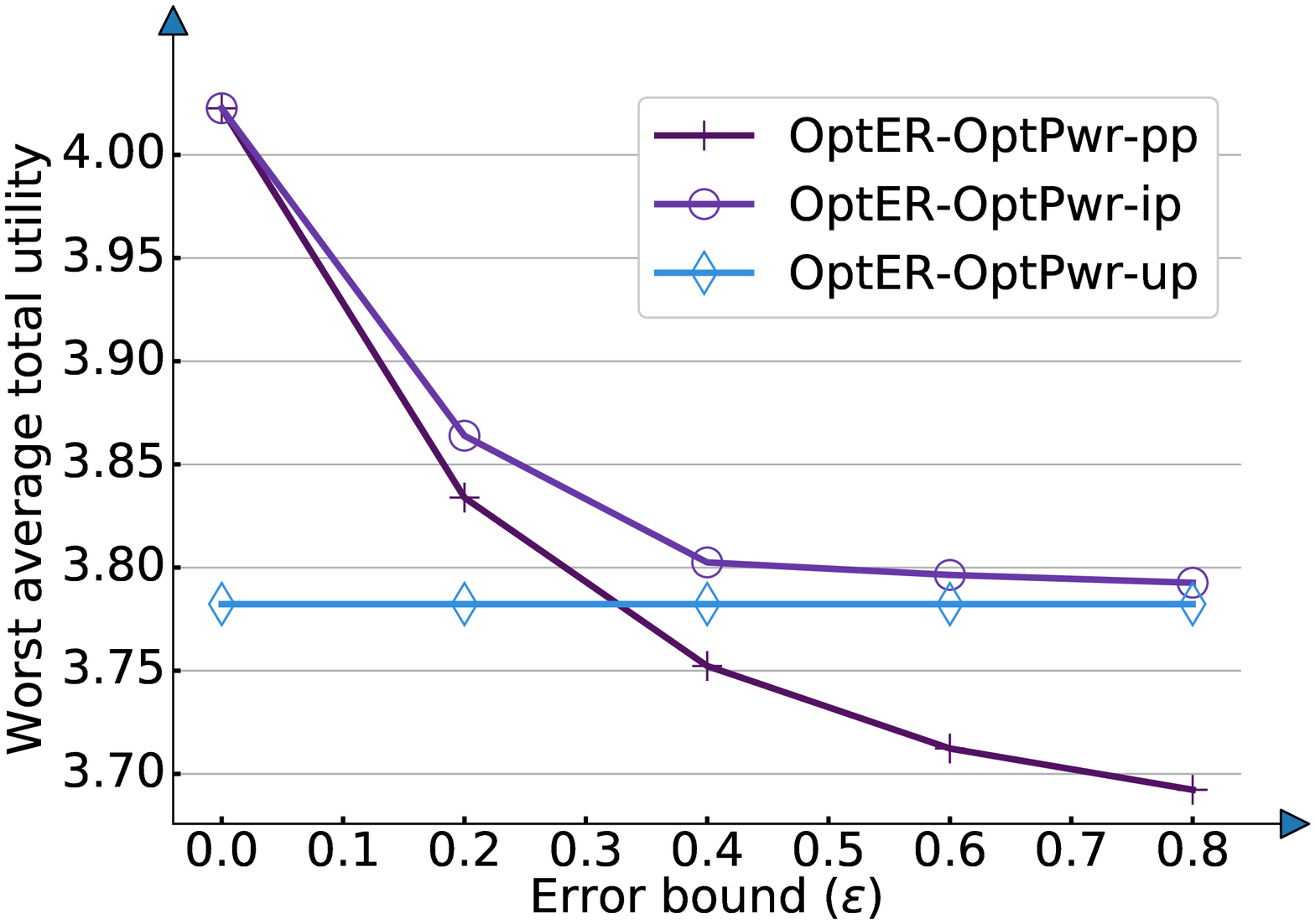}}}
 \subfigure[\small{Total utility versus $M$ at $\varepsilon$ = 0.4, $P$ = 30 dBm.}]
 {\resizebox{4.3cm}{!}{\includegraphics{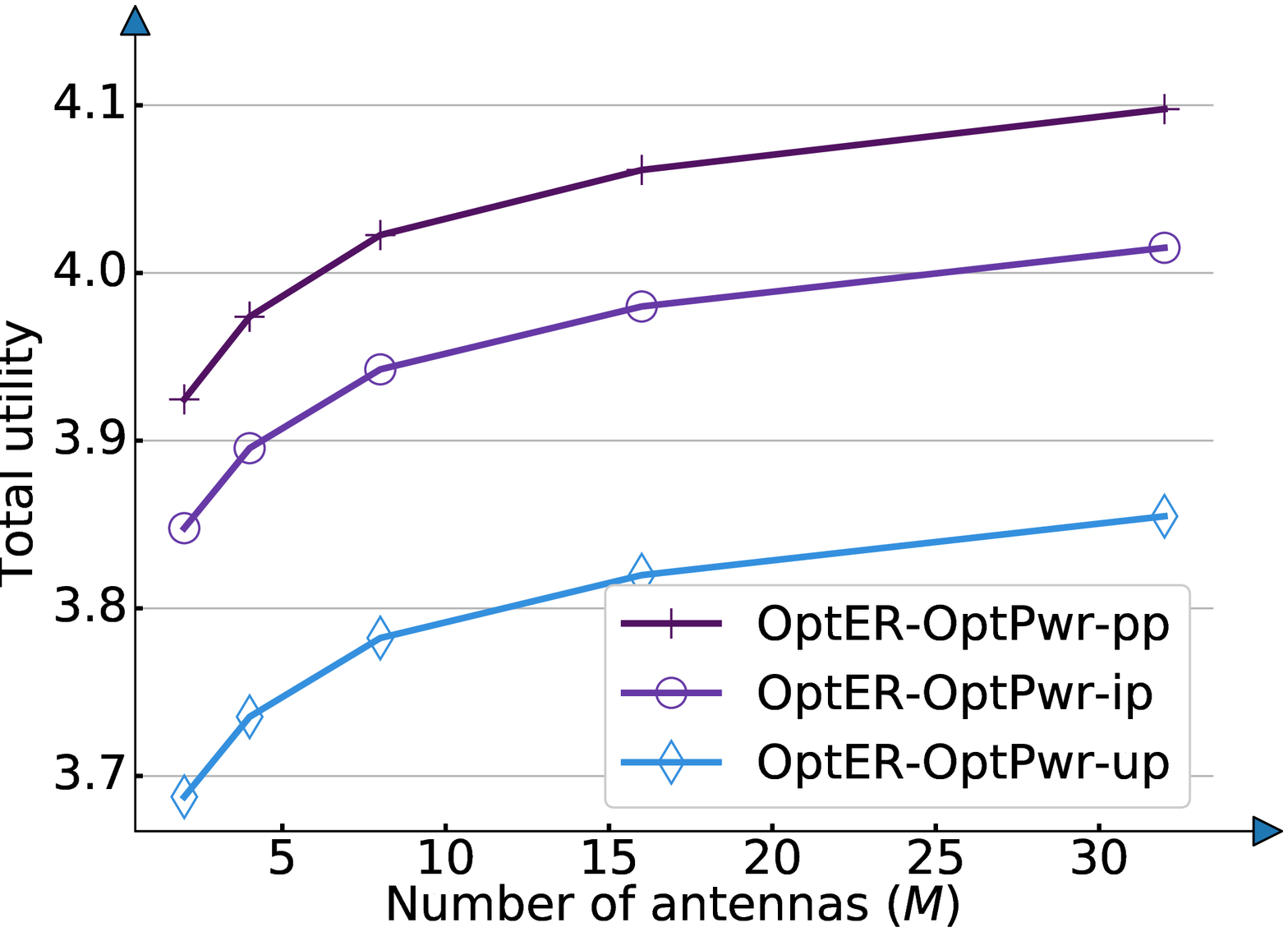}}}
  \subfigure[\small{Total utility versus $P$ at $\varepsilon$ = 0.4, $M$ = 8.}]
 {\resizebox{4.3cm}{!}{\includegraphics{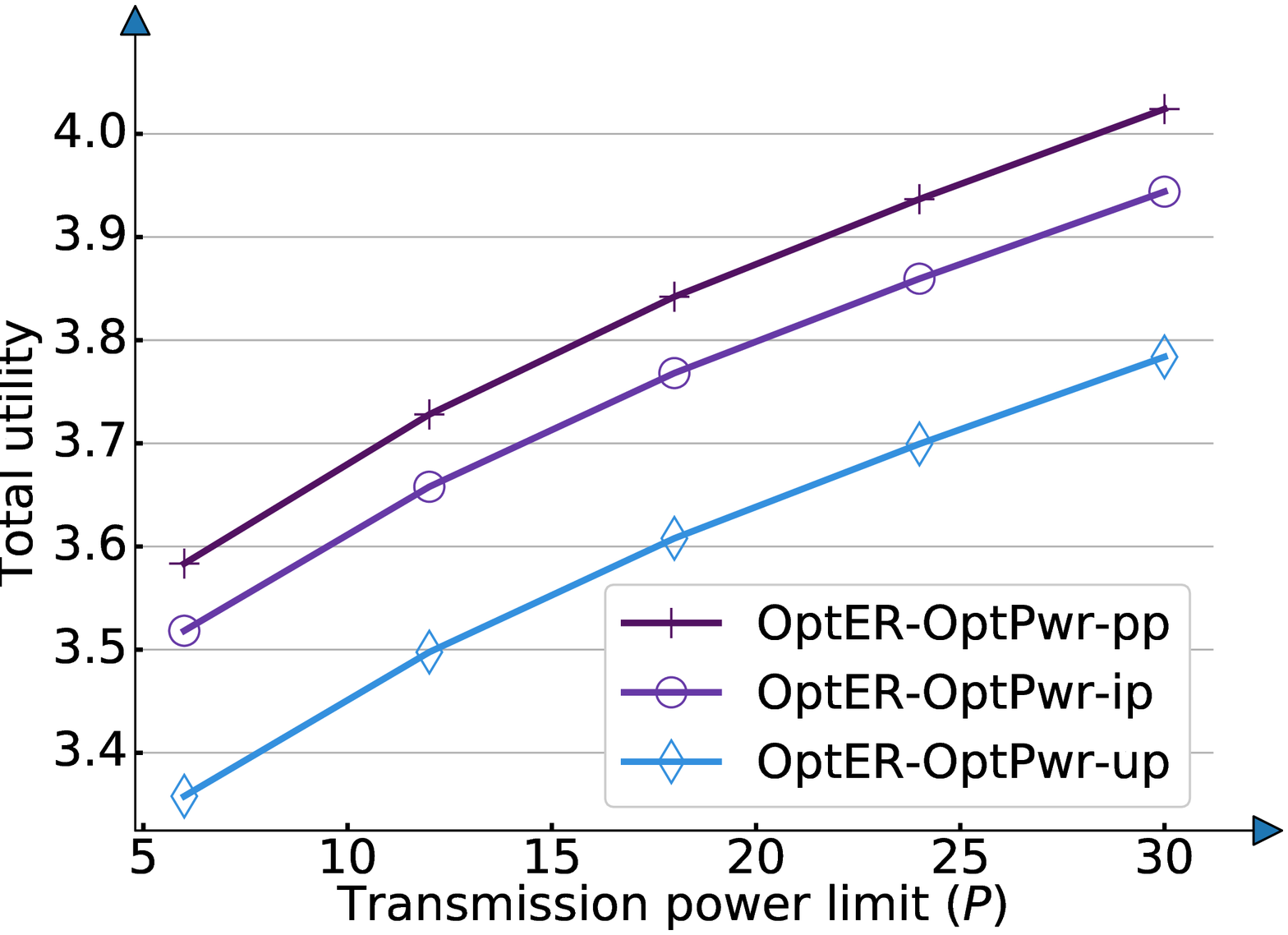}}}
 \end{center}
   \caption{\small{Total utility for the 3-rd GOP of the proposed solutions in the single-user secnario ($K$ = 1).}}
   \label{single_quality_error}
\end{figure*}

\begin{figure}[h]
\begin{center}
 \subfigure[\small{OptER-OptPwr-pp}]
 {\resizebox{2.8cm}{!}{\includegraphics{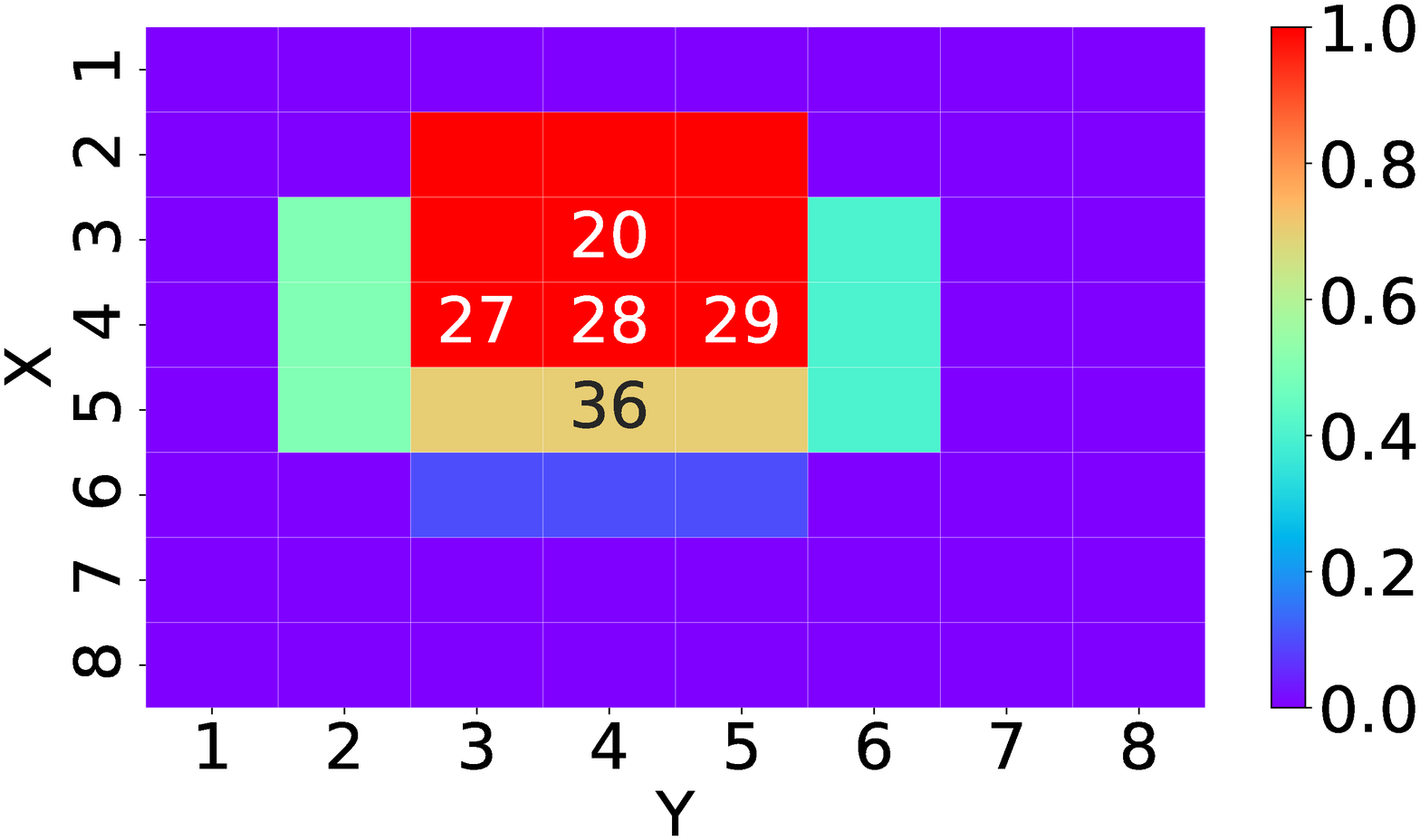}}}
  \subfigure[\small{OptER-OptPwr-ip}]
 {\resizebox{2.8cm}{!}{\includegraphics{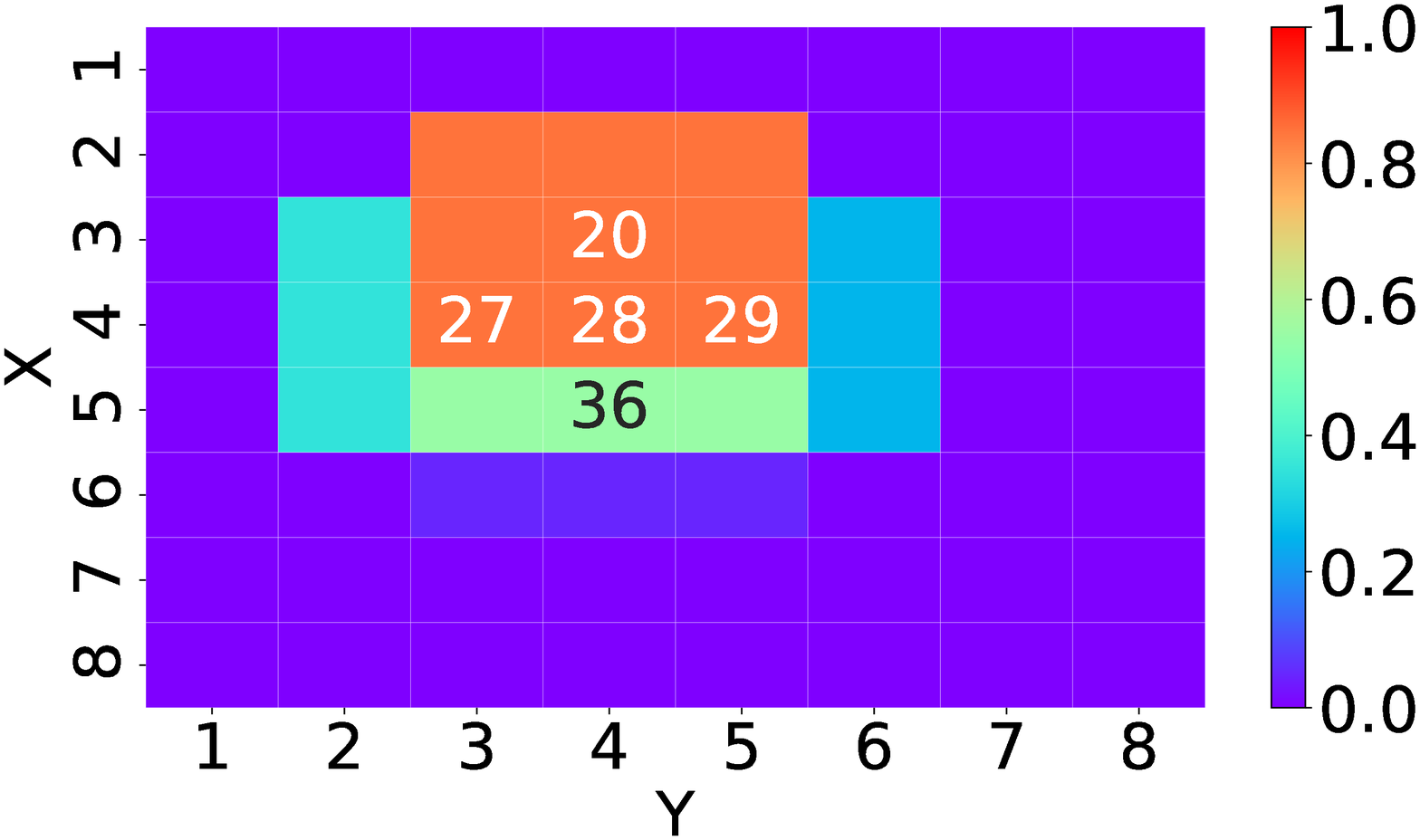}}}
   \subfigure[\small{OptER-OptPwr-up}]
   {\resizebox{2.8cm}{!}{\includegraphics{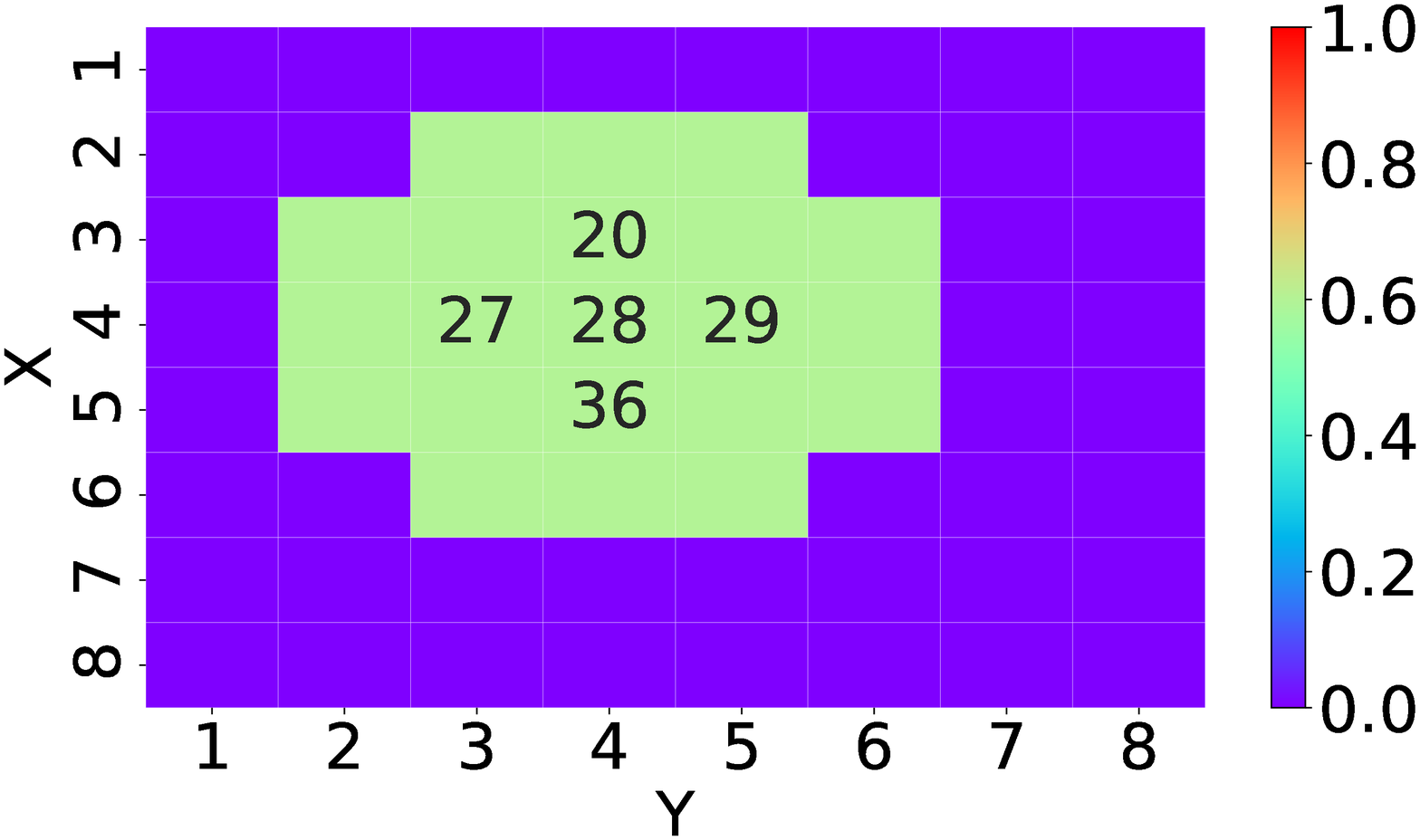}}}
 \end{center}
   \caption{\small{Encoding rates of the tiles for the 3-rd GOP given by OptER-OptPwr-up, OptER-OptPwr-ip, and OptER-OptPwr-up in the single-user scenario ($K = 1$) at $M$ = 8, $P$ = 30 dBm, and $\varepsilon = 0.4$.}}
   \label{result_single}
\end{figure}

In the simulation, we consider adaptive streaming of five 360 video sequences, i.e., \textit{Diving}, \textit{Rollercoaster}, \textit{Timelapse}, \textit{Venice}, \textit{Paris}, provided by \cite{inproceedings}. They are indexed by 1, 2, 3, 4, and 5, respectively. Each video sequence lasts 60 s. As illustrated in Fig. \ref{example}, we divide each 360 video into $8 \times 8$ tiles, i.e., $X = 8$, $Y = 8$, and choose $\overline{I} = 64$ FoVs, each of size $3\times 3$ (in the number of tiles). We use \textit{Kvazaar} as the 360 video encoder and set $D_{l},l\in\mathcal{L}$ according to Table \ref{table2}. Each video sequence is encoded into 60 GOPs, each of 1 s. We set the slot duration as 5 ms. That is to say, each GOP contains 200 slots. For each video sequence, based on the viewpoint data of 59 users provided by \cite{inproceedings}, we obtain a viewpoint sequence for each user, with one viewpoint for each GOP. We view users 2, 8, 24, 32, and 40 in \cite{inproceedings} as the users who request videos 1, 2, 3, 4, and 5, i.e., users 1, 2, 3, 4, and 5, respectively. The 59 viewpoint sequences for video $k$ are used for FoV prediction for user $k$. Let $i_{k}$ denote the index of the FoV of user $k$ corresponding to the current GOP in his viewpoint sequence. Set $\mathcal{I}_{k} = \{i_{k}-8,i_{k}-1,i_{k},i_{k}+1,i_{k}+8\}$, which contains $i_{k}$ and the indices of the neighbouring FoVs of FoV $i_{k}$. Let $n_{i_{k},i}$ denote the number of users with the current GOP and the next GOP in his viewpoint sequence for video $k$ being $i_{k}$ and $i$, respectively, where $i\in\mathcal{I}_{k}$. Then, we calculate the FoV viewing probabilities according to $p_{i,k} = \frac{n_{i_{k},i}}{\sum\nolimits_{i\in\mathcal{I}_{k}}n_{i_{k},i}},~i\in\mathcal{I}_{k},k\in\mathcal{K}.$ For example, the values of $p_{i,k},i\in\mathcal{I}_{k},k\in\mathcal{K}$ for the 3-rd GOP are given in Table \ref{table}, which will be used for plotting Fig. \ref{single_quality_error}, Fig. \ref{result_single}, Fig. \ref{multi_quality_error} and Fig. \ref{result_multi}. We set $\hat{p}_{i,k} = p_{i,k},\varepsilon_{i,k} =  \varepsilon, i\in\mathcal{I}_{k},k\in\mathcal{K}$, $B$ = 39 kHz, $N$ = 128, and $\sigma^{2} = 10^{-9}$ W. We consider the spatially
correlated Rayleigh-fading channel model with the correlation following the one-ring scattering model as in \cite{JSAC}.
\begin{table}[t]  
\caption{\small{Encoding rates (in kbit/s) for $L = 3,5,7$.}}
\begin{center}
\resizebox{8cm}{!}{  
\begin{tabular}{|c|c|} 
\hline  
$L$& $D_{l}$, $l\in\mathcal{L}$ \\ \hline  
3 & $D_{1} = 500,~D_{2} = 3000,~D_{3} = 8000$   \\ \hline 
5 & $D_{1} = 500,~D_{2} = 1000,~D_{3} = 3000,~D_{4} = 6000,~D_{5} = 8000$\\ \hline
7 &$D_{1} = 500,~D_{2} = 1000,~D_{3} = 2000,~D_{4} = 3000,~D_{5} = 4000,~D_{6} = 6000,~D_{7} = 8000$\\ \hline
\end{tabular}
} \label{table2}
\end{center}
\end{table}

\begin{table}\large
\caption{\small{Prediction parameters.}}
\begin{center}
\resizebox{8.5cm}{!}{  
\begin{tabular}{|p{1cm}|p{2.2cm}|p{1.7cm}|p{1.7cm}|p{4.2cm}|p{7cm}|} 
\hline  
$k$& Video sequence& User&Current FoV& Predicted FoVs& FoV viewing probability distributions  \\ \hline  
1 & \textit{Diving}  &2& 28 &$\mathcal{I}_{1} = \{20,27,28,29,36\}$& \tabincell{l}{ $(p_{20,1},p_{27,1},p_{28,1},p_{29,1},p_{36,1})$ \\$= (0.4138,0.1724,0.2414,0.1667,0.0417)$ }  \\ \hline  
2 & \textit{Rollercoaster} &8& 21 &$\mathcal{I}_{2} = \{13,20,21,22,29\}$&  \tabincell{l}{$(p_{13,2},p_{20,2},p_{21,2},p_{22,2},p_{29,2})$\\$ = (0,0.4615,0.3077,0.0769,0.1538)$}  \\ \hline  
3 & \textit{Timelapse} &24& 24 &  $\mathcal{I}_{3} = \{16,23,24,17,32\}$&\tabincell{l}{$(p_{16,3},p_{23,3},p_{24,3},p_{17,3},p_{32,3})$\\$ = (0.1481,0.037,0.2963,0.5185,0)$}\\ \hline
4 & \textit{Venice} & 32&29 &$\mathcal{I}_{4} = \{21,28,29,30,37\}$&\tabincell{l}{$(p_{21,4},p_{28,4},p_{29,4},p_{30,4},p_{37,4}) $\\$= (0.25,0.375,0.25,0.0625,0.0625)$} \\ \hline
5 & \textit{Paris} & 40&18 & $\mathcal{I}_{5} = \{10,17,18,19,26\}$&\tabincell{l}{$(p_{10,5},p_{17,5},p_{18,5},p_{19,5},p_{26,5}) $\\$= (0.375,0.5,0.125,0,0)$} \\ \hline
\end{tabular}
} \label{table}
\end{center}
\end{table}

We adopt the utility function in \cite{Zhang2013QoE}, i.e., $U(r) = 0.6 \log(1000\frac{r}{D_{L}})$. For ease of presentation, in the following, $Q^{(\phi)}(\mathbf{r}),\phi =$ pp, ip, up are referred to as total utility. We evaluate the average total utility of the 3-rd GOP over 100 random realizations of $\mathbf{h}_{{k,n}},k\in\mathcal{K},n\in\mathcal{N}$ for the first slot of the 3-rd GOP. We evaluate the Cumulative Distribution Function (CDF) of the total utility, the CDF of the total utility variation (i.e., the difference of the total utilities in two adjacent GOPs), and the means and variances of the rebuffering time for the transmitted FoVs over the 60 GOPs. We also evaluate the means and variances of the total PSNR, PSNR variation, SSIM, and SSIM variation for the viewing FoVs over the 60 GOPs.

\subsection{Single-user Scenario}\label{simulation_single}
In the single-user scenario, we consider adaptive streaming of \textit{Diving} to user 1. In case-$\phi$, the proposed solution is referred to as OptER-OptPwr-$\phi$.
\subsubsection{Properties of Proposed Solutions}
Fig. \ref{single_quality_error} (a) illustrates the total utility for the 3-rd GOP versus the number of quality levels $L$. In Fig. \ref{single_quality_error} (a), OptER-OptPwr-Disc-$\phi$ represents the discrete solution constructed based on OptER-OptPwr-$\phi$ (as illustrated in Section \ref{s3}). Notice that the total utility of OptER-OptPwr-$\phi$ does not change with $L$, for all $\phi$ = pp, ip, up. Fig. \ref{single_quality_error} (a) shows that in case-$\phi$, the gap between the total utilities of OptER-OptPwr-$\phi$ and OptER-OptPwr-Disc-$\phi$ decreases with $L$. Furthermore, the gap is small when $L$ is large, implying that the performance loss due to continuous relaxation is negligible at a large $L$. 

Fig. \ref{single_quality_error} (b) shows the worst average total utility for the 3-rd GOP versus the estimation error bound $\varepsilon$. Note that the worst average total utility of OptER-OptPwr-up is irrelevant to $\varepsilon$. From Fig. \ref{single_quality_error} (b), we can see that in the case of an imperfect FoV viewing probability distribution, the worst average total utility of OptER-OptPwr-ip is greater than those of OptER-OptPwr-pp and OptER-OptPwr-up, which reveals the importance of explicitly considering the imperfectness of the predicted FoV viewing probability distribution in this case; and the worst-case average total utility of OptER-OptPwr-up is greater than that of OptER-OptPwr-pp when $\varepsilon$ is large, as OptER-OptPwr-up is designed to maximize the worst-case total utility and does not depend on any information of the FoV viewing probability distribution. Furthermore, the gain of OptER-OptPwr-ip over OptER-OptPwr-pp increases with $\varepsilon$, as it is more important to take into account of FoV prediction error when $\varepsilon$ is large; and the gain of OptER-OptPwr-ip over OptER-OptPwr-up decreases with $\varepsilon$, as the imperfect FoV viewing probability distribution becomes less important when $\varepsilon$ is large. 

Fig. \ref{single_quality_error} (c) and (d) demonstrate the total utility for the 3-rd GOP versus the number of transmit antennas $M$ and the total transmission power budget $P$, respectively. Fig. \ref{single_quality_error} (c) and (d) show that the total utility of each scheme increases with $M$ and $P$. Besides, Fig. \ref{single_quality_error} (c) and (d) show that $U^{(\text{pp})\star} > U^{(\text{ip})\star} > U^{(\text{up})\star}$, where $U^{(\phi)\star}$ represents the total utility of OptER-OptPwr-$\phi$ in the single-user scenario. Such observation coincides with the optimality properties in Statement (\romannumeral3) of Theorem \ref{single_theorem}. 

Fig. \ref{result_single} illustrates the heatmap of the encoding rates of all tiles of the 3-rd GOP given by the proposed solutions in the three cases. From Fig. \ref{result_single} (a) and (b), we can see that the encoding rates of the tiles in an FoV with a larger viewing probability are higher. From Fig. \ref{result_single} (c), we can tell that the encoding rates of the tiles given by OptER-OptPwr-up are identical. Such observations are in accordance with the optimality properties in Statement (\romannumeral2) of Theorem \ref{single_theorem}.
\begin{figure*}[t]
\begin{center}
 \subfigure[\small{Total utility}]
 {\resizebox{6cm}{!}{\includegraphics{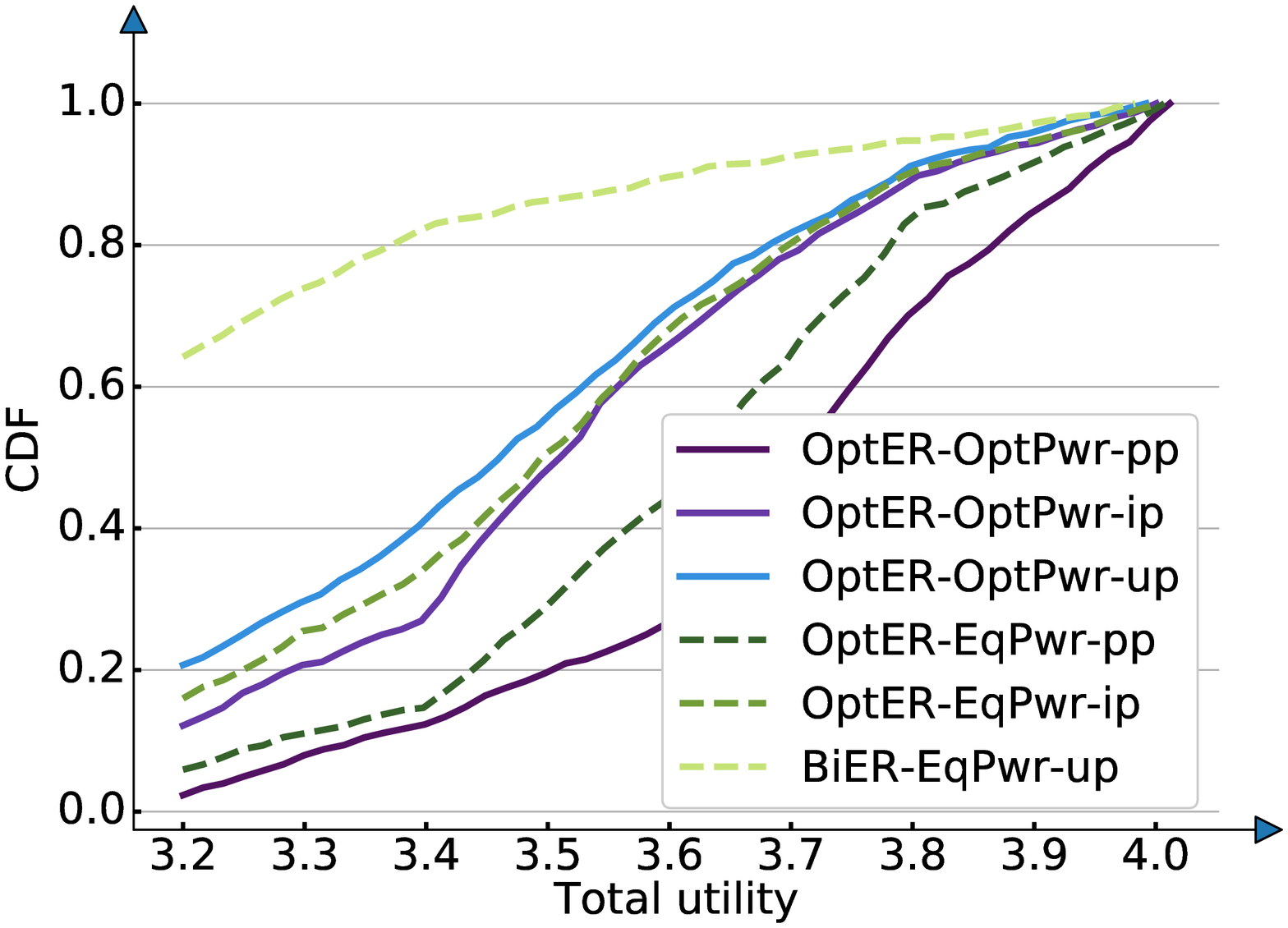}}}
 \subfigure[\small{Total utility variation}]
 {\resizebox{6cm}{!}{\includegraphics{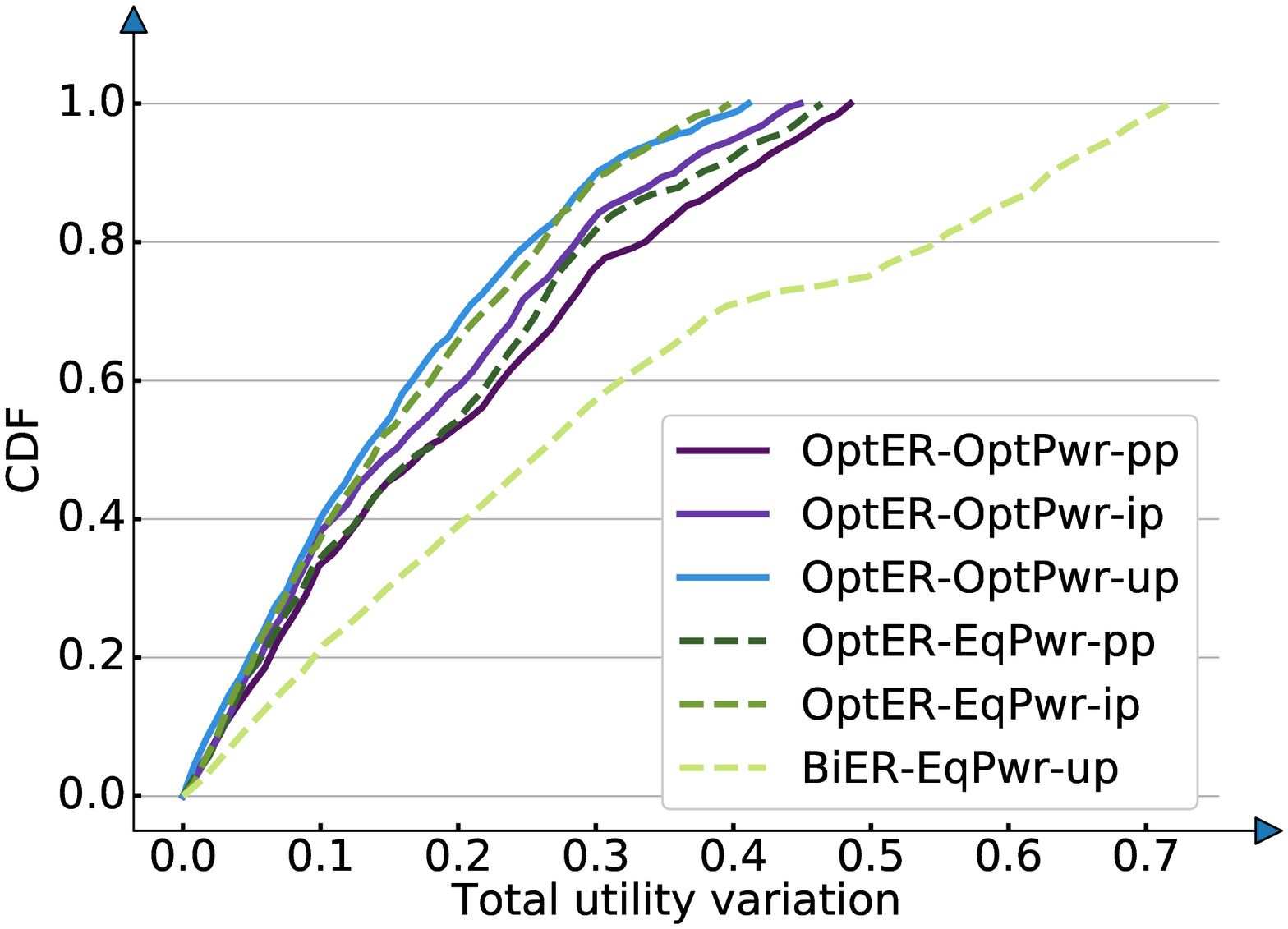}}}
 \end{center}
   \caption{\small{Comparing the proposed solutions with baseline schemes on the total utility and the total utility variation over 60 GOPs in the single-user scenario ($K = 1$) at $M$ = 8, $P$ = 30 dBm, and $\varepsilon = 0.4$.}}
   \label{single_overtime}
\end{figure*}

\begin{table*}
\centering
\caption{\small{Comparing the proposed solutions with baseline schemes on PSNR, PSNR variation, SSIM, SSIM variation, and rebuffering time over 60 GOPs in the single-user scenario ($K = 1$) at $M$ = 8, $P$ = 30 dBm, and $\varepsilon = 0.4$.}}
\resizebox{14cm}{!}{ 
\begin{tabular}{llllll} 
\toprule
   &    PSNR & PSNR variation & SSIM & SSIM variation & Rebuffering time (s) \\
  \midrule
  OptER-OptPwr-pp & $29.61\pm 1.07$  & $1.13 \pm 0.22 $& $0.907\pm 0.015$ &$0.0172\pm 0.0003$& $0.25\pm 0.04$  \\  
OptER-OptPwr-ip & $28.68\pm 1.06$ & $0.91 \pm 0.19$& $0.878 \pm 0.015$ &$0.0158\pm 0.0004$&  $0.23\pm 0.03$  \\ 
OptER-OptPwr-up & $28.27 \pm 0.90$ &$0.79 \pm 0.14$& $0.864 \pm 0.016$ &  $0.0147\pm 0.0004$& $0.22 \pm 0.02$\\ 
OptER-EqPwr-pp & $29.11\pm 1.02$ & $1.02 \pm 0.21$&$0.866 \pm 0.015$ &$0.0163\pm 0.0003$& $0.28\pm 0.04$ \\ 
OptER-EqPwr-ip & $28.21 \pm 0.93$ & $0.81 \pm 0.10$&$0.857\pm 0.015$ & $0.0153\pm 0.0003$& $0.25\pm 0.03$ \\ 
BiER-EqPwr-up & $27.13 \pm 2.33$ & $3.89 \pm 0.58$ &$0.843 \pm 0.050$ & $0.0492\pm 0.0014$&$0.21\pm 0.02$ \\ 
  \bottomrule
  \end{tabular}
  }
  \label{table_single}
\end{table*}

\begin{figure*}[t]
\begin{center}
 \subfigure[\small{Total utility versus $L$ at $\varepsilon = 0.4$, $M$ = 64, $P$ = 30 dBm.}]
 {\resizebox{4.3cm}{!}{\includegraphics{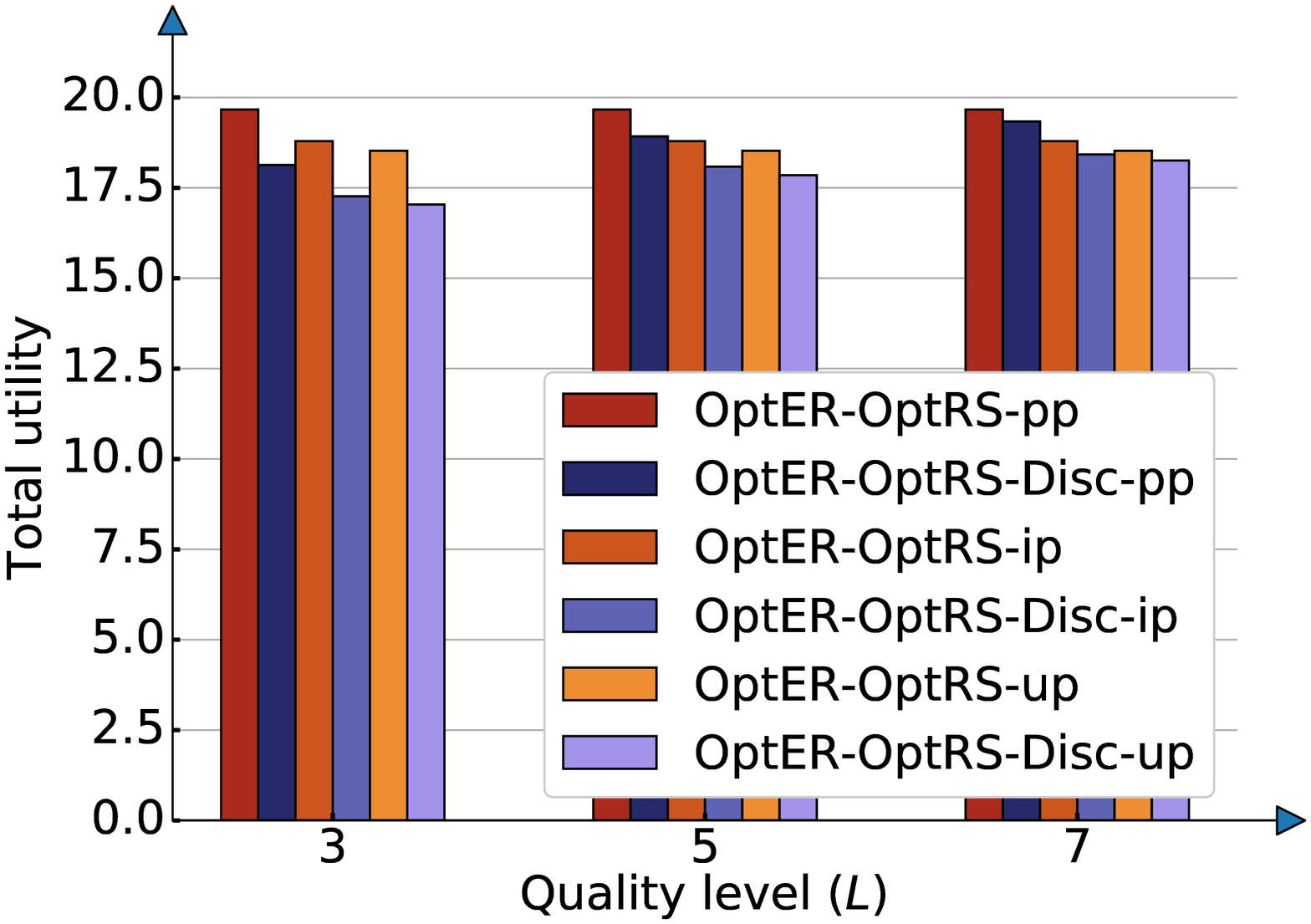}}}
  \subfigure[\small{Worst average total utility versus $\varepsilon$ at $M$ = 64, $P$ = 30 dBm.}]
 {\resizebox{4.3cm}{!}{\includegraphics{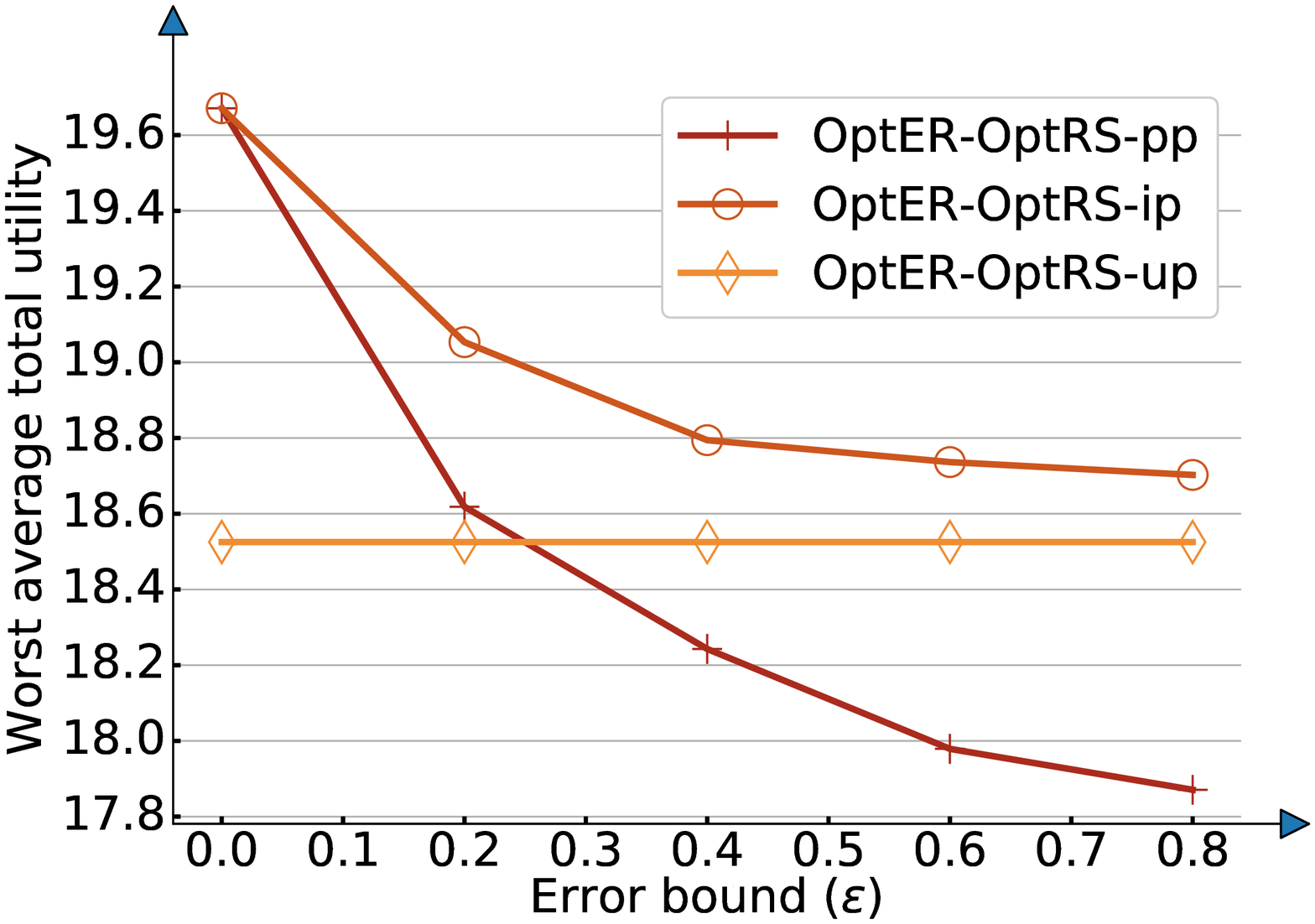}}}
 \subfigure[\small{Total utility versus $M$ at $\varepsilon = 0.4$, $P$ = 30 dBm.}]
   {\resizebox{4.3cm}{!}{\includegraphics{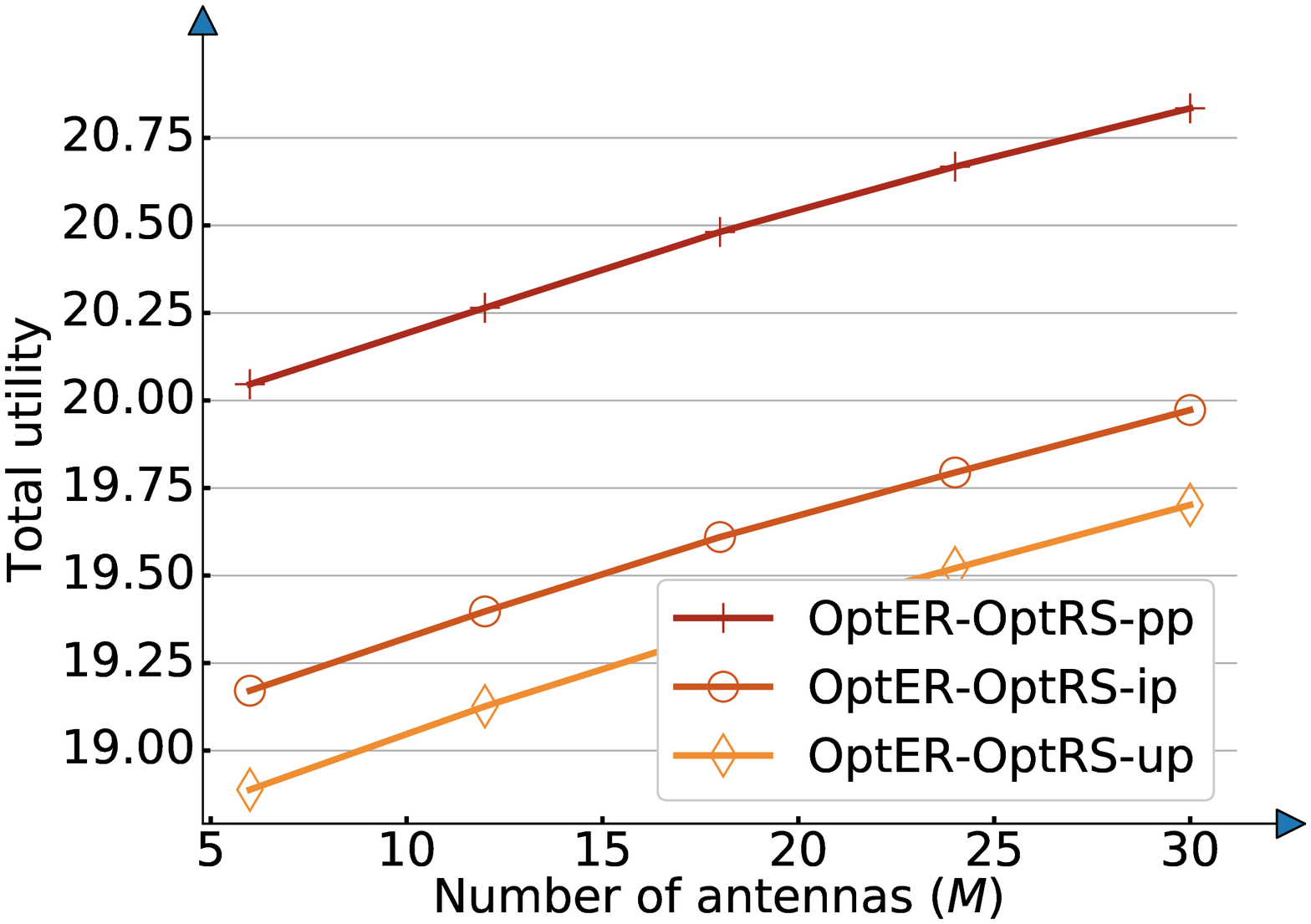}}}
   \subfigure[\small{Total utility versus $P$ at $\varepsilon = 0.4$, $M$ = 64.}]
 {\resizebox{4.3cm}{!}{\includegraphics{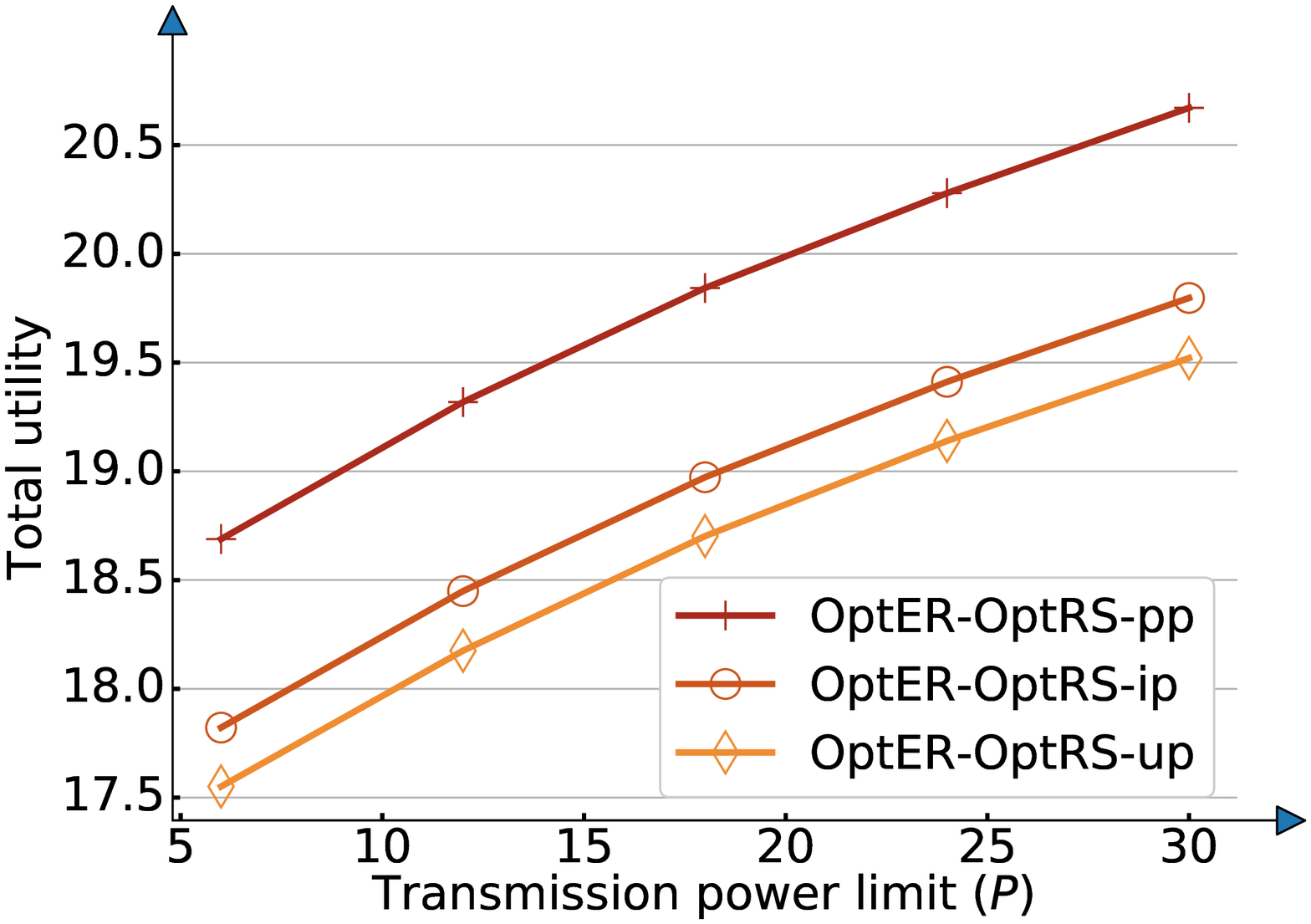}}}
 \end{center}
   \caption{\small{Total utility for the 3-rd GOP of the proposed solutions in the multi-user scenario ($K$ = 5). }}
   \label{multi_quality_error}
\end{figure*}

\begin{figure}[t]
\begin{center}
 \subfigure[\small{OptER-OptRS-pp}]
 {\resizebox{2.8cm}{!}{\includegraphics{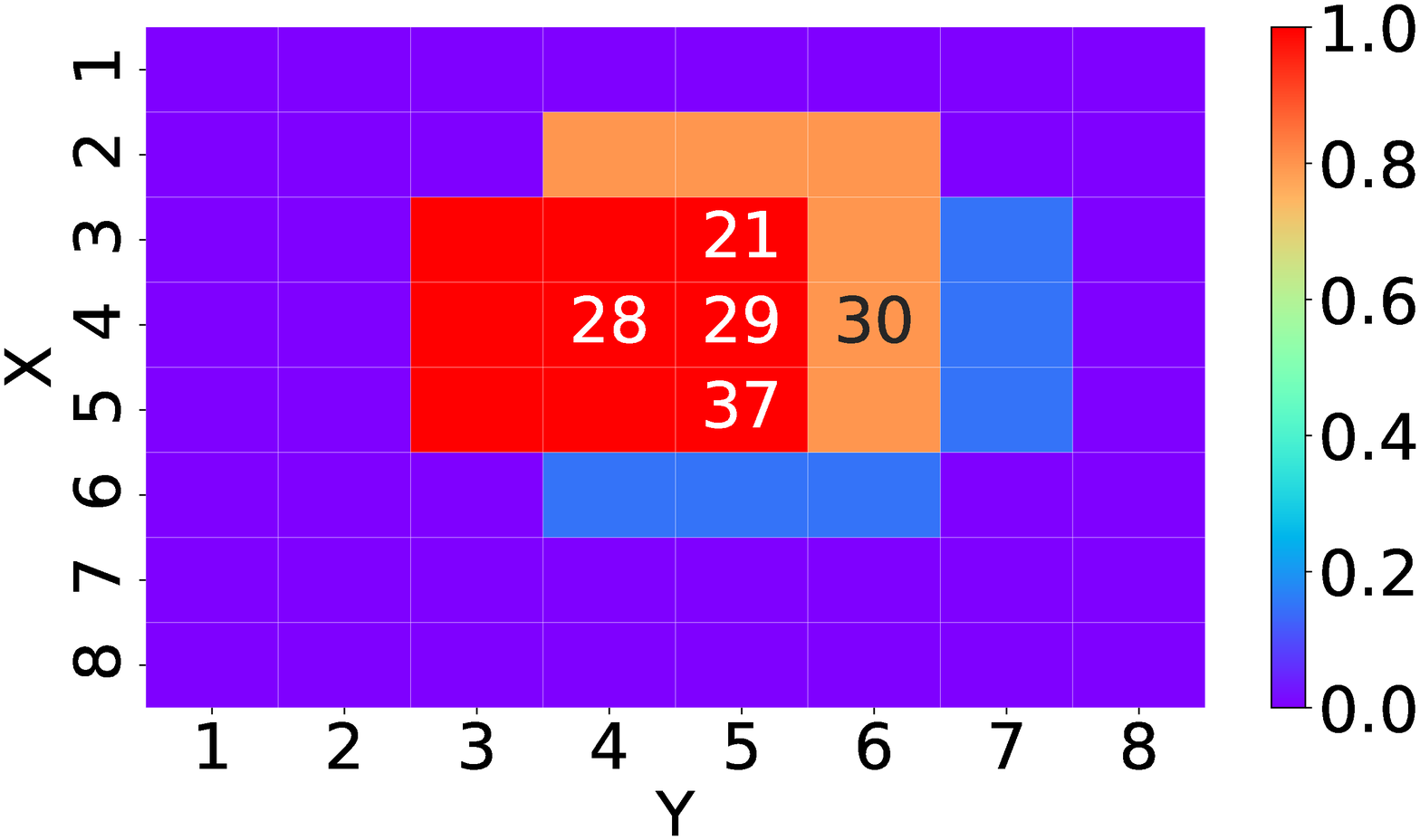}}}
  \subfigure[\small{OptER-OptRS-ip}]
 {\resizebox{2.8cm}{!}{\includegraphics{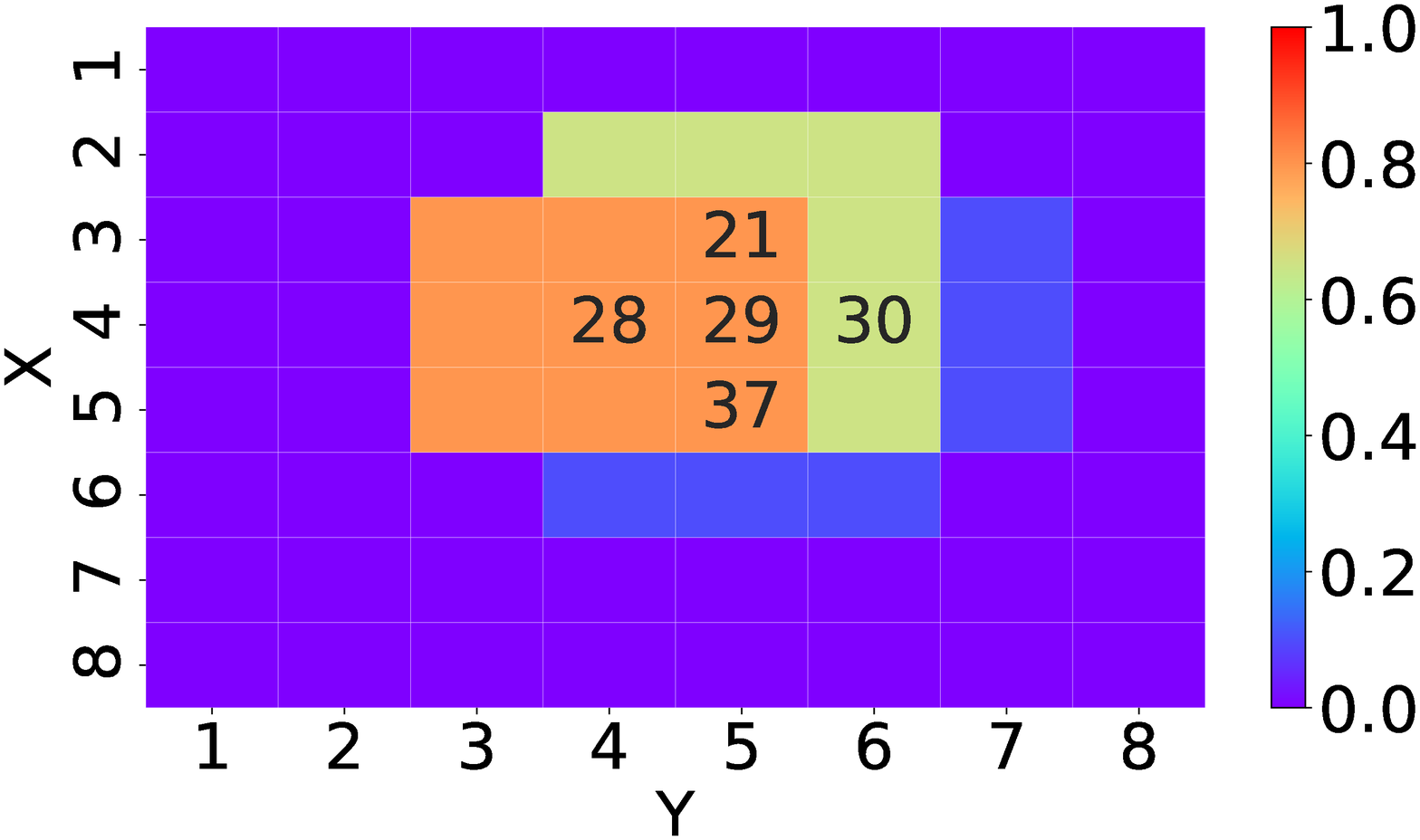}}}
   \subfigure[\small{OptER-OptRS-up}]
   {\resizebox{2.8cm}{!}{\includegraphics{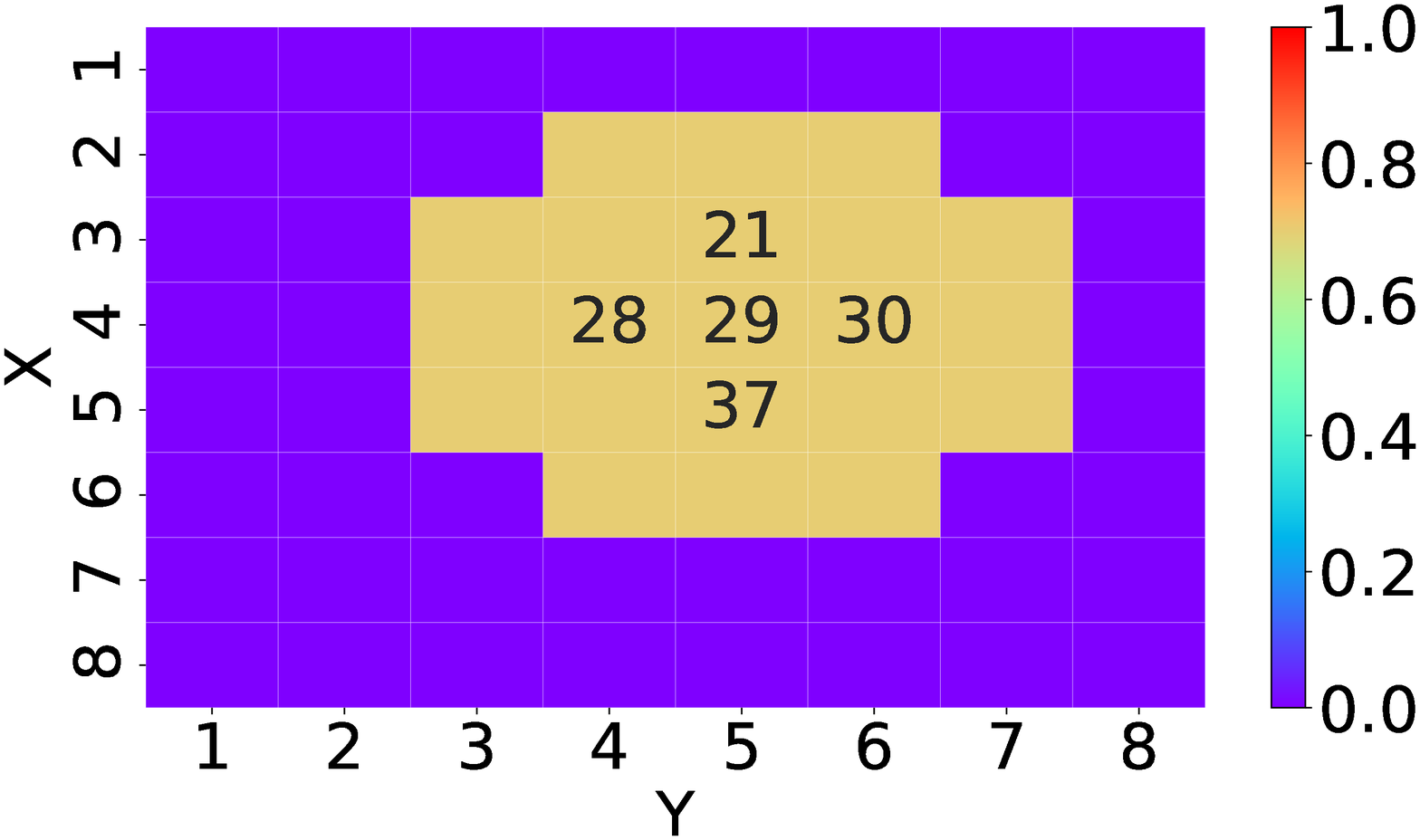}}}
 \end{center}
   \caption{\small{Encoding rates of the tiles for the 3-rd GOP of user 4 given by OptER-OptRS-pp, OptER-OptRS-ip and OptER-OptRS-up in the multi-user scenario ($K = 5$) at $M$ = 64, $P$ = 30 dBm, and $\varepsilon = 0.4$.}}
   \label{result_multi}
\end{figure}

\subsubsection{Comparisions with Baseline Schemes}We consider three baseline schemes, namely OptER-EqPwr-pp, OptER-EqPwr-ip, and BiER-EqPwr-up, for the three cases, respectively. All three baseline schemes adopt equal power allocation (i.e., $v_{n}(t) = \frac{P}{N}, n\in\mathcal{N},t\in\mathcal{T}$)\cite{eqpower} and determine the transmission rate for each slot based on the instantaneous channel conditions accordingly. OptER-EqPwr-pp and OptER-EqPwr-ip adopt the same encoding rate adaptation mechanism as the proposed one. But the encoding rate adaptation of OptER-EqPwr-ip is based on $\hat{p}_{i},i\in\mathcal{I}$ (rather than $p_{i},i\in\mathcal{I}$) without considering possible estimation errors for the FoV viewing probability distribution. BiER-EqPwr-up adopts the encoding rate adaptation mechanism in\cite{VR/AR18}. Specifically, the encoding rate adaptation of OptER-EqPwr-pp is obtained by solving Problem \ref{single_case_general_equal} with $\phi =$ pp and $C^{\dag}(1)$ given by the transmission rate at slot 1 under equal power allocation. The encoding rate adaptation of OptER-EqPwr-ip is obtained by solving Problem \ref{single_case_general_equal} with $\phi =$ ip, $\hat{p}_{i},i\in\mathcal{I}$ (rather than $p_{i},i\in\mathcal{I}$), and $C^{\dag}(1)$ given by the transmission rate at slot 1 under equal power allocation. The encoding rate adaptation of BiER-EqPwr-up chooses $D_{1}$ as the encoding rate of each FoV in $\mathcal{I}\backslash \{i\}$ and optimizes the encoding rate for the current FoV $i$ to maximize the total utility\cite{VR/AR18}.

Fig. \ref{single_overtime} shows the CDF of the total utility and the CDF of the total utility variation over the 60 GOPs. Table \ref{table_single} demonstrates the means and variances of the PSNR, PSNR variation, SSIM, and SSIM variation for the viewing FoVs over the 60 GOPs, and the means and variances of the rebuffering time for the transmitted FoVs over the 60 GOPs. Fig. \ref{single_overtime} (a) and Table \ref{table_single} demonstrate that in case-pp or case-ip, the proposed solution outperforms the baseline scheme in the average total utility, PSNR, SSIM, and rebuffering time, and the proposed solution and the baseline scheme have similar average total utility variations, PSNR variations, and SSIM variations.
In case-up, the proposed solution outperforms the baseline scheme in the average total utility, total utility variation, PSNR, PSNR variation, SSIM, and SSIM variation, and the proposed solution and the baseline scheme have similar rebuffering times. Besides, Table \ref{table_single} demonstrates that the proposed approach achieves a tradeoff among quality, quality variation, and rebuffering time. For example, OptER-OptPwr-pp achieves the highest PSNR (SSIM) and PSNR (SSIM) variation and the longest rebuffering time.

The gains of OptER-OptPwr-pp over OptER-EqPwr-pp in the average total utility, PSNR, and SSIM are due to the fact that the transmission rate adaptation is determined by the optimal power allocation; the gains of OptER-OptPwr-ip and OptER-OptPwr-up over OptER-EqPwr-ip and BiER-EqPwr-up, respectively, in the average total utility, PSNR, and SSIM arise from the fact that the transmission rate adaptation is determined by the optimal power allocation and robust encoding rate optimization is considered.

\begin{figure*}[t]
\begin{center}
   \subfigure[\small{Total utility}]
   {\resizebox{6cm}{!}{\includegraphics{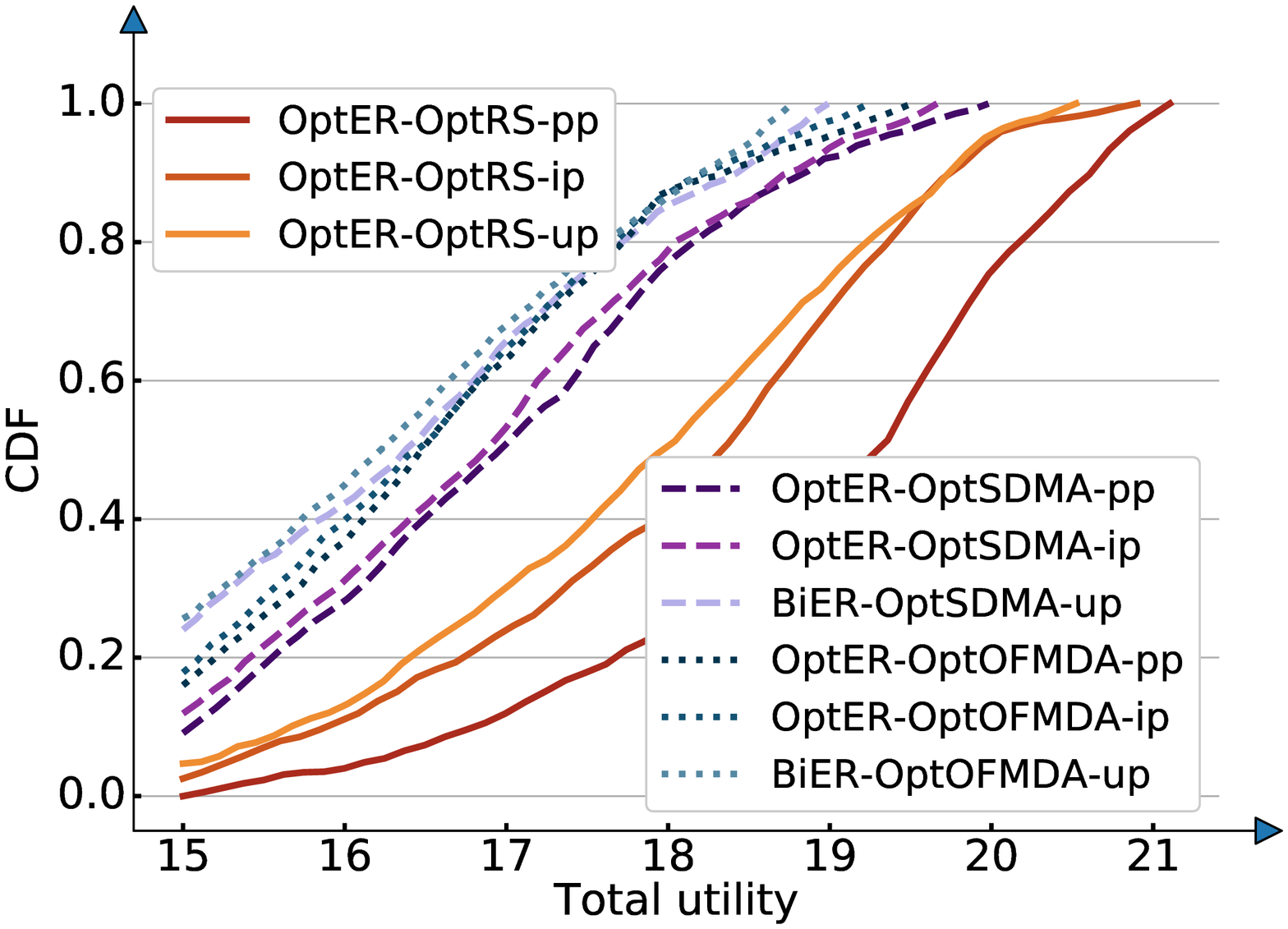}}}
  \subfigure[\small{Total utility variation}]
   {\resizebox{6cm}{!}{\includegraphics{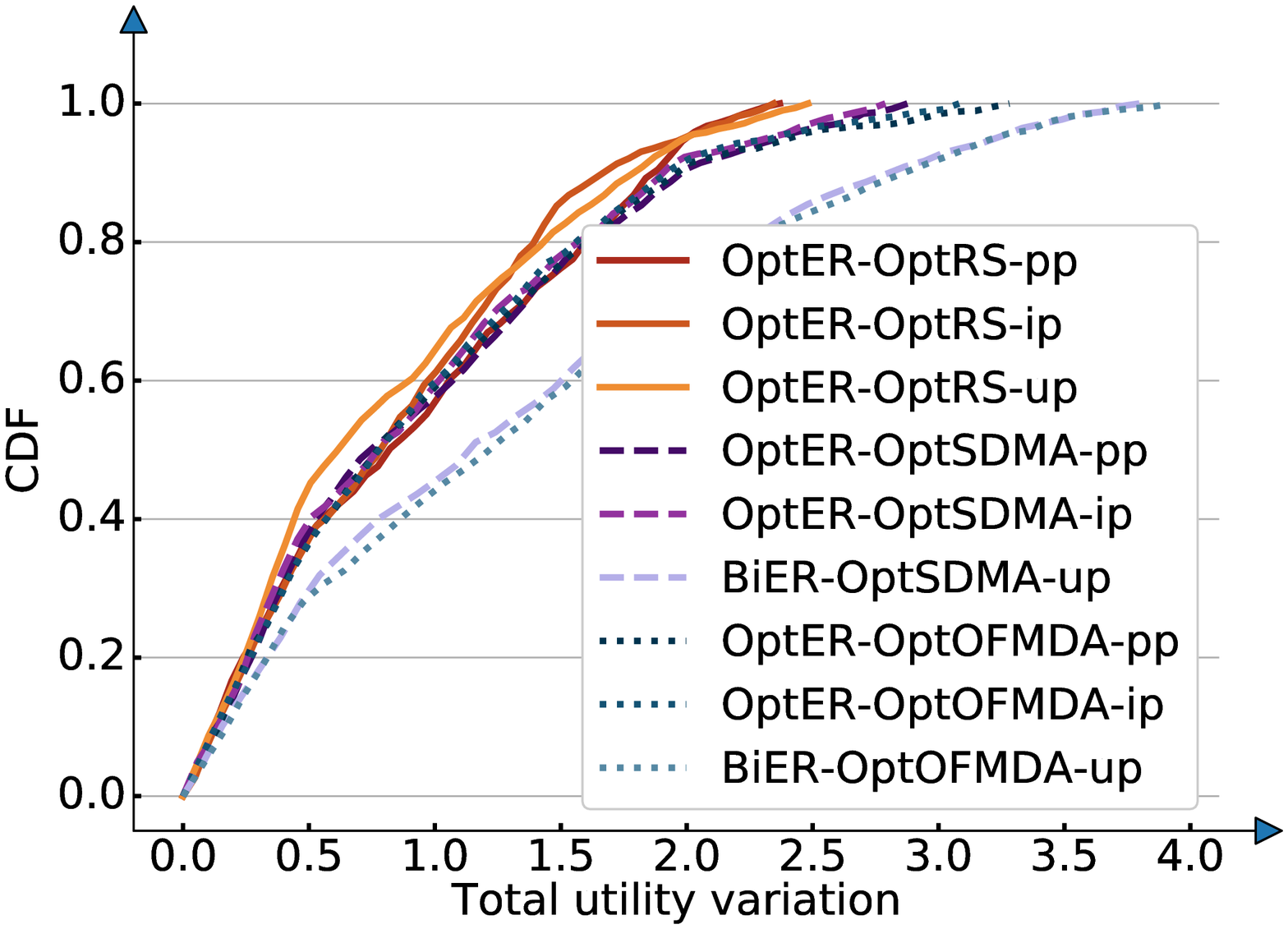}}}
 \end{center}
   \caption{\small{Comparing the proposed solutions with baseline schemes on the total utility and the total utility variation over 60 GOPs in the multi-user scenario ($K = 5$) at $M$ = 64, $P$ = 30 dBm, and $\varepsilon = 0.4$.}}
   \label{multi_overtime}
\end{figure*}

\begin{table*}
\centering
\caption{\small{Comparing proposed solutions with baseline schemes on PSNR, PSNR variation, SSIM, SSIM variation, and rebuffering time over 60 GOPs in the multi-user scenario ($K = 5$) at $M$ = 64, $P$ = 30 dBm, and $\varepsilon = 0.4$.}}
\resizebox{14cm}{!}{ 
\begin{tabular}{llllll} 
\toprule
   &     PSNR & PSNR variation & SSIM & SSIM variation & Rebuffering time (s) \\
  \midrule
  OptER-OptRS-pp & $169.30\pm 6.67$  & $5.61 \pm 0.36 $& $4.717\pm 0.048$ &$0.055\pm 0.004$& $1.263\pm 0.082$  \\  
OptER-OptRS-ip & $165.30\pm 4.58$ & $5.21 \pm 0.58$& $4.662 \pm 0.043$ &$0.052\pm 0.007$&  $1.213\pm 0.079$  \\ 
OptER-OptRS-up & $164.57 \pm 3.87$ &$4.58 \pm 0.44$& $4.646 \pm 0.050$ &  $0.047\pm 0.003$& $1.082 \pm 0.074$\\ 
OptER-OptSDMA-pp & $160.95\pm 8.73$ & $5.69 \pm 0.63$&$4.683 \pm 0.084$ &$0.055\pm 0.007$& $1.478\pm 0.121$ \\ 
OptER-OptSDMA-ip & $155.95 \pm 6.02$ & $5.48 \pm 0.56$&$4.601\pm 0.048$ & $0.054\pm 0.007$& $1.393\pm 0.106$ \\ 
BiER-OptSDMA-up & $152.87 \pm 16.14$ & $8.08 \pm 1.11$ &$4.405 \pm 0.161$ & $0.171\pm 0.016$&$1.310\pm 0.094$ \\ 
OptER-OptOFDMA-pp & $155.94\pm 8.61$ & $5.59 \pm 0.67$&$4.664 \pm 0.084$ &$0.054\pm 0.008$& $1.612\pm 0.113$ \\ 
OptER-OptOFDMA-ip & $150.94 \pm 4.70$ & $5.29 \pm 0.37$&$4.583\pm 0.077$ & $0.053\pm 0.007$& $1.415\pm 0.095$ \\ 
BiER-OptOFDMA-up & $145.94 \pm 16.24$ & $8.04 \pm 1.13$ &$4.377 \pm 0.158$ & $0.171\pm 0.013$&$1.211\pm 0.088$ \\ 
  \bottomrule
  \end{tabular}
  }
  \label{table_multi}
\end{table*}

\subsection{Multi-user Scenario}\label{simulation_multi}
In the multi-user scenario, we consider adaptive streaming of the five 360 videos given in Table \ref{table} to the five users, respectively, i.e., $K$ = 5. In case-$\phi$, the proposed solution is referred to as OptER-OptRS-$\phi$.
\subsubsection{Properties of Proposed Solutions}
Fig. \ref{multi_quality_error} (a), (c), (d) illustrate the total utility for the 3-rd GOP versus the number of quality levels $L$, the number of transmit antennas $M$, and the transmission power budget $P$, respectively. In Fig. \ref{multi_quality_error} (a), OptER-OptRS-Disc-$\phi$ represents the discrete solution which is constructed based on OptER-OptRS-$\phi$ (as illustrated in Section \ref{section3}). Fig. \ref{multi_quality_error} (b) illustrates the worst average total utility for the 3-rd GOP versus the estimation error bound $\varepsilon$. The results are the same as those in the single-user scenario. Besides, Fig. \ref{multi_quality_error} (a), (c), and (d) show that $U^{(\text{pp})\dag} > U^{(\text{ip})\dag} > U^{(\text{up})\dag}$, where $U^{(\phi)\dag}$ represents the total utility of OptER-OptRS-$\phi$ in the multi-user scenario. The relationship among the total utilities of the suboptimal solutions of Problem \ref{rs_case_general} in the three cases is the same as that of the optimal solutions in the three cases (which is shown in Statement (\romannumeral3) of Theorem \ref{theorem}). Fig. \ref{result_multi} illustrates the heatmap of the encoding rates of all tiles of the 3-rd GOP given by the proposed solutions in the three cases. Similarly, we see that the optimality properties in Statement (\romannumeral2) of Theorem \ref{theorem} hold.

\subsubsection{Comparision with Baseline Schemes}
We consider six baseline
schemes, namely OptER-OptSDMA-pp, OptER-OptSDMA-ip, BiER-OptSDMA-up, OptER-OptOFDMA-pp, OptER-OptOFDMA-ip, and BiER-OptOFDMA-up. OptER-OptSDMA-pp, OptER-OptSDMA-ip, and BiER-OptSDMA-up adopt SDMA and optimize the beamforming vector on each subcarrier\cite{5G}. OptER-OptOFDMA-pp, OptER-OptOFDMA-ip, and BiER-OptOFDMA-up adopt OFDMA, consider the maximum ratio transmission (MRT) on each subcarrier and optimize the subcarrier and power allocation\cite[pp. 39]{5G}. OptER-OptSDMA-pp, OptER-OptSDMA-ip, OptER-OptOFDMA-pp, and OptER-OptOFDMA-ip adopt the same encoding rate adaptation mechanism as the proposed one. The difference is that the encoding rate adaptation of OptER-OptSDMA-ip and OptER-OptOFDMA-ip is based on $\hat{p}_{i,k},i\in\mathcal{I}_{k},k\in\mathcal{K}$ (rather than $p_{i,k},i\in\mathcal{I}_{k},k\in\mathcal{K}$) without considering possible estimation errors for the FoV viewing probability distributions. BiER-OptSDMA-up and BiER-OptOFDMA-up adopt the encoding rate adaptation mechanism in \cite{VR/AR18}. The total utility maximization problems of these six baseline schemes are formulated similarly to Problem~\ref{rs_case_general_new} and solved using a similar separate approach. The only difference is that the objective function for the transmission rate optimization for slot $2,\ldots, T$ is the sum transmission rate. The separate optimization problems of OptER-OptSDMA-pp, OptER-OptSDMA-ip, and BiER-OptSDMA-up are solved similarly using CCCP. The separate optimization problems of OptER-OptOFDMA-pp, OptER-OptOFDMA-ip, and BiER-OptOFDMA-up are solved by continuous relaxation and the KKT conditions.

Fig. \ref{multi_overtime} shows the CDF of the total utility and the CDF of the total utility variation over the 60 GOPs. Table \ref{table_multi} demonstrates the means and variances of the total PSNR, PSNR variation, SSIM, and SSIM variation for the viewing FoVs over the 60 GOPs and the means and variances of the total rebuffering time for the transmitted FoVs over the 60 GOPs. From Fig. \ref{multi_overtime} and Table \ref{table_multi}, we can tell that in each case, the proposed solution outperforms the two baseline schemes in all considered performance metrics. Similarly, from Table \ref{table_multi}, we can see that the proposed approach achieves a tradeoff among the quality, quality variation, and rebuffering time.

The gains of OptER-OptRS-pp, OptER-OptRS-ip, and OptER-OptRS-up over OptER-OptSDMA-pp, OptER-OptSDMA-ip, and BiER-OptSDMA-up, respectively, in the average total utility, PSNR, and SSIM are due to the fact that the cost for SDMA to suppress interference can be high, while rate splitting together with SIC partially decodes interference and partially treats interference as noise. The gains of OptER-OptRS-pp, OptER-OptRS-ip, and OptER-OptRS-up over OptER-OptOFDMA-pp, OptER-OptOFDMA-ip, and BiER-OptOFDMA-up, respectively, in the average total utility, PSNR, and SSIM come from effective nonorthogonal transmission design. Besides, the gains of OptER-OptRS-ip and OptER-OptRS-up over OptER-OptSDMA-ip (OptER-OptOFDMA-ip) and BiER-OptSDMA-up (BiER-OptOFDMA-up), respectively, in the average total utility, PSNR, and SSIM also derive from the robust encoding rate optimizations. Moreover, the gains of OptER-OptRS-pp, OptER-OptRS-ip, and OptER-OptRS-up over OptER-OptSDMA-pp (OptER-OptOFDMA-pp), OptER-OptSDMA-ip (OptER-OptOFDMA-ip), and BiER-OptSDMA-up (BiER-OptOFDMA-up), respectively, in the average total rebuffering time arise from the fact that the proposed separate approach can reduce the sum of infeasibilities of the encoding rate constraints in \eqref{sum_rate_constraint} for each GOP.

\section{Conclusion}
In this paper, we investigated adaptive streaming of one or multiple tiled 360 videos from a multi-antenna BS to one or multiple single-antenna users, respectively, in a multi-carrier wireless system. We considered three cases of FoV viewing probability distributions and introduced a total utility metric for each case. In the single-user scenario, we optimized the encoding rate adaptation of each GOP and the transmission adaptation of each slot to maximize the total utility in each case. In the multi-user scenario, we adopted rate splitting with successive decoding and optimized the encoding rate adaptation of each GOP and the transmission adaptation of each slot to maximize the total utility in each case. We separated the challenging optimization problem into multiple tractable problems in each scenario. We obtained a globally optimal solution of each problem in the single-user scenario and a KKT point of each problem in the multi-user scenario. Finally, we evaluated the quality, quality variation, and rebuffering time of the proposed solutions. Numerical results demonstrated notable gains of the proposed solutions over existing schemes and revealed the impact of FoV prediction on adaptive streaming of tiled 360 videos.

\section*{Appendix A: Proof of Theorem \ref{lemma_decouple}}\label{app_decouple}
First, we obtain an equivalent problem of Problem \ref{single_case_general}. By introducing an auxiliary variable and an extra constraint, and by contradiction, Problem \ref{single_case_general} can be equivalently transformed to:
\begin{align}
\max_{\mathbf{R},\mathbf{r}}\quad &Q^{(\phi)}(\mathbf{r})\label{prove1_p3}\\
    \mathrm{s.t.}\quad&\eqref{rate_smooth},~\eqref{single_tile_relax},~\eqref{single_fov_relax},\nonumber\\
    &\sum\nolimits_{(x,y)\in\overline{\mathcal{F}}}R_{x,y} \leq C^{\ddag}(1),\nonumber
\end{align}
where
\begin{align}
C^{\ddag}(1) \triangleq \max_{\mathbf{w}(1)}\quad &\sum\limits_{n\in\mathcal{N}}B\log_{2}\left(1+\frac{|\mathbf{h}^{H}_{n}(1)\mathbf{w}_{n}(1)|^{2}}{\sigma^{2}}\right)\label{prove1_p4}\\
\mathrm{s.t.}\quad &\eqref{single_power_allocation_constraint}\nonumber.
\end{align}
Therefore, $(\mathbf{R}^{(\phi)\dag},\mathbf{r}^{(\phi)\dag})$ and $\mathbf{w}^{\star}(1)$ are the optimal solutions of the problem in \eqref{prove1_p3} and the problem in \eqref{prove1_p4}, respectively.
Next, we show that the problem in \eqref{water} is equivalent to the problem in \eqref{prove1_p4}. By the \textit{Cauchy-Schwartz inequality}, we have $|\mathbf{h}^{H}_{n}(1)\mathbf{w}_{n}(1)|^{2} \leq \|\mathbf{h}_{n}(1)\|^{2}_{2}\|\mathbf{w}_{n}(1)\|^{2}_{2}$, where the equality holds if and only if $\frac{\mathbf{h}_{n}(1)}{\|\mathbf{h}_{n}(1)\|_{2}} = \frac{\mathbf{w}_{n}(1)}{\|\mathbf{w}_{n}(1)\|_{2}},n\in\mathcal{N}.$ Thus, without loss of optimality, the problem in \eqref{prove1_p4} can be equivalently transformed into the problem in (\ref{water}), $C^{\ddag}(1) = C^{\dag}(1)$, and $\mathbf{w}^{\star}_{n}(1) = \frac{\mathbf{h}_{n}(1)}{\|\mathbf{h}_{n}(1)\|_{2}}\sqrt{v^{\dag}_{n}(1)},~n\in\mathcal{N}$.
Finally, we show that $\mathbf{R}^{(\phi)\star} = \mathbf{R}^{(\phi)\dag}$ and $\mathbf{r}^{(\phi)\star} = \mathbf{r}^{(\phi)\dag}$. Since $C^{\ddag}(1) = C^{\dag}(1)$, the problem in \eqref{prove1_p3} is equivalent to Problem \ref{single_case_general_equal}. Thus, $\mathbf{R}^{(\phi)\star} = \mathbf{R}^{(\phi)\dag}$, and $\mathbf{r}^{(\phi)\star} = \mathbf{r}^{(\phi)\dag}$.

\section*{Appendix B: Proof of Theorem \ref{lemma_single_case2}}\label{app_dual}
The inner problem of Problem \ref{single_case_general} with $\phi$ = ip is a linear program (LP) with respect to $\mathbf{p}$ for any given $\mathbf{r}$, satisfying \eqref{rate_smooth},~\eqref{single_fov_relax}. Strong duality holds for this LP and its dual problem can be readily obtained\cite[pp. 225]{boyd2004convex}. Thus, we can show that the max-min problem in Problem \ref{single_case_general} with $\phi$ = ip can be equivalently transformed to Problem \ref{single_case_two_equal}, by replacing the inner problem with its dual problem.

\section*{Appendix C: Proof of Theorem \ref{single_theorem}}\label{app_theo}
\subsection{Proof of Statement (\romannumeral1) of Theorem \ref{single_theorem}}
We rewrite the optimal value of Problem \ref{single_case_general} with transmission power budget $P$ as $U^{(\phi)\star}(P)$. By contradiction, we can easily show that the optimal value of Problem \ref{single_case_general_equal} strictly increases with $C^{\dag}(1)$, and $C^{\dag}(1)$ strictly increases with $P$. Thus, by Theorem 1, for all $P > P'$, we have: 
\begin{equation}
U^{(\phi)\star}(P) > U^{(\phi)\star}(P').\label{prove31_c1}
\end{equation}
According to the first inequality in \eqref{rate_smooth}, we have $R^{(\phi)\star}_{x,y} \geq \max_{i\in\mathcal{I}: (x,y)\in\mathcal{F}_{i}}r^{(\phi)\star}_{i},~(x,y)\in\overline{\mathcal{F}}.$ Suppose that there exists $(x^{(\phi)'},y^{(\phi)'})\in\overline{\mathcal{F}}$ such that $R^{(\phi)\star}_{x^{(\phi)'},y^{(\phi)'}} > \max_{i\in\mathcal{I}: (x^{(\phi)'},y^{(\phi)'})\in\mathcal{F}_{i}}r^{(\phi)\star}_{i}$. We construct a feasible solution $(\mathbf{R}^{(\phi)'},\mathbf{r}^{(\phi)\star},\mathbf{w}^{'}(1))$ of Problem \ref{single_case_general}. Specifically, $R^{(\phi)'}_{x^{(\phi)'},y^{(\phi)'}} = \max_{i\in\mathcal{I}: (x^{(\phi)'},y^{(\phi)'})\in\mathcal{F}_{i}}r^{(\phi)\star}_{i}, R^{(\phi)'}_{x,y} = R^{(\phi)\star}_{x,y},(x,y)\in\overline{\mathcal{F}}\backslash\{(x^{(\phi)'},y^{(\phi)'})\}$, $\mathbf{w}^{'}_{n}(1) = \alpha \mathbf{w}^{\star}_{n}(1),n\in\mathcal{N}$, where $\alpha \in (0,1)$ satisfies $R^{(\phi)\star}_{x^{(\phi)'},y^{(\phi)'}} - R^{(\phi)'}_{x^{(\phi)'},y^{(\phi)'}} = \sum_{n\in\mathcal{N}}B\log_{2}\left(1 + \frac{|\mathbf{h}^{H}_{n}(1)\mathbf{w}^{\star}_{n}(1)|^{2}}{\sigma^{2}}\right) - \sum_{n\in\mathcal{N}}B\log_{2}\left(1 + \frac{\alpha^{2}|\mathbf{h}^{H}_{n}(1)\mathbf{w}^{\star}_{n}(1)|^{2}}{\sigma^{2}}\right)$. By the construction, $\sum_{(x,y)\in\overline{\mathcal{F}}} R^{(\phi)'}_{x,y} = \sum\nolimits_{n\in\mathcal{N}}B\log_{2}\left(1 + \frac{\alpha^{2}|\mathbf{h}^{H}_{n}(1)\mathbf{w}^{\star}_{n}(1)|^{2}}{\sigma^{2}}\right)$, implying that $(\mathbf{R}^{(\phi)'},\mathbf{r}^{(\phi)\star},\mathbf{w}^{'}(1))$ satisfies \eqref{single_successful_transmit}. In addition, it is obvious that $(\mathbf{R}^{(\phi)'},\mathbf{r}^{(\phi)\star},\mathbf{w}^{'}(1))$ satisfies the constraints in \eqref{rate_smooth}, \eqref{single_power_allocation_constraint}, \eqref{single_tile_relax}, \eqref{single_fov_relax}. Thus, $(\mathbf{R}^{(\phi)'},\mathbf{r}^{(\phi)\star},\mathbf{w}^{'}(1))$ is a feasible solution of Problem \ref{single_case_general} with transmission power budget $P$, and achieves $U^{(\phi)\star}(P)$. Let $P^{'} \triangleq \sum_{n\in\mathcal{N}}\|\mathbf{w}^{'}_{n}(1)\|^{2}_{2} = \alpha^{2} P < P$. It is clear that $(\mathbf{R}^{(\phi)'},\mathbf{r}^{(\phi)\star},\mathbf{w}^{'}(1))$ is also a feasible solution of Problem \ref{single_case_general} with transmission power budget $P'$. Thus, we have $U^{(\phi)\star}(P') \geq U^{(\phi)\star}(P)$, which contradicts with \eqref{prove31_c1}. Thus, by contradiction, we can show Statement (\romannumeral1).

\subsection{Proof of Statement (\romannumeral2) of Theorem \ref{single_theorem}}
For $\phi$ = pp and ip, we construct $(\mathbf{R}^{(\phi)'},\mathbf{r}^{(\phi)'},\mathbf{w}^{\star}(1))$. Specifically, let
\begin{align}
\Psi^{(\phi)} \triangleq \left\{ \begin{array}{ll}
 \min\left\{\frac{|\mathcal{T}_{m}| - 1}{|\mathcal{T}_{m}| - |\mathcal{T}_{j}| - 1}\delta, (|\mathcal{T}_{j}| - 1)\delta, \frac{r^{(\phi)\star}_{m} - r^{(\phi)\star}_{j}}{2}\right\},\\
 ~~~~~~~~~~~~~~~~~~~~~~~~~~~~~~~~~~|\mathcal{T}_{m}| > |\mathcal{T}_{j}| + 1,\\
 \min\left\{(|\mathcal{T}_{j}| - 1)\delta, \frac{r^{(\phi)\star}_{m} - r^{(\phi)\star}_{j}}{2}\right\},\\
 ~~~~~~~~~~~~~~~~~~~~~~~~~~~~~~~~~~|\mathcal{T}_{m}| = |\mathcal{T}_{j}| + 1.
  \end{array} \right.\nonumber
\end{align}
Set 
\begin{align}
&r^{(\phi)'}_{h} = r^{(\phi)\star}_{h},~h\in\mathcal{I}\backslash\{m,j\},\label{t3s2_0}\\
&r^{(\phi)'}_{m} = r^{(\phi)\star}_{m} - \Delta,\quad r^{(\phi)'}_{j} = r^{(\phi)\star}_{j} +  \Delta, \label{t3s2_1}\\
&R^{(\phi)'}_{x,y} = R^{(\phi)\star}_{x,y},~(x,y)\in\overline{\mathcal{F}}\backslash(\mathcal{T}_{m}\cup\mathcal{T}_{j}). \label{t3s2_2}
\end{align}
Choose any $(a,b)\in\mathcal{T}_{m}$ and any $(c,d)\in\mathcal{T}_{j}$, and set 
\begin{align}
&R^{(\phi)'}_{a,b} = r^{(\phi)\star}_{m} -  \Delta, R^{(\phi)'}_{x,y} = r^{(\phi)\star}_{m} - \frac{|\mathcal{T}_{j}|}{|\mathcal{T}_{m}| - 1} \Delta,\nonumber\\
&~~~~~~~~~~~~~~~~~~~~~~~~~~~~~~~~~~~~(x,y)\in\mathcal{T}_{m}\backslash\{(a,b)\}\label{t3s2_3},\\
&R^{(\phi)'}_{c,d} = r^{(\phi)\star}_{j} +  \Delta, R^{(\phi)'}_{x,y} = r^{(\phi)\star}_{j} + \frac{|\mathcal{T}_{j}|}{|\mathcal{T}_{j}| - 1} \Delta,\nonumber\\
&~~~~~~~~~~~~~~~~~~~~~~~~~~~~~~~~~~~~~(x,y)\in\mathcal{T}_{j}\backslash\{(c,d)\},\label{t3s2_4}
\end{align}
where $\Delta > 0$. Thus, $\sum_{(x,y)\in\overline{\mathcal{F}}}R^{(\phi)'}_{x,y} - \sum_{(x,y)\in\overline{\mathcal{F}}}R^{(\phi)\star}_{x,y} \overset{(a)}= \sum_{(x,y)\in\mathcal{T}_{j}}\left(R^{(\phi)'}_{x,y} - R^{(\phi)\star}_{x,y}\right) + \sum_{(x,y)\in\mathcal{T}_{m}}\left(R^{(\phi)'}_{x,y} - R^{(\phi)\star}_{x,y}\right)  \overset{(b)} = (|\mathcal{T}_{j}| + 1) \Delta - (|\mathcal{T}_{j}| + 1) \Delta  = 0$, where $(a)$ is due to \eqref{t3s2_2}, and $(b)$ is due to \eqref{t3s2_3} and \eqref{t3s2_4}, $R^{(\phi)\star}_{x,y} = \max_{i\in\mathcal{I}:(x,y)\in\mathcal{F}_{i}}r^{(\phi)\star}_{i} = r^{(\phi)\star}_{m},(x,y)\in\mathcal{T}_{m}$, and $R^{(\phi)\star}_{x,y} = \max_{i\in\mathcal{I}:(x,y)\in\mathcal{F}_{i}}r^{(\phi)\star}_{i} = r^{(\phi)\star}_{j},(x,y)\in\mathcal{T}_{j}$. As $\sum_{(x,y)\in\overline{\mathcal{F}}}R^{(\phi)\star}_{x,y} = C^{\dag}(1)$, we have $\sum_{(x,y)\in\overline{\mathcal{F}}}R^{(\phi)'}_{x,y} = C^{\dag}(1)$, i.e., $(\mathbf{R}^{(\phi)'},\mathbf{r}^{(\phi)'},\mathbf{w}^{\star}(1))$ satisfies \eqref{single_successful_transmit}. It is also obvious that $(\mathbf{R}^{(\phi)'},\mathbf{r}^{(\phi)'},\mathbf{w}^{\star}(1))$ satisfies the constraint in \eqref{single_power_allocation_constraint}.

For $\phi$ = pp, suppose that there exist $m,j\in\mathcal{I},m\not=j$ such that $p_{m} \leq p_{j},|\mathcal{T}_{m}| > |\mathcal{T}_{j}| > 1$ and $ r^{(\text{pp})\star}_{m} > r^{(\text{pp})\star}_{j}$. As  $r^{(\text{pp})\star}_{m} > r^{(\text{pp})\star}_{j}$ and $|\mathcal{T}_{j}| > 1$, we have $\Psi^{(\text{pp})} > 0$. Choose $\Delta \in (0, \Psi^{(\text{pp})})$. It is obvious that $(\mathbf{R}^{(\text{pp})'},\mathbf{r}^{(\text{pp})'},\mathbf{w}^{\star}(1))$ satisfies the constraints in \eqref{rate_smooth}, \eqref{single_tile_relax}, \eqref{single_fov_relax}. Thus, $(\mathbf{R}^{(\text{pp})'},\mathbf{r}^{(\text{pp})'},\mathbf{w}^{\star}(1))$ with $\Delta \in (0, \Psi^{(\text{pp})})$ is a feasible solution of Problem \ref{single_case_general} with $\phi$ = pp. In addition, we have: 
\begin{align}
&\sum_{i\in\mathcal{I}}p_{i}U(r^{(\text{pp})'}_{i}) - \sum_{i\in\mathcal{I}}p_{i}U(r^{(\text{pp})\star}_{i})\nonumber\\ 
&\overset{(c)}{=} p_{m}U(r^{(\text{pp})'}_{m}) + p_{j}U(r^{(\text{pp})'}_{j}) - p_{m}U(r^{(\text{pp})\star}_{m}) - p_{j}U(r^{(\text{pp})\star}_{j}) \nonumber\\
&\overset{(d)}\geq p_{m}\left(U(r^{(\text{pp})'}_{m}) + U(r^{(\text{pp})'}_{j}) - U(r^{(\text{pp})\star}_{m}) - U(r^{(\text{pp})\star}_{j})\right),\label{t3s2_5}
\end{align}
where $(c)$ is due to \eqref{t3s2_0}, $(d)$ is due to $\frac{\mathrm{d}U(x)}{\mathrm{d}x} > 0$ and $p_{j}-p_{m} \geq 0$. As 
$\frac{\mathrm{d}^{2}U(x)}{\mathrm{d}^{2}x} < 0$, $\frac{\mathrm{d}U(x)}{\mathrm{d}x}$ is strictly decreasing. Thus, $\frac{\mathrm{d}U(x)}{\mathrm{d}x} < \frac{\mathrm{d}U(A - x)}{\mathrm{d}x}$ for all $x\in[\frac{A}{2}, A].$ Thus, $U(x) + U(A-x)$ is a strictly increasing function of $x$ when $x\in[\frac{A}{2}, A]$. By \eqref{t3s2_1}, we have $r^{(\text{pp})'}_{m} + r^{(\text{pp})'}_{j} = r^{(\text{pp})\star}_{m} + r^{(\text{pp})\star}_{j}$. As $r^{(\text{pp})'}_{m} > r^{(\text{pp})'}_{j} > 0$, we have $r^{(\text{pp})'}_{m}\in\left(\frac{r^{(\text{pp})\star}_{m} + r^{(\text{pp})\star}_{j}}{2}, r^{(\text{pp})\star}_{m} + r^{(\text{pp})\star}_{j}\right)$. Thus, we can show:
\begin{equation}
U(r^{(\text{pp})'}_{m}) + U(r^{(\text{pp})'}_{j}) - U(r^{(\text{pp})\star}_{m}) - U(r^{(\text{pp})\star}_{j}) > 0. \label{t3s2_6}
\end{equation} 
By \eqref{t3s2_5} and \eqref{t3s2_6}, we have $\sum_{i\in\mathcal{I}}p_{i}U(r^{(\text{pp})'}_{i}) - \sum_{i\in\mathcal{I}}p_{i}U(r^{(\text{pp})\star}_{i}) > 0$, which contradicts with the optimality of $(\mathbf{R}^{(\text{pp})\star},\mathbf{r}^{(\text{pp})\star},\mathbf{w}^{\star}(1))$. Therefore, by contradiction, we can show Statement (\romannumeral2) for $\phi$ = pp.

For $\phi$ = ip, suppose that there exist $m,j\in\mathcal{I},m\not=j$ such that $\overline{p}_{m} \leq \underline{p}_{j}$, $|\mathcal{T}_{m}| > |\mathcal{T}_{j}| > 1$, $r^{(\text{ip})\star}_{m} > r^{(\text{ip})\star}_{j}$. As $r^{(\text{ip})\star}_{m} > r^{(\text{ip})\star}_{j}$ and $|\mathcal{T}_{j}| > 1$, we have $\Psi^{(\text{ip})} > 0$. Choose $\Delta \in (0, \Psi^{(\text{ip})})$. It is obvious that $(\mathbf{R}^{(\text{ip})'},\mathbf{r}^{(\text{ip})'},\mathbf{w}^{(\text{ip})\star}(1))$ satisfies the constraints in \eqref{rate_smooth}, \eqref{single_tile_relax}, \eqref{single_fov_relax}. Thus, $(\mathbf{R}^{(\text{ip})'},\mathbf{r}^{(\text{ip})'},\mathbf{w}^{\star}(1))$ with $\Delta \in (0, \Psi^{(\text{ip})})$ is a feasible solution of Problem \ref{single_case_general} with $\phi$ = ip. In addition, we have:
\begin{align}
&\min_{\mathbf{p}\in\mathcal{P}}\sum_{i\in\mathcal{I}}p_{i}U(r^{(\text{ip})'}_{i}) - \min_{\mathbf{p}\in\mathcal{P}}\sum_{i\in\mathcal{I}}p_{i}U(r^{(\text{ip})\star}_{i}) \nonumber\\
& \overset{(e)}\geq \sum_{i\in\mathcal{I}}p'_{i}U(r^{(\text{ip})'}_{i}) - \sum_{i\in\mathcal{I}}p'_{i}U(r^{(\text{ip})\star}_{i})\nonumber\\
&\overset{(f)}= p'_{m}U(r^{(\text{ip})'}_{m}) + p'_{j}U(r^{(\text{ip})'}_{j}) - p'_{m}U(r^{(\text{ip})\star}_{m}) - p'_{j}U(r^{(\text{ip})\star}_{j})\nonumber\\
& \overset{(g)}\geq p'_{m}\left(U(r^{(\text{ip})'}_{m}) + U(r^{(\text{ip})'}_{j}) - U(r^{(\text{ip})\star}_{m}) - U(r^{(\text{ip})\star}_{j})\right),\nonumber 
\end{align}
where $\mathbf{p}'$ denotes the optimal solution of $\min_{\mathbf{p}\in\mathcal{P}}\sum_{i\in\mathcal{I}}p_{i}U(r^{(\text{ip})'}_{i})$, $(e)$ is due to the fact that $\mathbf{p}'$ is a feasible solution of $\min_{\mathbf{p}\in\mathcal{P}}\sum_{i\in\mathcal{I}}p_{i}U(r^{(\text{ip})\star}_{i})$, $(f)$ is due to \eqref{t3s2_0}, $(g)$ is due to $\frac{\mathrm{d}U(x)}{\mathrm{d}x} > 0$ and $p'_{j}-p'_{m} \geq 0$. Following the proof for $U(r^{(\text{pp})'}_{m}) + U(r^{(\text{pp})'}_{j}) - U(r^{(\text{pp})\star}_{m}) - U(r^{(\text{pp})\star}_{j}) > 0$, we can show $U(r^{(\text{ip})'}_{m}) + U(r^{(\text{ip})'}_{j}) - U(r^{(\text{ip})\star}_{m}) - U(r^{(\text{ip})\star}_{j}) > 0$. Thus, $\min_{\mathbf{p}\in\mathcal{P}}\sum_{i\in\mathcal{I}}p_{i}U(r^{(\text{ip})'}_{i}) - \min_{\mathbf{p}\in\mathcal{P}}\sum_{i\in\mathcal{I}}p_{i}U(r^{(\text{ip})\star}_{i}) > 0$, which contradicts with the optimality of $(\mathbf{R}^{(\text{ip})\star},\mathbf{r}^{(\text{ip})\star},\mathbf{w}^{\star}(1))$. Thus, by contradiction, we can show Statement (\romannumeral2) for $\phi$ = ip.
  
For $\phi$ = up, suppose that $\min_{i\in\mathcal{I}}U(r^{(\text{up})\star}_{i}) < \max_{i\in\mathcal{I}}U(r^{(\text{up})\star}_{i})$. Let $\mathcal{I}_{\max} \triangleq \{h\in\mathcal{I}: U(r^{(\text{up})\star}_{h}) = \max_{i\in\mathcal{I}}U(r^{(\text{up})\star}_{i})\}$ and $\mathcal{I}_{\min} \triangleq \{h\in\mathcal{I}: U(r^{(\text{up})\star}_{h}) = \min_{i\in\mathcal{I}}U(r^{(\text{up})\star}_{i})\}$. We construct a feasible solution $(\mathbf{R}^{(\text{up})'},\mathbf{r}^{(\text{up})'},\mathbf{w}^{\star}(1))$ of Problem \ref{single_case_general} with $\phi$ = up. Let $\mathcal{S}_{\max} \triangleq \cup_{i\in\mathcal{I}_{\max}}\mathcal{T}_{i}$ and $\mathcal{S}_{\min} \triangleq \cup_{i\in\mathcal{I}_{\min}}\mathcal{T}_{i}$. Set 
\begin{align}
&r^{(\text{up})'}_{i} = r^{(\text{up})\star}_{i} - \Delta, i\in\mathcal{I}_{\max}, \label{t3s2_700}\\
&r^{(\text{up})'}_{i} = r^{(\text{up})\star}_{i} + \frac{|\mathcal{S}_{\max}|}{|\mathcal{S}_{\min}|}\Delta,i\in\mathcal{I}_{\min},\label{t3s2_70}\\
&r^{(\text{up})'}_{h} = r^{(\text{up})\star}_{h},h\in\mathcal{I}\backslash(\mathcal{I}_{\min}\cup\mathcal{I}_{\max}), \label{t3s2_7}
\end{align}
and set $R^{(\text{up})'}_{x,y} = \max_{i\in\mathcal{I}}r^{(\text{up})\star}_{i} - \Delta,(x,y)\in\mathcal{S}_{\max}, R^{(\text{up})'}_{x,y} =  \min_{i\in\mathcal{I}}r^{(\text{up})\star}_{i} + \frac{|\mathcal{S}_{\max}|}{|\mathcal{S}_{\min}|}\Delta,(x,y)\in\mathcal{S}_{\min}, R^{(\text{up})'}_{x,y} = R^{(\text{up})\star}_{x,y},(x,y)\in\overline{\mathcal{F}}\backslash(\mathcal{S}_{\max}\cup\mathcal{S}_{\min})$, where $\Delta \in \left(0, \frac{|\mathcal{S}_{\max}|}{|\mathcal{S}_{\max}| + |\mathcal{S}_{\min}|}(\max_{i\in\mathcal{I}}r^{(\text{up})\star}_{i} - \min_{i\in\mathcal{I}}r^{(\text{up})\star}_{i})\right)$. Similarly, by the construction, we have $\sum_{(x,y)\in\overline{\mathcal{F}}}R^{(\text{up})'}_{x,y} - \sum_{(x,y)\in\overline{\mathcal{F}}}R^{(\text{up})\star}_{x,y} = 0$, implying that $(\mathbf{R}^{(\text{up})'},\mathbf{r}^{(\text{up})'},\mathbf{w}^{\star}(1))$ satisfies \eqref{single_successful_transmit}. It is obvious that $(\mathbf{R}^{(\text{up})'},\mathbf{r}^{(\text{up})'},\mathbf{w}^{\star}(1))$ satisfies the constraint in \eqref{single_power_allocation_constraint}. As $\Delta \in \left(0, \frac{|\mathcal{S}_{\max}|}{|\mathcal{S}_{\max}| + |\mathcal{S}_{\min}|}(\max_{i\in\mathcal{I}}r^{(\text{up})\star}_{i} - \min_{i\in\mathcal{I}}r^{(\text{up})\star}_{i})\right)$, it is also obvious that $(\mathbf{R}^{(\text{up})'},\mathbf{r}^{(\text{up})'},\mathbf{w}^{\star}(1))$ satisfies the constraints in \eqref{rate_smooth}, \eqref{single_tile_relax}, \eqref{single_fov_relax}. Thus, $(\mathbf{R}^{(\text{up})'},\mathbf{r}^{(\text{up})'},\mathbf{w}^{\star}(1))$ is a feasible solution of Problem \ref{single_case_general} with $\phi$ = up. In addition, by \eqref{t3s2_700}, \eqref{t3s2_70}, \eqref{t3s2_7} and $\frac{\mathrm{d}U(x)}{\mathrm{d}x} > 0$, we have $\min_{i\in\mathcal{I}}U(r^{(\text{up})\star}_{i}) < \min_{i\in\mathcal{I}}U(r^{(\text{up})'}_{i})$, which contradicts with the optimality of $(\mathbf{R}^{(\text{up})\star},\mathbf{r}^{(\text{up})\star},\mathbf{w}^{\star}(1))$. Thus, by contradiction, we show Statement (\romannumeral2) for $\phi$ = up.
\subsection{Proof of Statement (\romannumeral3) of Theorem \ref{single_theorem}}
Note that the feasible sets of Problem \ref{single_case_general} with $\phi$ = pp, ip and up are identical. For any $\mathbf{p}$ satisfying $p_{i} \geq 0, i\in\mathcal{I}$, $\sum_{i\in\mathcal{I}}p_{i} = 1$ and for all $\mathbf{r} \succeq 0$, we have:
\begin{equation}
\sum\nolimits_{i\in\mathcal{I}}p_{i}U(r_{i}) \geq \min\nolimits_{i\in\mathcal{I}}U(r_{i}).\label{prove3_c1}
\end{equation}
For any $\hat{\mathbf{p}}$ satisfying $\hat{p}_{i} \geq 0, i\in\mathcal{I}$, $\sum_{i\in\mathcal{I}}\hat{p}_{i} = 1$ and for all $\mathbf{r} \succeq 0$, we have:
\begin{equation}
\sum\nolimits_{i\in\mathcal{I}}p_{i}U(r_{i}) \geq \min\nolimits_{\mathbf{p}\in\mathcal{P}}\sum\nolimits_{i\in\mathcal{I}}p_{i}U(r_{i}),~\mathbf{p} \in\mathcal{P}.\label{prove3_c2}
\end{equation}
First, we have $U^{(\text{pp})\star} = \sum_{i\in\mathcal{I}}p_{i}U(r^{(\text{pp})\star}_{i}) \overset{(a)}{\geq} \sum_{i\in\mathcal{I}}p_{i}U(r^{(\text{up})\star}_{i}) \overset{(b)}{\geq} \min_{i\in\mathcal{I}}U(r^{(\text{up})\star}_{i}) = U^{(\text{up})\star}$, where ($a$) is due to the fact that $r^{(\text{pp})\star}_{i}$ is the optimal solution of Problem \ref{single_case_general} with $\phi$ = pp, and ($b$) is due to \eqref{prove3_c1}. Next, we have $U^{(\text{ip})\star} = \sum_{i\in\mathcal{I}}p^{\star}_{i}U(r^{(\text{ip})\star}_{i}) \overset{(c)}{\geq} \sum_{i\in\mathcal{I}}p^{\star}_{i}U(r^{(\text{up})\star}_{i}) \overset{(d)}{\geq} \min_{i\in\mathcal{I}}U(r^{(\text{up})\star}_{i}) = U^{(\text{up})\star},$ where ($c$) is due to the fact that $r^{(\text{ip})\star}_{i}$ is the optimal solution of Problem \ref{single_case_general} with $\phi$ = ip, and ($d$) is due to \eqref{prove3_c1}. Finally, we have $U^{(\text{pp})\star} = \sum_{i\in\mathcal{I}}p_{i}U(r^{(\text{pp})\star}_{i}) \overset{(e)}{\geq} \sum_{i\in\mathcal{I}}p_{i}U(r^{(\text{ip})\star}_{i}) \overset{(f)}{\geq} \min_{\mathbf{p}\in\mathcal{P}}\sum_{i\in\mathcal{I}}p_{i}U(r^{(\text{ip})\star}_{i}) =  U^{(\text{ip})\star}, \mathbf{p}\in\mathcal{P}$, where ($e$) is due to the fact that $r^{(\text{pp})\star}_{i}$ is the optimal solution of Problem \ref{single_case_general} with $\phi$ = pp, and ($f$) is due to \eqref{prove3_c2}. Therefore, we can show Statement (\romannumeral3).

\end{document}